\documentclass[12pt]{article}
\usepackage{graphicx,psfrag,epsf}
\usepackage{url} 

\usepackage[english]{babel}
\usepackage{amsmath,amssymb,enumerate,rotate,multirow}
\usepackage{bm,latexsym,mathrsfs}
\usepackage[usenames]{color}
\usepackage{subfig}
\usepackage[authoryear]{natbib}
\usepackage{booktabs}
\usepackage{colortbl}
\usepackage{xcolor}
\usepackage{xfrac}
\usepackage{xspace}
\usepackage{amsthm}
\usepackage{bbm}
\usepackage{verbatim}
\usepackage{multirow}

\usepackage{sidecap} 
\sidecaptionvpos{figure}{c}

\newcommand{\calN}{\mathcal{N}}

\newcommand{\be}{\begin{equation}}
\newcommand{\ee}{\end{equation}}
\newcommand{\iid}{\stackrel{\mathrm{iid}}{\sim}}
\newcommand{\ind}{\stackrel{\mathrm{ind}}{\sim}}

\newcommand{\e}{\mathrm{e}}
\newcommand{\rt}{\rightarrow}

\newcommand{\Rea}{{\mathbb{R}}}

\newcommand{\Rd}{\mathbb{R}{\scriptscriptstyle{^{^d}}} }
\newcommand{\Z}{{\mathbb{Z}}}
\newcommand{\intS}{\int_S}

\newcommand{\rmd}{\mathrm{d}}
\newtheorem*{remark}{Remark}
\newtheorem*{definition}{Definition}
\DeclareMathOperator{\Var}{Var}

\DeclareMathOperator{\dir}{Dirichlet}

\newcommand{\Prob}{\mathbb{P}}
\newcommand{\E}{\mathbb{E}}

\newcommand{\virgolette}[1]{``#1"}

\newcommand{\blind}{0}

\begin{document}

\def\spacingset#1{\renewcommand{\baselinestretch}%
{#1}\small\normalsize} \spacingset{1}


\if0\blind
{
  \title{\bf Determinantal point process mixtures\\
 via spectral density approach}
  \author{Ilaria Bianchini,
    Alessandra Guglielmi \thanks{
    The authors gratefully acknowledge grant FONDECYT 1141057; the first two authors
thank people from the Departamento de Estad\'{\i}stica at PUC, Chile, for
their kind hospitality.   }\\
    Politecnico di Milano (Italy)\\
    and \\
  Fernando A. Quintana \\
   Pontificia Universidad Cat\'olica de Chile (Chile)\\    }
  \maketitle
} \fi

\if1\blind
{
  \bigskip
  \bigskip
  \bigskip
  \begin{center}
    {\LARGE\bf Title}
\end{center}
  \medskip
} \fi

\bigskip
\begin{abstract}
We consider mixture models where location parameters are a priori
encouraged to be well separated. We explore a class of determinantal point
process (DPP) mixture models, which provide the desired notion of
separation or repulsion.  Instead of using the rather restrictive case
where analytical results are available, we adopt a spectral representation
from which approximations to the DPP intensity functions can be readily
computed. For the sake of concreteness the presentation focuses on a power
exponential spectral density, but the proposed approach is in fact quite
general. We later extend our model to incorporate covariate information in
the likelihood and also in the assignment to mixture components, yielding
a trade-off between repulsiveness of locations in the mixtures and
attraction among subjects with similar covariates. We develop full
Bayesian inference, and explore model properties and posterior behavior
using several simulation scenarios and data illustrations. Supplementary
material for this article can be found at the end of this document.
\end{abstract}

\noindent%
  \vfill

\newpage
\spacingset{1.45} 
\section{Introduction}

Mixture models are an extremely popular class of models, that have been
successfully used in many applications. For a review, see, e.g.
\cite{fruhwirth:mixt}. Such models are typically stated as
\begin{equation}\label{eq:mixture}
y_i\mid k,\bm{\theta},\bm{\pi}\ind \sum_{j=1}^k \pi_j f(y_i\mid\theta_j),\quad i=1,\ldots,n,
\end{equation}
where $\bm{\pi}=(\pi_1,\ldots,\pi_k)$ are constrained to be nonnegative
and sum up to 1, $\bm{\theta}=(\theta_1,\ldots,\theta_k)$, and $1\le k\le
\infty$, with $k=\infty$ corresponding to a nonparametric model. A common
prior assumption is that $\bm{\pi}\sim\dir(\delta_1,\ldots,\delta_k)$ and
that the components of $\bm{\theta}$ are drawn i.i.d. from some suitable
prior $p_0$. However, the weights $\bm{\pi}$ may be constructed
differently, e.g. using a stick-breaking representation (finite or
infinite), which poses a well-known connection with more general models,
including nonparametric ones. See, e.g., \cite{ishwaran&james:01} and
\cite{MilHar2017}. A popular class of Bayesian nonparametric models is the
Dirichlet process mixture (DPM) model, introduced in \cite{Ferguson1983}
and \cite{Lo1984}. It is well-known that this class of mixtures usually
overestimates the number of clusters, mainly because of the
\virgolette{rich gets richer} property of the Dirichlet process. By this
we mean that both prior and posterior distributions are concentrated on a
relatively large number of clusters, but a few are very large, and the
rest of them have very small sample sizes. Mixture models may even
be inconsistent; see \cite{RouMen2011}, where concerns about over-fitted
mixtures are illustrated, and \cite{MilHar2013}, for inconsistency
features of DPMs.

Despite their success, mixture models like~\eqref{eq:mixture} tend to use
excessively many mixture components. As pointed out in \cite{Xu_etal2016},
this is due to the fact that the component-specific parameters are a
priori i.i.d., and therefore, free to move. This motivated
\cite{PetraliRaoDunson:12}, \cite{FuqueSteelRoss:2016} and
\cite{QuQuPa:17} to explicitly define joint distributions for
$\bm{\theta}$ having the property of {\em repulsion} among its components,
i.e. that $p(\theta_1,\ldots,\theta_k)$ puts higher mass on configurations
such that components are well separated. For a different approach, via
sparsity in the prior, see \cite{malsiner_etal2016}.

\cite{Xu_etal2016} explored a similar way to accomplish separation of
mixture components, by means of a Determinantal Point Process (DPP) acting
on the parameter space. DPPs have recently received increased attention in
the statistical literature~\citep{Lav_etal2015}. DPPs are point processes
having a product density function expressed as the determinant of a
certain matrix constructed using a covariance function evaluated at the
pairwise distances among points, in such a way that higher mass is
assigned to configurations of well-separated points. We give details
below. DPPs have been used in a number of different modeling efforts.
\cite{bardenet2015inference} and \cite{affandi2014learning} applied DPPs
to model spatial patterns of nerve fibers in diabetic patients, a basic
motivation being that such fibers become more clustered as diabetes
progresses. The latter discussed also applications to image search,
showing how such processes could be used to study human perception of
diversity in different image categories. Similarly,
\cite{kulesza2012determinantal} show how DPPs can be applied to various
problems such as finding diverse sets of high-quality search results,
building informative summaries by selecting diverse sentences from
documents, modeling non-overlapping human poses in images or video, and
automatically building timelines of important news stories.

We discuss full Bayesian inference for a class of mixture densities where
the locations follow stationary DPPs. The approach is based on the
spectral representation of the covariance function defining the
determinant as the joint distribution of component-specific parameters.
While our methods can be used with any such valid spectral representation,
we focus for the sake of concreteness on the power exponential spectral
case; see examples with different spectral densities in the on-line
Supplementary Material. This particular specification allows for flexible
repulsion patterns, and we discuss how to set up different types of prior
behavior, shedding light on the practical use of our approach in that
particular scenario. We discuss applications in the context of synthetic
and real data applications. We later extend our model to incorporate
covariate information in the likelihood and also in the assignment to
mixture components. In particular, subjects with similar covariates are a
priori more likely to co-cluster, just as in mixtures of experts
models~\citep[see, e.g.,][]{McLachlan:2005}, where weights are defined as
normalized exponential functions.

We implement a reversible jump (RJ) MCMC posterior simulation scheme for
each of the models we propose. In all cases, we estimate clusters for the
subjects in the sample, by considering the partition that minimizes the
posterior expectation of Binder's loss function \citep{binder1978bayesian}
under equal misclassification costs. This is a common choice in the
applied Bayesian nonparametric literature \citep{lau2007bayesian}.

The first mixture model we propose is as in \cite{Xu_etal2016}. They
consider, as a prior for the location points, a particular case of DPPs,
called L-ensambles. These are defined as a direct generalization from the
same definition but for the finite state space case. Our proposal does not
use L-ensembles. Instead, we stick to the more general case, following the
spectral approach discussed in \cite{Lav_etal2015}. Although we limit
ourselves to the case of isotropic DPPs, inhomogeneous DPPs can be
obtained by transforming or thinning a stationary process; see
\cite{Lav_etal2015}, Section~3 and Appendix~A. A crucial point in our
models and algorithms is the DPP density expression, which is only defined
for DPPs restricted to compact subsets $S$ of the state space, with
respect to the unit rate Poisson process. When this density exists, it
explicitly depends on $S$. A sufficient condition for the existence of the
density is that all the eigenvalues of the covariance function, restricted
to $S$, are smaller than 1. We follow the spectral approach and assume
that the covariance function defining the DPP has a spectral
representation. A basic motivation for our choice is that conditions for
the existence of a density become easier to check. We review here the
basic theory on DPPs, making an effort to be as clear and concise as
possible in the presentation of our subsequent models.

We acknowledge that the RJ scheme for our first mixture model is as in
\cite{Xu_etal2016}. In both cases the algorithm requires computing the DPP
density with respect to the unit rate Poisson process. We explain how to
adapt the calculations to our case, and discuss the need to restrict the
process to (any) compact subset. When extending the model to incorporate
covariate information in both, likelihood and prior assignment to mixture
components, the RJ MCMC algorithm requires modifications, as discussed
below. 

The rest of this article is organized as follows. Section~\ref{sec:2}
presents notation and theoretical background necessary to understand how
DPP mixture models are constructed. Section~\ref{sec:3} illustrates the
covariate-dependent extension. Here we build on regular mixture models,
incorporating covariate dependence in the mixture weights and optionally
in the likelihood, which still allows for repulsion among components after
correcting for the regression effect. Section~\ref{sec:4} presents results
from a simulation study and for the Galaxy dataset, while data
applications are discussed in Section~\ref{sec:biopic}. We conclude in
Section~\ref{sect:disc} with final comments and discussion. The on-line
Supplementary Material contains a description of the two RJ MCMC
algorithms, additional illustrative examples on simulated and real data,
and supplemental figures.


\section{Using DPPs to induce repulsion}\label{sec:2}

We review here the basic theory on DPPs to the extent required to explain
our mixture model. We use the same notation as in \cite{Lav_etal2015},
where further details on this theory may be found.

\subsection{Basic theory on DPPs}
\label{subsec:DPP}

Let $B\subseteq \Rd$; we mainly consider the cases $B=\Rd$ and
$B=S$, a compact subset in $\Rd$. By $X$ we denote a simple locally finite
spatial point process defined on $B$, i.e. the number of points of the
process in any bounded region is a finite random variable, and there is at
most one point at any location. See \citeauthor{DalVJ_volI}
(\citeyear{DalVJ_volI,DalVJ_volII}) for a general presentation on point
processes. The class of determinantal point processes we consider 
is defined in terms of their moments, expressed by their product density
functions $\rho^{(n)}:B^n\rt [0,+\infty)$, $n=1,2,\ldots$. Intuitively,
for any pairwise distinct points $x_1,\ldots,x_n \in B$,
$\rho^{(n)}(x_1,\ldots,x_n) dx_1\cdots d x_n$  is the probability that $X$
has a point in an infinitesimal small region around $x_i$ of volume
$dx_i$, for each $i=1,\ldots,n$. More formally, $X$ has $n$-th order
product density function $\rho^{(n)}:B^n\rt [0,+\infty)$ if this function
is locally integrable (i.e. $\intS |\rho^{(n)}(x) | \rmd x< +\infty$ for
any compact $S$) and, for any Borel-measurable function $h:B^n\rt
[0,+\infty)$,
$$ \E\left( \sum_{x_1,\ldots,x_n\in X}^{\neq} h(x_1,\ldots,x_n)\right)=\int_{B^n} \rho^{(n)}(x_1,\ldots,x_n) h(x_1,\ldots,x_n) dx_1\cdots d x_n,$$
where the $\neq$ sign over the summation means that $x_1,\ldots,x_n$ are
pairwise distinct. See also \cite{MolWaa2007}. Let us consider a
covariance function $C:B\times B\rt \Rea$. The theory recalled here is
valid also for complex-valued covariance functions, but such extensions
are not needed in what follows. 

\begin{definition}\normalfont
A simple locally finite spatial point process $X$ on $B$ is called a
determinantal point process with kernel $C$ if its product density
functions are \be \label{eq:determinant} \rho^{(n)}(x_1,\ldots,x_n) =
det[C](x_1,\ldots,x_n), \quad (x_1,\ldots,x_n)\in B^n, \quad n=1,2,\ldots,
\ee where $[C](x_1,\ldots,x_n)$ is the $n\times n$ matrix with entries
$C(x_i,x_j)$. We write $X\sim DPP_B(C)$; when $B=\Rd$ we write $X\sim
DPP(C)$.
\end{definition}

\begin{remark}\normalfont
If $A$ is a Borel subset of $B$, then the restriction $X_A:=X\cap A$ of
$X$ to $A$ is a DPP with kernel given by the restriction of $C$ to
$A\times A$.
\end{remark}

By Theorem~2.3 in \cite{Lav_etal2015}, first proved by \cite{Macchi1975},
such DPP's exist under the two following conditions:
\begin{itemize}
\item $C$ is a continuous covariance function; hence, by Mercer's
    Theorem,
$$ C(x,y)=\sum_{k=1}^{+\infty} \lambda_k^S \phi_k(x)\phi_k(y), \quad (x,y)\in S\times S, \quad S \textrm{ compact subset},$$
    where $\{ \lambda_k^S  \}$ and $\{ \phi_k(x)  \}$ are the
    eigenvalues and eigenfunctions of $C$ restricted to $S\times S$,
    respectively.

\item $\lambda_k^S \leq 1$ for all compact $S$ in $\Rd$ and all $k$.
\end{itemize}

Formula (2.9) in \cite{Lav_etal2015} reports the distribution of the
number $N(S)$ of points of $X$ in $S$, for any compact $S$:
\begin{equation}\label{eq:distrib_N_S}
N(S) \stackrel{d} = \sum_{k=1}^{+\infty} B_k, \quad
\E(N(S))=\sum_{k=1}^{+\infty} \lambda_k^S, \quad
\Var(N(S))=\sum_{k=1}^{+\infty} \lambda_k^S (1- \lambda_k^S),
\end{equation}
where $B_k\ind Be(\lambda_k^S)$, i.e. the Bernoulli random variable with
mean $\lambda_k^S$. When restricted to any compact subset $S$, the DPP has
a density with respect to the unit rate Poisson process which, when
$\lambda_k^S<1$ for all $k=1,2,\ldots$, has the following expression:
\begin{equation}\label{eq:density_DPP}
f(\{x_1,\ldots,x_n\})=
\e^{|S|-D_S}det[\widetilde C](x_1,\ldots,x_n), \quad n=1,2,\ldots,
\end{equation}
where $|S|=\intS \rmd x$, $D_S=-\sum_1^{+\infty}\log(1-\lambda_k^S)$  and
\begin{equation}\label{eq:Ctilde} \widetilde
C(x,y)=\sum_1^{+\infty}\frac{\lambda_k^S}{1-\lambda_k^S}\phi_k(x)\phi_k(y),
\qquad x, y\in S.
\end{equation}
When $n=0$ the density (as well as the determinant) is defined to be equal
to 0. See \cite{MolWaa2007} for a thorough definition of absolute
continuity of a spatial process with respect to the unit rate Poisson
process. However, note that from the first part of \eqref{eq:distrib_N_S}
we have $\Prob(N(S)=0)=\prod_{k=1}^{+\infty}(1-\lambda_k^S)$;  this
probability could be positive due to the assumption $\lambda_k^S<1$ for
all $k=1,2,\ldots$.

From now on we restrict our attention to stationary DPP's, that is,
when $C(x,y)=C_0(x-y)$, where $C_0\in L^2(\Rd)$ is such that its spectral
density $\varphi$ exists, i.e. 
$$C_0(x)=\int_{\Rd} \varphi(y) \cos( 2\pi  x \cdot y)\rmd y, \quad x \in\Rd$$
and $x \cdot y$ is the scalar product in $\Rd$. If $\varphi\in L^1(\Rd)$
and $0\leq \varphi \leq 1$, then the $DPP(C)$ process exists. Summing up,
the distribution of a stationary DPP can be assigned by its spectral
density; see Corollary~3.3 in \cite{Lav_etal2015}.

To explicitly evaluate \eqref{eq:density_DPP} over ${\displaystyle
S=\left[-\dfrac{1}{2}, \dfrac{1}{2}\right]^d }$, we approximate
$\widetilde C$ as suggested in \cite{Lav_etal2015}. In other words, we
approximate the density of $X$ on $S$
by
\begin{equation}\label{eq:approx_dens}
f^{app}(\{x_1,\ldots,x_n\})= \e^{|S|-D_{app}}det[\widetilde
C_{app}](x_1,\ldots,x_n), \quad \{x_1,\ldots,x_n\} \subset S,
\end{equation}
where
\begin{equation}\label{eq:Ct_app}
\widetilde C_{app}(x,y)=  C_{app,0}(x-y)=
\sum_{k\in \Z^d} \left[\dfrac{\varphi(k)}{1-\varphi(k)}\right]
\cos(2\pi k\cdot(x-y)), \qquad x,y \in S,
\end{equation}
\begin{equation} \label{eq:D_app}
D_{app}= \sum_{k\in \Z^d}\log\left(1+
\dfrac{\varphi(k)}{1-\varphi(k)}\right).
\end{equation}
The approximation $C(x,y)\approx C_{app,0}(x-y)$ ($x-y\in S$) follows
because the exact Fourier expansion of $C_0(x-y)$ in $S$ is as in
\eqref{eq:Ct_app} with  the real part of $\int_S C_0(y) \e^{-2\pi i k\cdot
y} \rmd y$ instead of $\varphi(k)$; if we assume $C_0$ such that
$C_0(t)\approx 0$ for $t\not\in S$, then
$$Re\left(\int_S C_0(y) \e^{-2\pi i k\cdot y} \rmd y\right)\approx \varphi(k):=Re\left( \int_{\Rd} C_0(y) \e^{-2\pi i k\cdot y} \rmd y\right).$$
See  also \cite{Lav_etal2015}, Section 4.1.
See  Figure~\ref{fig:C0} in the Supplementary Material, where we display
an example of the function $C_0$.

When $S=R$ is a rectangle in $\Rd$, we can always find an affine
transformation $T$ such that $T(R)=S=\left[-\dfrac{1}{2},
\dfrac{1}{2}\right]^d$. Define $Y=T(X)$. If $f^{app}_Y$ is the approximate
density of $Y$ as in \eqref{eq:approx_dens}, we can then approximate the
density of $X_R$ by
\begin{equation} \label{eq:approx_densR}
f^{app}(\{x_1,\ldots,x_n\})= |R|^{-n}\e^{|R|-|S|}
f^{app}_Y(T(\{x_1,\ldots,x_n\})), \quad \{x_1,\ldots,x_n\} \subset R.
\end{equation}
In practice, the summation over $\Z^d$ in
\eqref{eq:Ct_app} above is truncated to ${\Z_N}^d$, where
$\Z_N:=\{ -N,-N+1,\ldots,0,\ldots,N-1,N\}$; see further details in
\cite{Lav_etal2015}, Section~4.3.

One particular example of spectral density that we found useful is
\begin{equation}\label{eq:spectral_dens}
\varphi(x;\rho,\nu)= s^d \exp\left\{-
\left(\frac{s}{\sqrt{\pi}}\right)^\nu
\left(\frac{\Gamma(\frac{d}{2}+1)}{\Gamma(\frac{d}{\nu}+1)}  \right)^{\nu
d} \rho^ {\nu d}  \| x\|^{\nu}\right\}, \qquad \rho,\nu >0,
\end{equation}
for fixed $s\in (0,1)$ (e.g. $s=\frac12$) and $\|x \|$ is the
Euclidean norm of $x\in\Rd$. This function is the spectral density of a
{\em power exponential spectral model} (see (3.22) in \cite{Lav_etal2015}
when $\alpha=s \  \alpha_{max} (\rho,\nu)$). In this case, we write $X\sim
PES-DPP(\rho,\nu)$. The corresponding spatial process is isotropic. When
$\nu=2$, the spectral density is
$$\varphi(x;\rho,\nu)= s^d \exp\left\{-  \frac{s^2\rho^{2d}}{\sqrt{\pi}}  \| x\|^{2} \right\}, \qquad \rho>0,$$
corresponding to the Gaussian spectral density.

\begin{remark}\normalfont
Note that $\varphi(x;\rho,\nu)<1$ when $0<s<1$ for any $x\in\Rd$,
$\rho,\nu>0$, so that $X_S$ has a density as described in
\eqref{eq:density_DPP}. Figure~\ref{fig:spectral_density} in the
Supplementary material shows a plot of the power exponential spectral
density \eqref{eq:spectral_dens} for different values of parameters
$\rho$, $\nu$. Note that $\nu$ controls the shape of
$\varphi(x;\rho,\nu)$, which ranges from a slowly decreasing function of
$x$ to an indicator function. On the other hand $\rho$ plays the role of a
centering parameter, with higher values retarding the decay speed of
$\varphi(x;\rho,\nu)$. As discussed above, knowledge about the spectral
density is all that is needed for the approximations to work. 
Moreover, even if the analytic expression of $C(x,y),
(x,y)\in\Rd\times\Rd$ is known, as in, e.g.~\eqref{eq:spectral_dens}, we
still need to compute the eigenvalues and eigenfunctions of $C$ restricted
to $S\times S$ for any compact $S$, and this may be analytically
impossible. A potential disadvantage derived from this is that parameter
interpretation  in the spectral domain becomes unclear. A possible way to
alleviate this difficulty consists of conducting extensive simulations
that help understanding the role of such parameters. We do that for the
case of \eqref{eq:spectral_dens}; see Section~\ref{sec:4}.
\end{remark}

Typically, DPPs have been used to make inference mostly on spatial data
(including images, videos and point patterns related to towns, trees and
cells); see, for instance, \cite{ShiGel2016} who describe an approximate
Bayesian computation method to fit DPPs to spatial point pattern data.
Historically, the first paper where DPPs were adopted as a prior for
statistical inference in mixture models is \cite{affandi2013approximate}.

The statistical literature includes a number of papers
illustrating theoretical properties for estimators of DPPs from a
non-Bayesian viewpoint. \cite{biscio2016contrast} study asymptotic
properties of minimum contrast estimators for parametric stationary
determinantal point processes. \cite{biscio2016quantifying} quantify the
repulsiveness of a DPP, based on its second-order properties, and address
the question of how repulsive a stationary DPP can be. Moreover,
\cite{bardenet2015inference} derive bounds on the likelihood of a DPP, and
\cite{kulesza2012determinantal} provide an introduction to DPPs, focusing
on the intuitions, algorithms, and extensions that are most relevant to
the machine learning community.

\subsection{The mixture model with repulsive means}\label{subsec:nocov_model}

To deal with limitations of model \eqref{eq:mixture} or DPMs, we consider
repulsive mixtures. Our aim is to estimate a random partition of the
available subjects, and we want to do so using \virgolette{few}
groups. By repulsion we mean that cluster locations are a priori
encouraged to be well separated, thus inducing fewer clusters than if they
were allowed to be independently selected. We start from parametric
densities $f(\cdot;\theta)$, which we take to be Gaussian,
and assume that the collection of location parameters follows a DPP.
We specify a hierarchical model that achieves the goals previously
described. Concretely, we propose:
\begin{align}
\label{eq:ver}
 y_i \mid s_i = k, \{ \mu_k\}, \{\sigma^2_k \}, K  &\ind \mathcal{N}\left(y_i; \mu_k, \sigma^2_k  \right)
 \quad i=1,\dots,n\\
\label{eq:X}
X=\{\mu_1, \mu_2, \dots, \mu_K, K\} &\sim PES-DPP(\rho, \nu)\\
\label{eq:rhonu}
 (\rho,\nu) &\sim \pi\\
\label{eq:cluster}
p(s_i=k) &= w_k, \quad k =1,\dots,K \ \textrm{ for each } i\\
\label{eq:w}
w_1, \dots, w_K \mid K &\sim Dirichlet(\delta, \delta, \dots, \delta)\\
\label{eq:sigma}
 \sigma^2_k \mid K &\iid inv-gamma(a_0, b_0),
\end{align}
where the PES-DPP$(\rho,\nu)$ assumption \eqref{eq:X} is regarded
as a default choice that could be replaced by any other valid DPP
alternative. We note that, as stated, the prior model may assign a
positive probability to the case $K=0$. This case of course makes no sense
from the viewpoint of the model described above. Nevertheless, we adopt
the working convention of redefining the prior to condition on $K\ge 1$,
i.e., truncating the DPP to having at least one point. In practice, the
posterior simulation scheme later described simply ignores the case $K=0$,
which produces the desired result. Note also that we have assumed prior
independence among blocks of parameters not involving the locations
$\mu_k$.

Model \eqref{eq:ver}-\eqref{eq:sigma} is a DPP mixture model along the
lines proposed in \cite{Xu_etal2016}. Indeed, we both use DPPs as priors
for location points in the mixture of parametric densities.
However, the specific DPP priors are different, as they restrict to a
particular case of DPPs (L-ensambles), and choose a Gaussian covariance
function for which eigenvalues and eigenfunctions are analytically
available. We adopt instead a spectral approach for assigning the
prior \eqref{eq:X}. Similar to \cite{Xu_etal2016}, we carry out posterior
simulation using a reversible jump step as part of the Gibbs sampler.
However, when updating the location points $\mu_1,\ldots,\mu_K$ we refer
to formulas \eqref{eq:approx_dens}-\eqref{eq:approx_densR}.
\cite{Xu_etal2016} take advantage of the analytical expressions that we do
not have for our case, and that are also unavailable in other possible
specific choices of the spectral density. The posterior simulation
algorithm we propose for our model is described in Section
\ref{supp:nocov_Gibbs} in the Supplementary material.

\section{Generalization to covariate-dependent models}
\label{sec:3}

The methods discussed in Section~\ref{sec:2} were devised for density
estimation-like problems. We now extend the previous modeling to
the case where $p$-dimensional covariates $x_1,\ldots,x_n$ are recorded as
well. We do so by allowing the mixture weights to depend on such
covariates. In this case, there is a trade-off between repulsiveness of
locations in the mixtures and attraction among subjects with similar
covariates. We also entertain the case where covariate dependence is added
to the likelihood part of the model. Our modeling choice here is akin to
mixtures of experts models~\citep[see, e.g.,][]{McLachlan:2005}, i.e., the
weights are defined by means of normalized exponential function.

Building on the model from Section~\ref{subsec:nocov_model}, we assume the
same likelihood \eqref{eq:ver} and the DPP prior for $X=\{\mu_1, \mu_2,
\dots, \mu_K, K\}$ in \eqref{eq:X}-\eqref{eq:rhonu}, but change
\eqref{eq:cluster} and \eqref{eq:w} to
\begin{align}
\label{eq:clusterx}
p(s_i=k) = w_k(x_i) &= \dfrac{\exp\left(\beta^T_k x_i \right)}{\sum_{l=1}^K \exp\left(\beta_l^T x_i \right)}, \quad k =1,\dots,K \\
\label{eq:beta}
\beta_2, \dots, \beta_K \mid K &\iid \calN_p\left(\beta_0, \Sigma_0\right), \quad \beta_1 = 0,
\end{align}
where the $\beta_1=0$ assumption is to ensure identifiability.  To
complete the model, we assume \eqref{eq:sigma} as the conditional marginal
for $\sigma^2_k$; the prior for $(\rho,\nu)$ in \eqref{eq:rhonu} is later
specified.  Here $\beta_0\in\Rea^p$, and to  choose $\Sigma_0$, we
use a g-prior approach, namely $\Sigma_0 = \phi \times \left( X^T
X\right)^{-1}$, where $\phi$ is fixed, typically of the same order of
magnitude of the sample size \citep[see][]{zellner1986assessing}.

Assuming now \eqref{eq:ver} on top of
\eqref{eq:clusterx}-\eqref{eq:beta} rules out the case of a likelihood
explicitly depending on covariates, which instead would generally achieve a better
fit than otherwise. Of course, there are many ways in which such
dependence may be added. For the sake of concreteness, we assume here a
Gaussian regression likelihood, where only the intercept parameters arise
from the DPP prior. More precisely, we assume
\begin{align}
\label{eq:verx}
 y_i \mid s_i=k, x_i,  \{ \mu_k\}, \{\sigma^2_k \}, K  &\ind \mathcal{N}\left(y_i; \mu_k + x_i^T \gamma_k, \sigma^2_k \right)
 \quad i=1,\dots,n\\
\label{eq:sigmagamma}
(\gamma_1, \sigma^2_1),\dots, (\gamma_K, \sigma^2_K)|K &\iid \ norm-invgamma(\gamma_0, \Lambda_0, a_0, b_0),
\end{align}
where the $\gamma_k$'s are $p$-dimensional regression coefficients. The
notation in \eqref{eq:sigmagamma} means that $\gamma_k \mid \sigma_k^2
\sim \calN_p(\gamma_0, \sigma_k^2 \Lambda_0)$, and $\sigma_k^2 \sim
inv-gamma(a_0,b_0)$, where $\gamma_0\in\Rea^p$ and $\Lambda_0$ is a
covariance matrix. The prior for $\{ s_i \}$ and $\beta_j$'s is
given in \eqref{eq:clusterx}-\eqref{eq:beta} as in the previous model.
Note that \eqref{eq:verx} implies that only the intercept term is
distributed according to the repulsive prior. Thus, we allow the
response mean to be corrected by a linear combination of the covariates
with cluster-specific coefficients, with the repulsion acting only on the
residual of this regression. The result is a more flexible model than the
repulsive mixture \eqref{eq:ver}-\eqref{eq:sigma}. Observe that there is
no need to assume the same covariate vector in \eqref{eq:verx} and
\eqref{eq:clusterx}, but we do so for illustration purposes only. 

The Gibbs sampler algorithm employed to carry out posterior inference for
this model is detailed in Section \ref{supp:cov_Gibbs} of the
Supplementary material. However, it is worth noting that the reversible
jump step related to updating the number of mixture components $K$ and the
update of the coefficients  $\left\{ \beta_2, \beta_3, \dots, \beta_K
\right\}$ are complicated by the presence of the covariates. For the
$\beta$ coefficients, we resort to a Metropolis-Hastings step, with a
multivariate Gaussian proposal centered in the current value. For $K$,
we employ an ad hoc Reversible Jump move.

\section{Simulated data and reference datasets}\label{sec:4}

Before illustrating the application of our models to specific datasets, we
discuss some general choices that apply to all examples. Every run of
the Gibbs sampler (implemented in \texttt{R}) produced a final sample
size of 5,000 or 10,000 iterations (unless otherwise specified), after a
thinning of 10 and initial burn-in of 5,000 iterations. In all cases,
convergence was checked using both visual inspection of the chains and
standard diagnostics available in the \textsc{CODA} package. 
Elicitation of the prior for $(\rho,\nu)$ requires some care, as the role
of these parameters is difficult to understand. Therefore, an extensive
robustness analysis with respect to $\pi(\rho,\nu)$ for those datasets was
carried out. See Sections \ref{subsec:galaxy} and \ref{supp:8mixt} of the
Supplementary material. We point out that an initial prior independence
assumption $\pi(\rho,\nu)= \pi(\rho) \ \pi(\nu)$ produced bad mixing of
the chain. In particular, when $\rho$ is small with respect to $\nu$, the
spectral function $\varphi(\cdot)$ has a very narrow support, concentrated
near the origin, forcing the covariance function
$\widetilde{C}_{app}(x,y)$ to become nearly constant for $x, y\in S$ and
thus producing nearly singular matrices.
We next investigated the case $\pi(\rho, \nu) = \pi(\rho\mid\nu)\times
\pi_\nu(\nu)$, where
$$\rho\mid\nu \stackrel{d}{=} M(s,\varepsilon, \nu) + \rho_0, \quad \rho_0
\sim gamma(a_\rho, b_\rho). $$ Here, $M(s,\varepsilon, \nu)$ is a constant
that is the minimum value of $\rho$ such that $\varphi(2)>\varepsilon$
(here $\varphi(2)$ is a reference value chosen to avoid a small support),
and $\varepsilon$ is a threshold value, assumed to be small (0.05, for
instance). From Figure~\ref{fig:spectral_density} in the Supplementary
Material, it is clear that $\varphi(\cdot;\rho,\nu)$ goes to $0$ too fast
when $\nu$ is small relative to $\rho$. It follows that
$$ M(s,\varepsilon, \nu) = \dfrac{2 s \Gamma(1/\nu+1)\pi^{1/2}}{\Gamma(3/2)
  \left(\log\left(\dfrac{s}{\varepsilon}\right) \right)^{1/\nu}}. $$
On the other hand, two different choices for $\pi_\nu$ were
considered: a gamma distribution, which gave a bad chain mixing, and a
discrete distribution on $\mathcal{V}_2=\{0.5, 1, 2,3, 5, 10, 15, 20, 30,
50 \}$ (or on one of its subsets). In this case, the mixing of the chain
was better, but the posterior for $\nu$ did not discriminate among the
values in the support. For this reason, in Sections 5.3, 6 and 7, we
assume $\nu=2$, $s=1/2$ and
\begin{equation}
\label{eq:prior_rho}
 \rho \stackrel{d}{=} \sqrt{\frac{\pi}{\log(\frac{1}{2\varepsilon})}} + \rho_0, \quad \rho_0 \sim gamma(a_\rho, b_\rho).
\end{equation}

\subsection{Galaxy data}

\label{subsec:galaxy} 
\begin{figure}[h!]
\begin{center}
       \includegraphics[width=0.45\textwidth,height=0.5\textwidth]{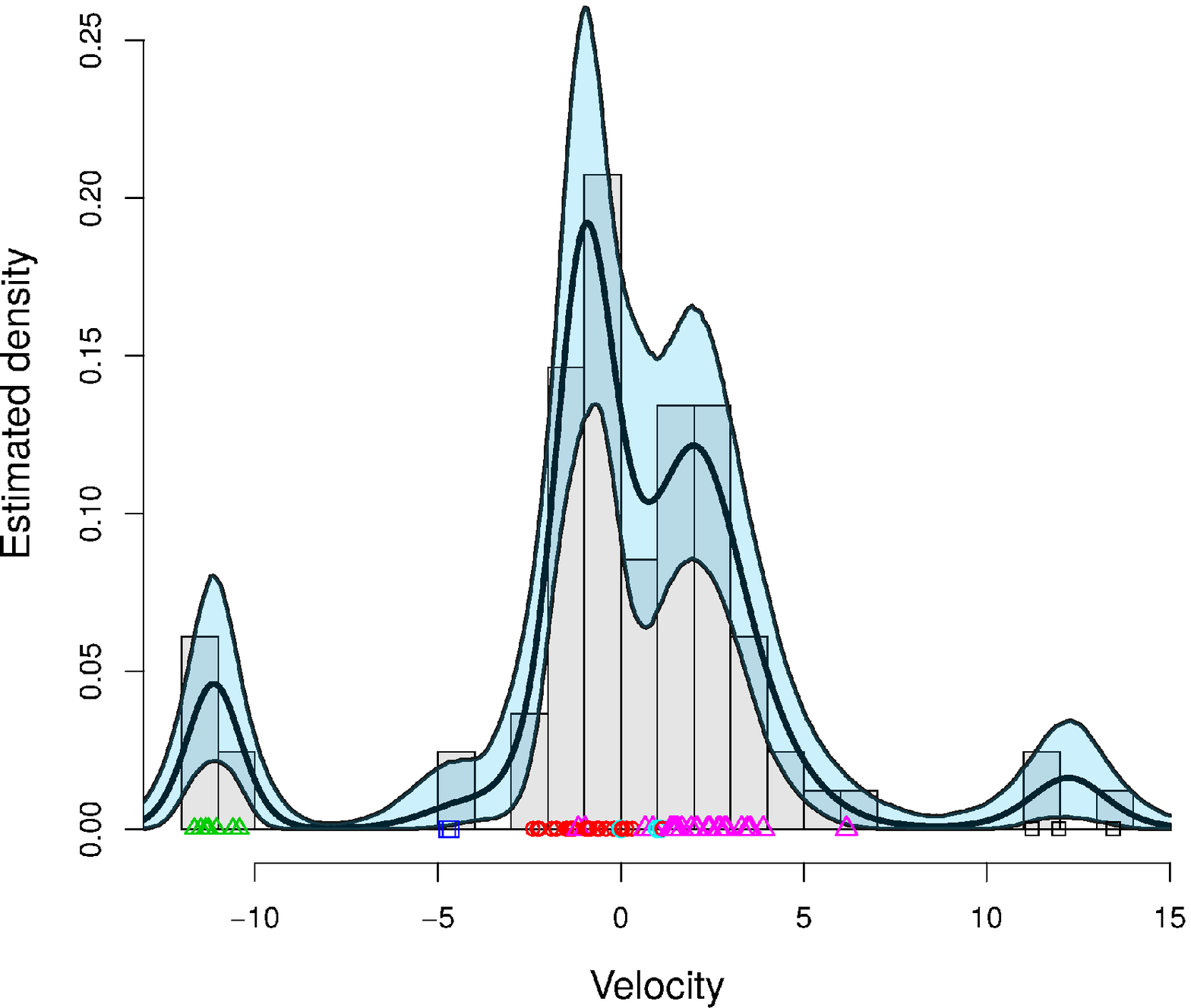}
     \includegraphics[width=0.45\textwidth,height=0.55\textwidth]{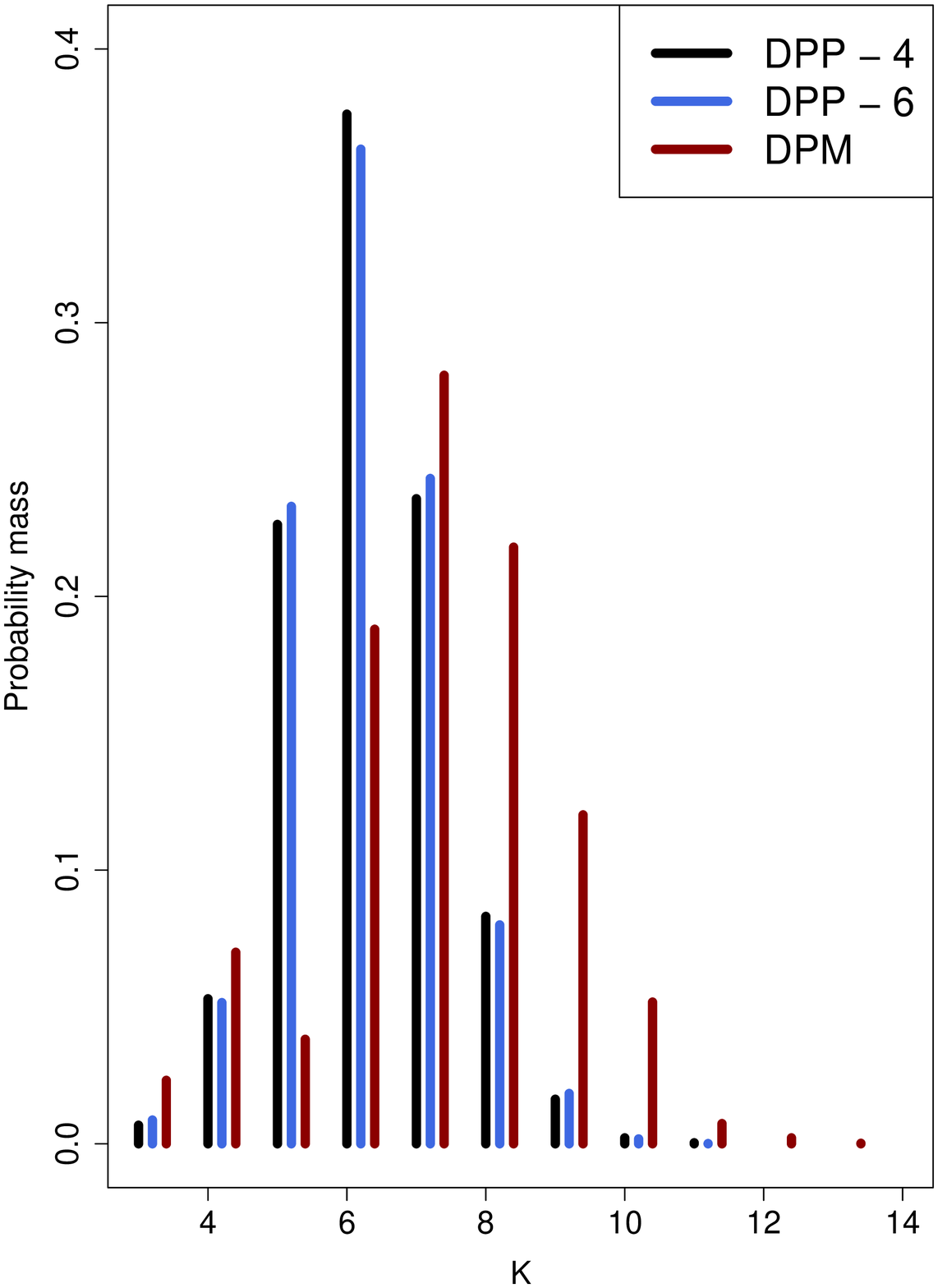}
\end{center}
\caption{Density estimates and estimated partition (left) for the Galaxy dataset
under Test~4 in Table~\ref{tab:test_galaxy}, including 90\% credibility
bands (light blue). Posterior distribution (right) of the number $K$ of
components under Test~4 (black) and 6 (blue) in
Table~\ref{tab:test_galaxy} and under the DPM model (red) as in Test~7 in
Table~\ref{tab:test_galaxy_dpm}.} \label{fig:gal_dens}
\end{figure}

This popular dataset contains $n = 82$ measured
velocities of different galaxies from six well-separated conic sections of
space. Values are expressed in Km/s, scaled by a factor of $10^{-3}$. We
set the hyperparameters in this way: for the variance $\sigma^2_k $ of the
components, $\left(a_0, b_0\right) = \left(3, 3\right)$ (such that the
mean is 1.5 and the variance is $9/4$) and for the weights $\{w_k\}$ the
Dirichlet has parameter $(1,1,\dots,1)$. The other hyperparameters are
modified in the tests, as in Table~\ref{tab:test_galaxy}, where we report
summaries of interest, such as the prior and posterior mean and variance
for the number of components $K$. In addition, we also display the mean
squared error (MSE) and the log-pseudo marginal likelihood (LPML) as
indices of goodness of fit, defined as $MSE = \sum_{i=1}^n(y_i -
\hat{y_i})^2$ and $LPML = \sum_{i=1}^n \log\left( f(y_i\mid y^{(-i)})
\right)$, where $\hat{y_i}$ is the posterior predictive mean and $
f(y_i\mid y^{(-i)}) $ is the $i-$th \textit{conditional predictive
ordinate}, that is the predictive distribution obtained using the dataset
without the $i-$th observation. Figure~\ref{fig:gal_dens} (left) shows
density estimates and the estimated partition of the data, obtained as the
partition that minimizes the posterior expectation of Binder's loss
function under equal misclassification costs
\citep[see][]{lau2007bayesian}. The points at the bottom of the plots
represent observations,  while colors refer to the corresponding cluster.
Figure~\ref{fig:gal_dens} (right) displays the posterior distribution of
$K$ for Test 4 and 6 in Table~\ref{tab:test_galaxy}.

\begin{table}[h!]
  \begin{center}\small
    \begin{tabular}{l  c  c  c  c  c  c c c}
    \toprule
  Test & $\rho$ & $\nu$ & $\mathbb{E}(K)$ & $Var(K)$ & $\mathbb{E}(K|data)$ & $Var(K|data)$ & MSE & LPML \\
  \midrule
\rowcolor{black!20}  1 & 2 & 2 & 2& 1.67& 6.09 & 1.10  & 78.95 &  -171.72  \\
  2 & 5 & 10  & 5.00&7.12 & 6.07 & 1.09  & 78.33 & -167.96  \\
\rowcolor{black!20}  3 & $a_\rho= 1, b_\rho=1$ & 2 &2.18 & 1.978 &  6.10  & 1.10 &  73.89 & -164.47 \\
  4 & $a_\rho= 1, b_\rho=1$ & 10 &  2.73 & 2.15 &  6.11 & 1.12 &  74.93 & -162.71 \\
\rowcolor{black!20}  5 &$a_\rho= 1, b_\rho=1$ & discr($\mathcal{V}_1$) & 2.47 & 2.21  & 6.06  & 1.08 & 74.02  & -172.54  \\
  6 & $a_\rho= 1, b_\rho=1$ & discr($\mathcal{V}_2)$ &  2.51 & 2.27 &  6.10  & 1.13  & 76.64 &  -170.94 \\
\bottomrule
    \end{tabular}
\end{center}
\caption{Prior specification for $(\rho,\nu)$ and $K$ and posterior
summaries for the Galaxy dataset; $(a_\rho, b_\rho)$ appear in
\eqref{eq:prior_rho}; here $\mathcal{V}_1$ is $\{1,2,5,10,20\}$  and
$\mathcal{V}_2 = \{0.5,1,2, 3, 5,10,15,20,30, 50 \}$. }
\label{tab:test_galaxy}
\end{table}

As a comparison, the same posterior quantities than in
Table~\ref{tab:test_galaxy} were computed using the DPM, fitted via
the function \texttt{DPdensity} available from \texttt{DPpackage}
\citep{jara2011dppackage}; see Table~\ref{tab:test_galaxy_dpm}. The
same prior for $\sigma_k^2$ as in our model was assumed. Note that
$\alpha$ is fixed in Test 8 in such a way that  a priori $\E(K)=2$, while
for Test 3, $a_0$ and $b_0$ are chosen so that $\mathbb{E}(\alpha)=2$ and
$\Var(\alpha)=1$. Clearly, Test 10 (DPM) shows the best indexes of
goodness of fit but at the cost of overestimating the number of clusters.
It is well-known that, in general, clustering in the context of
nonparametric mixture models as DPMs is strongly affected by the base
measure \citep[see, e.g.][]{MilHar2017}. Our model, on the other hand,
avoids the delicate choice of the base measure leading to more robust
estimates of $K$. See also Figure~\ref{fig:gal_dens} (right) which
displays the posterior distribution of $K$ under the DPM mixture and our
models. 
\begin{table}[h]
  \begin{center}\small
    \begin{tabular}{ l  c c  c  c  c  c }
    \toprule
  Test & $\alpha$ & $\mathbb{E}(K|data)$ & $Var(K|data)$ & MSE & LPML\\
  \midrule
 \rowcolor{black!20}7 & $gamma(a_0=0.5, b_0=1)$ & 6.166 & 1.549 & 62.703 & -151.797  \\
 8 & $0.8$ & 5.936 & 1.25 & 61.255  & -151.146\\
\rowcolor{black!20} 9 & 0.45 & 4.371 & 1.142         & 139.659      & -169.978  \\
 10 &$ gamma(4,2)$ &  7.271 & 1.594 & 36.708   & -149.258     \\
 \bottomrule
    \end{tabular}
\end{center}
\caption{Prior specification for $\alpha$  and posterior summaries for the
Galaxy dataset using the function DPdensity in \texttt{DPpackage}.}
\label{tab:test_galaxy_dpm}
\end{table}

Finally, we recall that in Section~\ref{supp:diff_spectral_dens} in the
Supplementary material we report some further tests on the Galaxy
dataset to show the influence of various choices of spectral density on
the inference. We conclude that there is evidence to robustness with
respect to the choice of spectral density.

\subsection{Simulated data with covariates}
\label{subsec:cov_simulated}

We consider the same simulated dataset  as in \cite{muller2011product},
Section~5.2; the simulation \virgolette{truth} consists of 12 different
distributions, corresponding to different covariate settings (see Figure~1
of that paper). Model \eqref{eq:ver}-\eqref{eq:rhonu},
\eqref{eq:sigma}-\eqref{eq:beta} was fitted to the dataset, assuming
$\beta_0 = 0$, $\Sigma_0 = 400\times \left( X^T X\right)^{-1}$, $a_\rho =
1, b_\rho = 1.2$, and $a_0, b_0$ such that the prior mean of $\sigma^2_k$
is 50 and variance is 300. Recall also that here we assume $\nu = 2$.

As an initial step, inference for the complete dataset (1000
observations) was carried out, yielding a posterior of $K$, not reported
here, mostly concentrated over the set $\{8,9,\dots,16\}$, with a mode
at 11. Figure~\ref{fig:pred_comp} in the Supplementary Material shows
posterior predictive distributions for the 12 different reference
covariate values, along with 90\% credibility intervals. These are in good
accordance with the simulation truth \citep[compare Figure 1
in][]{muller2011product}.

To replicate the tests in \cite{muller2011product}, a total of
$M=100$ datasets of size 200 were generated by randomly subsampling 200
out of the 1000 available observations. Computational burden over multiple
repetitions was controlled by limiting the posterior sample sizes to
2,000. Table~\ref{tab:mse_12cov} displays the root MSE for estimating
$\mathbb{E}(y\mid x_1 , x_2 , x_3 )$ for each of the 12 covariate
combinations defining the different clusters for our model and for the
PPM$_x$, as in Table~1 of \cite{muller2011product}. The computations
also include evaluation of the root MSE and LPML for all the 100
datasets for estimating the data used to train the model, with $
MSE_{train} = \sum_{i=1}^{n} \left( y_i - \hat{y}_i \right)^2,$ where
$\hat{y}_i$ is the expected value of the estimated predictive
distribution, and for a test dataset of 100 new data, $ MSE_{test} =
\sum_{i=1}^{n} \left( y^{test}_i - \hat{y}_i \right)^2$. In addition,
we report $LPML_{train}$, value of the Log Pseudo Marginal Likelihood
for the training dataset. Table~\ref{tab:cfr} shows the values compared to
other competitor models, i.e the linear dependent Dirichlet process
mixture (LDDP) defined in \cite{de2004anova}, the product partition model
(PPM$_x$) in \cite{muller2011product} and the linear dependent tailfree
process model (LDTFP) in \cite{jara2011class}. The best values are in
bold: our model performs well according to the LPML, while the MSE
suggests to use PPM$_x$ or LDTFP. In general, our model is competitive
with respect to other popular models in the literature. Moreover, in the
LDDP case, we have that the average number of clusters is 20.6 with
variance 2.266, thus indicating a less parsimonious model compared to
ours.

\begin{table}[h!]
  \begin{center}\small
    \begin{tabular}{c c  c  c c }
    \hline
  $x_1$ & $x_2$ & $x_3$ & DPP & $PPM_X$ \\ \hline
\rowcolor{black!20}  -1 & 0 & 0 & 6.1 & 7.9 \\
  0 & 0 & 0 & 6.7 & 3.9 \\
\rowcolor{black!20}  1 & 0 & 0 & 7.2 & 2.8 \\
  -1 & 1 & 0 & 6.5 & 5.4 \\
 \rowcolor{black!20} 0 & 1 & 0 & 6.5 & 4.6 \\
  1 & 1 & 0 & 6.8 &  4.0\\
\rowcolor{black!20}  -1 & 0 & 1 & 6.8 & 6.1 \\
  0 & 0 & 1 & 6.1 & 4.2 \\
 \rowcolor{black!20} 1 & 0 & 1 & 5.7 &  4.5\\
  -1 & 1 & 1 & 5.9 &  9.5\\
 \rowcolor{black!20} 0 & 1 & 1 & 6.6 & 8.3 \\
  1 & 1 & 1 & 5.8 & 6.2 \\ \hline
 \rowcolor{black!50} \textbf{avg} & & &\textbf{6.4} & \textbf{5.6} \\
  \hline
    \end{tabular}
\end{center}
\caption{Root MSE for estimating $\mathbb{E}(y\mid x_1 , x_2 , x_3 )$ for
12 combinations of covariates $\left(x_1 , x_2 , x_3 \right)$ and $PPM_x$
as competing model of reference (compare also the results in Table~1 of
\citealp{muller2011product}).
}
\label{tab:mse_12cov}
\end{table}

\begin{table}[h!]
  \begin{center}\small
    \begin{tabular}{ c  c c c c  }
    \toprule
~ & DPPx & LDDP & $PPM_X$ & LDTFP \\
\midrule
\rowcolor{black!20} Root $MSE_{train}$ & 324.531 & 304.742 & \textbf{278.395} & 304.374 \\
Root $ MSE_{test}$ & 216.675 & 215.1694 & 217.2459 & \textbf{212.761} \\
\rowcolor{black!20} $LPML_{train}$ & \textbf{-871.8} &      -902.2295 & -873.1671 &  -901.465\\
\bottomrule
    \end{tabular}
\end{center}
\caption{Comparison with competitors for the simulated dataset with
covariates: best values according to each index are in bold. DPPx denotes
our model, while LDDP is the linear dependent Dirichlet process mixture,
$PPM_x$ is the product partition model with covariates, and LDTFP is the
linear dependent tailfree process model.} \label{tab:cfr}
\end{table}

In summary, our extensive simulations suggest that the proposed approach
tends to require less mixture components than other well-known alternative
models, while still providing a reasonably good fit to the data.

\section{Biopic movies dataset}
\label{sec:biopic}

For this illustrative example we consider the Biopics data
available in the R package \texttt{fivethirtyeight}
\citep{fivethirtyeight}. This dataset is based on the IMDB database,
related to biographical films released from 1915 through 2014. An
interesting explorative analysis of the data can be found in
\url{https://fivethirtyeight.com/features/straight-outta-compton-is-the-rare-biopic-not-about-white-dudes/}.

We consider the logarithm of the gross earnings at US box office  as a
response variable, with  the following covariates: (i) year of release of
the movie (in a suitable scale, continuous); (ii) a binary variable that
indicates whether the main character is a person of color; and (iii) a
categorical variable that considers if the country of the movie is US, UK
or other. After removing the missing data from the dataset, we  were left
with $n=437$ observations and $p=4$. We note that 76 biopics have a person
of color as a subject and the frequencies of the category ``origin'' are
$\left(256 ,79  ,  64\right)$ for US, UK and ``other'', respectively;
``other'' means mixed productions (e.g. US and Canada, or US and UK). 
In what follows, the hyperparameters in model
\eqref{eq:verx}-\eqref{eq:sigmagamma}, \eqref{eq:X}-\eqref{eq:rhonu},
\eqref{eq:clusterx}-\eqref{eq:beta} are chosen as $\beta_0=0$,
$(a_\rho, b_\rho)=(1,1)$. The prior mean and variance of $K$ induced by
these hyperparameters are 2.162 and 1.978, respectively. The scale
hyperparameter $\phi$ in the $g$-prior for $\beta$ and  $(a_0,b_0)$ vary
as determined in Table~\ref{tab:par_biopics}, where $m$ and $v$ denote the
prior mean ${\displaystyle b_0/(a_0-1)}$ and variance ${\displaystyle
b_0^2/((a_0-1)^2 (a_0-2))}$, respectively, of the inverse gamma
distribution for $\sigma_k^2$ as in \eqref{eq:sigmagamma}. We also assume
$\gamma_0$ equal to the vector of all 0's, while $\Lambda_0$ is such that
the marginal a priori variance of $\gamma_k$ is equal to $diag (0.01,
0.1,0.1,0.1)$, in accordance to the variances of the corresponding
frequentist estimators.

 \begin{table}[h!]
  \begin{center}
    \begin{tabular}{ l  c  c   c c  c  c c }
    \hline
  Test  & $\phi$ & $m$ & $v$  &${\mathbb E}(K\mid data)$ & sd($K\mid$data) & MSE & LPML\\
  \hline
\rowcolor{black!20}  $A$ &  50 & 5    &1    & 4.49 & 1.10  & 1126.32 &  -960.89\\
                     $B$ &  200&5     &10   & 4.45 & 1.19  & 983.55 &  -954.55  \\
\rowcolor{black!20}  $C$ &  50 & 3&$+\infty$&  5.66 & 1.27 & 501.22 & -918.74\\
                     $D$ &  200& 10  & 5    & 4.21 & 1.33  &  1805.83  & -980.61 \\
\rowcolor{black!20}  $E$ &  100 & 2  &1     & 5.31  & 1.21 & 564.26 & -935.56\\
                     $F$ &  200& 2  & 10    & 5.51 & 1.26  & 557.44 &  -925.22 \\
  \hline
    \end{tabular}
\end{center}
\caption{Prior specification for $\beta_k$'s and $\sigma_k^2$'s parameters
and posterior summaries for the Biopics dataset; $m$ and $v$ are
prior mean and variance, respectively, of $\sigma_k^2$. Posterior mean and
variance of the number $K$ of mixture components are in the fifth and
sixth columns, respectively, while the last two columns report MSE and
LPML, respectively.}
\label{tab:par_biopics}
\end{table}

The posterior of $K$ is robust with respect to the choice of prior
hyperparameters; on the other hand, our results show that by not
including covariates in the likelihood, i.e. setting all $\gamma_k$'s
are equal to 0, inference on $K$ is much more sensitive to the choice of
$(a_0, b_0)$ (results not shown here).

Predictive inference was also considered, by evaluating the posterior
predictive distribution at the following combinations of covariate values:
$(i)$ (mean value for covariate year, US, white); $(ii)$ (mean value
for covariate year, US, color); $(iii)$ (mean value for covariate year,
UK, white); $(iv)$ (mean value for covariate year, UK, color); $(v)$ (mean
value for covariate year, ``other'', white); and $(vi)$ (mean value for
covariate year, ``other'', color). Corresponding plots are shown in
Figure~\ref{suppfig:pred_bio} in the Supplementary material. These
distributions appear to be quite different in the six cases: in
particular, we can observe that in cases $(i)$ and $(ii)$, the posterior
is shifted towards higher values. This is quite easy to interpret, since
the measurements are given by the earnings in the US box offices;
therefore, we expect that in general US movies will be more profitable in
that market. The difference due to the race is, on the other hand, less
evident. However, the predictive densities show slightly higher earnings
for movies where the subject is a person of color, if the origin is
``other'' ($(v)$ and $(vi)$). Movies from the UK, on the other hand,
exhibit the opposite behavior ($(iii)$ and $(iv)$).

\begin{figure}
\centering
 \includegraphics[height=0.5\textwidth, width = 0.4\textwidth]{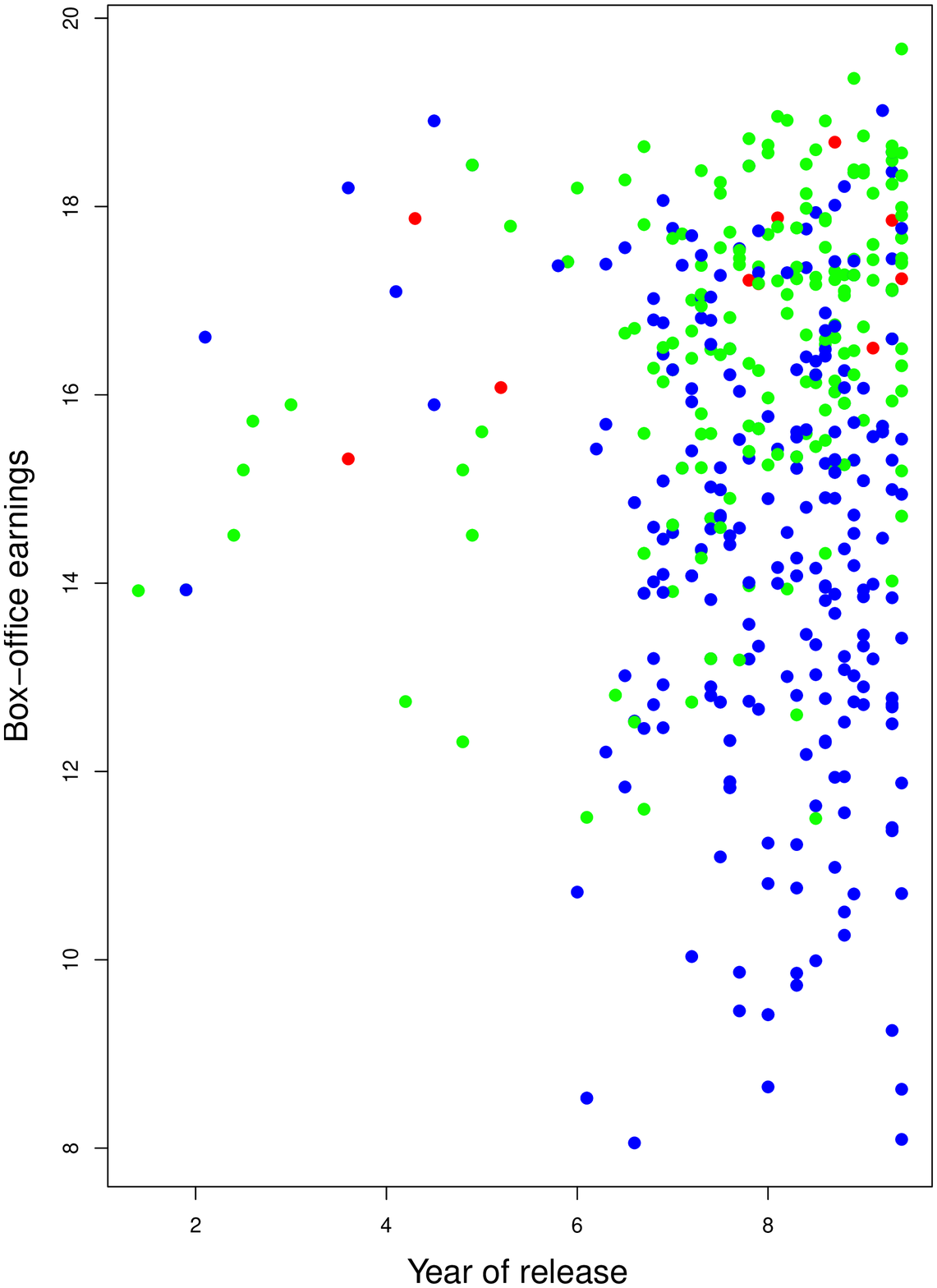}
 \includegraphics[height=0.5\textwidth, width = 0.4\textwidth]{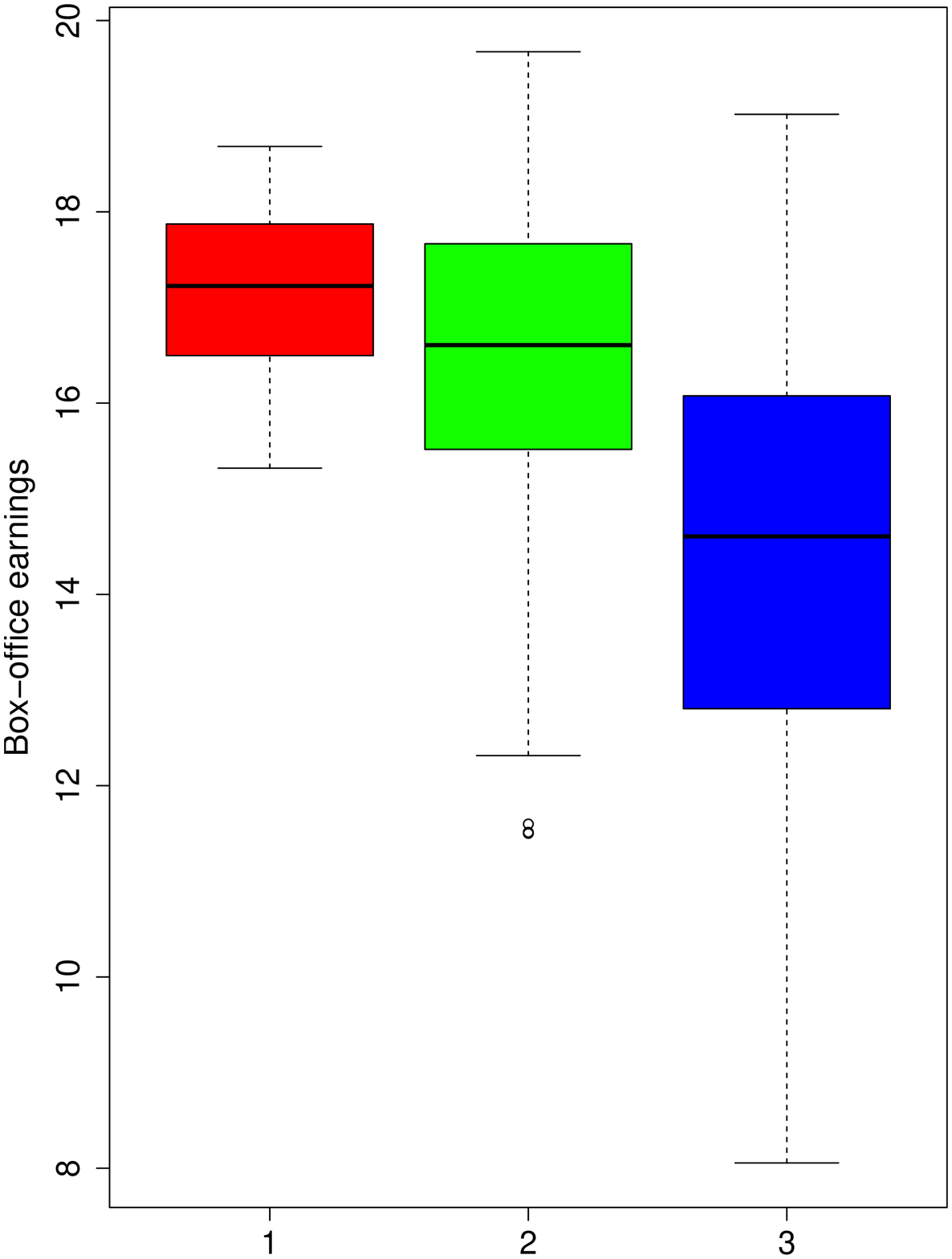}
\caption{Cluster estimate (left) under our model (Test B in
Table~\ref{tab:par_biopics}) for the Biopics dataset. Each color
represents one of the three estimated clusters. Coordinate $y$ is the
response, i.e. box-office earning, while coordinate $x$ is the covariate
year of release. The boxplot of the response per group is in the right
panel.}\label{fig:clu_est_bio}
\end{figure}

We report here the posterior cluster estimate for Test B in
Table~\ref{tab:par_biopics}. We found three groups, with sizes 10, 193,
234, respectively; see Figure~\ref{fig:clu_est_bio} for the estimated
clusters and boxplots of the response. As a comparison, it can be useful
to report the total average values for the response, 15.36, and for the
covariates: 7.89 (year), 0.18 (UK), 0.15 (``other''), 0.83 (white). These
3 groups have a nice interpretation in terms of covariates: group 1 is the
smallest, with a high average response (17.18), and it is characterized by
a high percentage of movies from ``other'' countries, with a person of
color as its subject. Group 2 corresponds also to a high average response
(16.42), but the average values of UK, ``other'' and person of color are
similar to the total averages (0.14, 0.09, 0.84, respectively). The
average response in group 3 is smaller (14.40) than the total sample mean,
while the average values of UK, ``other'' and person of color are 0.22,
0.17, 0.84, respectively.

To assess effectiveness of the proposed model,  we compare the results
with the linear dependent Dirichlet process mixture model introduced in
\cite{de2004anova} and implemented in the \texttt{LDDPdensity} function of
\texttt{DPpackage} \citep{jara2011dppackage}. Prior information has been
fixed as follows: for Test G the mass parameter of the Dirichlet process
$\alpha$ is set equal to 0.3 such that $\mathbb{E}(K) = 2.87$ and
$Var(K)=1.81$, that approximately match the prior information we gave on
the parameter $K$. Similarly, under Test H, $\alpha$ is distributed
according to the  $gamma(1/4,1/2)$, such that the prior mean on $K$ is 3.6
and variance 22.18. The normal baseline distribution is a multivariate
Gaussian with mean vector 0 and a random covariance matrix which is given
a non-informative prior and the inverse-gamma distribution for the
variances of the mixture components has parameters such that mean and
variance are equal to 5, 1, respectively similarly as in Table
\ref{tab:par_biopics}. Posterior summaries can be found in Table
\ref{tab:biopics_lddp}.

As a comparison between  the estimated partitions under our model (Figure
\ref{fig:clu_est_bio}) and the LDDP mixture model,
Figure~\ref{fig:clu_est_bio_lddp} in the Supplementary Material displays
the estimated partition obtained under the LDDP model under Test $G$, that
has 3 groups with sizes $\{ 300, 127,  10  \}$.

\begin{table}[h!]\centering \small
\begin{tabular}{ l c c c c  }
\hline
Case& $\mathbb{E}(K\mid data)$ &sd$(K\mid data)$ & MSE & LPML \\
\hline \rowcolor{black!20}$G$ & 2.95  & 1.03 & 1282.49 & -937.51 \\
$H$ & 3.56 & 2.36 & 682.98  &  -914.00\\
\hline
\end{tabular}
\caption{Posterior summaries for the tests on the Biopics dataset under a
linear dependent Dirichlet process mixture.}\label{tab:biopics_lddp}
\end{table}


An additional illustrative application, concerning data on air quality in
different locations in North and South America is reported in   Section
\ref{supp:airquality} of the Supplementary material.

\section{Conclusion}\label{sect:disc}
\label{sec:conc}

This work deals with mixture models where the prior has the property
of repulsion across location parameters. Specifically, the discussion is
centered on mixtures built on determinantal point processes (DPPs),
that can be constructed using a general spectral representation. The
methods work with any valid spectral density, but for the sake of
concreteness, illustrations were discussed in the context of the power
exponential case.

Though we limit ourselves to the case of isotropic  DPPs, inhomogeneous
DPPs can be obtained by transforming or thinning a stationary process.
However, we  believe that this case is not very interesting, unless there
is a strong reason to assume non-homogeneous locations a priori.

Our computational experiments and data illustrations show that the
repulsion induced by the DPP priors indeed tends to eliminate the annoying
case of very small clusters that commonly arises when using models that do
not constrain location/centering parameters. This happens with very small
sacrifice of model fit compared to the usual mixture models.

Another advantage of our model over DPMs is that we avoid the delicate
choice of the base measure of the Dirichlet process, leading to more
robust estimates on the number $K$ of components in the mixture.


\newpage

\begin{center}
 \textbf{\Large{Supplementary Material}}
\end{center}
\renewcommand{\thefigure}{S.\arabic{figure}}
\renewcommand{\thetable}{S.\arabic{table}}
\renewcommand{\thesection}{S.\arabic{section}}
\renewcommand{\theequation}{S.\arabic{equation}}
\renewcommand{\thesection}{S.\arabic{section}}

\setcounter {section} {0}
\section{Gibbs sampler for model  \eqref{eq:ver}-\eqref{eq:sigma}
}
\label{supp:nocov_Gibbs}

Posterior inference for our DPP mixture model as in
\eqref{eq:ver}-\eqref{eq:sigma} is carried out using a Gibbs sampler
algorithm. The full-conditionals are outlined below: we provide the
details of the computation only when the conditional posterior
distribution is not straightforward. In what follows, $rest$ refers to the
data and all parameters except for the one to the left of ``$\mid$''.
\begin{itemize}
\item The labels $\{ s_1, \dots, s_n\}$ are independently distributed
    according to a discrete distribution whose support is
    $\{1,2,\dots,K\}$:
    \begin{equation}\label{eq:sample_s}
    p(s_i=k\mid rest)\propto w_k \mathcal{N}\left(y_i; \mu_k, \sigma^2_k \right).
    \end{equation}
\item The distribution of the weights $\{w_1, \dots, w_K\}$ is
    conjugate: the conditional distribution is still a Dirichlet
    distribution, where the parameters are $\delta + n_k$,
    $k=1,\dots,K$.
\item The variances in each component of the mixture $\{\sigma^2_1,
    \dots, \sigma^2_K\}$ are generated independently according to the
    following distribution:
 $$ \sigma^2_k  \mid rest \sim inv-gamma\left(a_0 + \dfrac{n_k}{2},
 b_0 + \dfrac{1}{2}\sum_{i:\, s_i=k}\left( y_i - \mu_k \right)^2 \right), \quad k=1,\dots,K. $$
\item Sampling the means $\{\mu_1, \dots, \mu_K\}$ needs more care:
    following the reasoning in \cite{Xu_etal2016},  this
    full-conditional can be written as
 $$ p(\mu_1, \dots, \mu_K\mid rest) \propto
 det[\tilde{C} ](\{\tilde{\mu}_1, \dots, \tilde{\mu}_K\}; \rho, \nu)\prod_{k=1}^K
 \prod_{i:\, s_i=k} \mathcal{N}\left(y_i; \mu_k, \sigma^2_k \right)$$
 $$ \hspace{3.3cm} \propto \prod_{k=1}^K
 \left\{ \left(  \tilde{C}(\tilde{\mu}_k, \tilde{\mu}_k) - b\tilde{C}_{-k}^{-1}b^T  \right)
  \prod_{i:\, s_i=k} \mathcal{N}\left(y_i; \mu_k, \sigma^2_k \right)
    \right\}, $$ thanks to the Schur determinant identity. Note that
    $det[\tilde{C}](\{\tilde{\mu}_1, \dots, \tilde{\mu}_K \}; \rho,
    \nu)$ in the above expression follows from the expression of the
    density of a DPP on a compact set; see \eqref{eq:approx_densR}.
    Then, $b$ is a vector defined as $b=  \tilde{C}(\tilde{\mu}_k,
    \tilde{\mu}_{-k})$, $\tilde{\mu}_{-k} = \{
    \tilde{\mu}_{j}\}_{j\neq k}$ and $ \tilde{C}_{-k}^{-1}$ is a
    matrix of dimension $(K-1)\times (K-1)$ defined as $\tilde{C}
    \left(\tilde{\mu}_{-k}, \rho, \nu\right)$. Moreover,
    $\tilde{\mu}_{k} = T(\mu_k) $ is the transformed variable that
    takes values  on the set  $S=[-1/2,1/2]^d$. Typically, the
    rectangle $R$ such that $T(R)=S$ is fixed in such a way that it is
    large and contains abundantly all the datapoints.

    We update each mean $\mu_k$ separately for $k=1,\dots,K$ using a
    Metropolis-Hastings update.
\item The full-conditional for the parameters $(\rho, \nu)$ is as
    follows
$$p(\rho, \nu\mid rest) \propto  det\left(\tilde{C} \right)[\tilde{\mu}_1, \dots, \tilde{\mu}_K, \rho, \nu]
\exp\left(-\sum_{k=-N}^N \log(1+\tilde{\varphi}(k; \rho, \nu))
    \right)\pi(\rho,\nu).$$ The adaptive Metropolis-Hastings algorithm
    of \cite{roberts2009examples} is employed in this case, in order
    obtain a better mixing of the chains and to avoid the annoying
    choice of the parameters for the proposal distribution.
\item In order to sample $K$ we need a Reversible Jump step: standard
    proposals to estimate mixtures of densities with a variable number
    of components are based on moment matching
    \citep{richardson1997bayesian} and have been relatively often used
    in the literature. The idea is to build a proposal that preserves
    the first two moments before and after the move, as in
    \cite{Xu_etal2016}. In particular, the only possible moves are the
    splitting move, passing from $K$ to $K+1$, and the combine move,
    from $K$ to $K-1$.
\begin{itemize}
\item[(i)] \textbf{Choose move type:} uniformly choose among split
    and combine move (however, if $K=1$ the only possibility is to
    split)
\item[(ii.a)] \textbf{Combine:} randomly select a pair $(j_1,
    j_2)$ to merge into a new parameter indexed with $j_1$. The
    following relations must hold:
 $$ w^{new}_{j_1} = w_{j_1} + w_{j_2} $$
 $$ \quad w^{new}_{j_1}\mu^{new}_{j_1} = w_{j_1}\mu_{j_1} + w_{j_2}\mu_{j_2} $$
 $$  \qquad \quad w^{new}_{j_1}\left(\mu^{new2}_{j_1} +
 \sigma^{new2}_{j_1}\right) = w_{j_1}\left(\mu^{2}_{j_1} + \sigma^{2}_{j_1}\right)
  + w_{j_2}\left(\mu^{2}_{j_2} + \sigma^{2}_{j_2}\right) $$
\item[(ii.b)] \textbf{Split:} randomly select a component $j$ to
    split into two new components. In this case, we need to impose
    the following relationships:
 $$  w^{new}_{j_1} = \alpha w_{j}, \quad w^{new}_{j_2} = (1-\alpha) w_{j}  $$
 $$  \mu^{new}_{j_1} = \mu_j - \sqrt{\dfrac{w^{new}_{j_2}}{w^{new}_{j_1}}}r \left( \sigma^2_j\right)^{1/2}, \quad
     \mu^{new}_{j_2} = \mu_j - \sqrt{\dfrac{w^{new}_{j_1}}{w^{new}_{j_2}} }r \left( \sigma^2_j\right)^{1/2}$$
 $$ \sigma^{new2}_{j_1}= \beta (1-r^2)\dfrac{w_j}{w^{new}_{j_1}}\sigma^2_j,
 \quad   \sigma^{new2}_{j_2}= (1-\beta) (1-r^2)\dfrac{w_j}{w^{new}_{j_2}}\sigma^2_j$$
    where $ \alpha \sim Beta(1,1)$, $\beta \sim Beta(1,1)$ and $r
    \sim Beta(2,2)$.
\item[(iii)] \textbf{Probability of acceptance:} the proposed move
    is accepted with probability $\alpha=\min\left(1,
    \dfrac{1}{q(\mbox{proposed, old})} \right)$ if we selected a
    combine step, $\min\left(1, q(\mbox{old, proposed)} \right)$
    in the split case. In particular,
  $$  q(\mbox{old, proposed}) = |det(J)| \dfrac{p(K+1, \mathbf{w}^{new}, \boldsymbol{\mu}^{new},
   \boldsymbol{\sigma}^{2new} \mid\mathbf{y})}{
  p(K, \mathbf{w}^{old}, \boldsymbol{\mu}^{old}, \boldsymbol{\sigma}^{2old} \mid\mathbf{y})}
  \dfrac{p_{K+1}^{split}\dfrac{1}{(K+1)}}{(K+1)p_K^{comb}
p(\alpha)p(\beta)p(r)} $$  where
  $$  |det(J)| = \dfrac{w_j^4}{\left(w^{new}_{j_1}w^{new}_{j_2}\right)^{3/2}} (\sigma^2_j)^{3/2}(1-r^2)$$
  and $$   \dfrac{p(K+1, \mathbf{w}^{new}, \boldsymbol{\mu}^{new},
  \boldsymbol{\sigma}^{2new} \mid\mathbf{y})}{ p(K, \mathbf{w}^{old},
  \boldsymbol{\mu}^{old}, \boldsymbol{\sigma}^{2old} \mid\mathbf{y})}
  = \dfrac{\mbox{likelihood}(\mathbf{w}^{new}, \boldsymbol{\mu}^{new},
  \boldsymbol{\sigma}^{2new})}{ \mbox{likelihood}(\mathbf{w}^{old},
  \boldsymbol{\mu}^{old},
  \boldsymbol{\sigma}^{2old})}\dfrac{\pi(\sigma^{2new}_{j_1})\pi(\sigma^{2new}_{j_2})
  }{\pi(\sigma^{2}_{j})}
 $$
  $$\hspace{8cm} \times\dfrac{Dirichlet_{K+1}(\mathbf{w}^{new})}{Dirichlet_{K}(\mathbf{w}^{old})}
  \dfrac{det(\tilde{C}_{K+1})}{det(\tilde{C}_{K})}.$$
  Moreover, $p_{K+1}^{split} = 0.5 $ if $K>1$, 1 otherwise;
  $p_K^{comb} = 0.5 $ if $K>1$, 0 otherwise.

\end{itemize}

\end{itemize}

\section{Tests on data from a mixture with 8 components}
\label{supp:8mixt}

We simulated a dataset with $n = 100$ observations from a mixture of 8
components. Each component is the Gaussian density with mean $\theta_k$
and $\sigma^2_k = \sigma^2=0.05$: the means $\{\theta_k\}$ are evenly
spaced in the interval $(-10,10)$. In the model
\eqref{eq:ver}-\eqref{eq:sigma},  we set $a_0= 2.0025 $, $b_0= 0.050125$
so that $\mathbb{E}(\rho_0)=0.05$ and $Var(\rho_0)=1$; again, $s=0.5$ and
$\delta=1$.

\begin{table}[h!]
  \begin{center}
    \begin{tabular}{l c c c  c }
    \hline
 ~&
\multicolumn{4}{ c}{Prior specification}\\
    \hline
  Test & $\rho$ &  $\nu$  & $\mathbb{E}(K)$ & $V(K)$\\
\hline\rowcolor{black!20}    $S_0$ & 9.00 & 1 & 8.98 & 45.12    \\
    $S_1$ & 9 & 10 & 9 & 23.05    \\
 \rowcolor{black!20}$S_2$ &  $\mbox{  } a_\rho= 1, b_\rho=1$ & 1 & 1.94 & 1.99  \\
  $S_3$ &   $\mbox{  } a_\rho= 1, b_\rho=1$ & 2 & 2.18 & 1.99  \\
\rowcolor{black!20}$S_4$ &   $\mbox{  } a_\rho= 1, b_\rho=1$ & 10 & 2.74 & 2.17    \\
$S_5$ &   $\mbox{  } a_\rho= 1, b_\rho=1$ & discr(2,5,20) & 2.52 & 2.11  \\
\rowcolor{black!20}$S_6$ &  $\mbox{  } a_\rho= 1, b_\rho=1$ & discr($\mathcal{V}_1$) &  2.45 & 2.18    \\
$S_7$ &  $\mbox{  } a_\rho= 1, b_\rho=1$ &discr($\mathcal{V}_2$) &2.5 & 2.25 \\
\hline
    \end{tabular}\vspace{0.3cm}
 \begin{tabular}{l   c c c c}
     \hline
 ~&
\multicolumn{4}{ c}{Posterior summaries}\\
    \hline
  Test & $\mathbb{E}(K\mid data)$ & $V(K\mid data)$ & MSE & LPML\\
\hline\rowcolor{black!20}    $S_0$ &  7.98 & 0.20 &  4.65 & 2.39    \\
    $S_1$ &    7.99 & 0.19&  4.62  &3.10   \\
 \rowcolor{black!20}$S_2$ &  8.00  &0.17  &  4.62 &  3.66  \\
  $S_3$ &   7.991  & 0.16 & 4.62  & 3.03   \\
\rowcolor{black!20}$S_4$ &      7.99 & 0.17  &  4.63  & 2.96    \\
$S_5$ &  7.99 & 0.16  & 4.63  & 3.61  \\
\rowcolor{black!20}$S_6$ &  7.99 &  0.17 & 4.65  & 3.42   \\
$S_7$ &    7.99 & 0.18 & 4.63 & 3.36 \\
\hline
    \end{tabular}
\end{center}
\caption{Prior specification for $(\rho,\nu)$  and the corresponding mean
and variance induced on $K$ (top). Hyperparameters $(a_\rho, b_\rho)$
appear in \eqref{eq:prior_rho}, while $\mathcal{V}_1=\{ 1, 2, 5, 10, 20\}$
and $\mathcal{V}_2 = \mathcal{V}_1 \cup \{0.5,3,15,30,50\}$. Posterior
summaries for the simulated dataset from a mixture with 8 components are
in the bottom subtable.} \label{tab:test_sim8}
\end{table}

\begin{figure}[h!]
  \begin{center}
       \includegraphics[width=0.4\textwidth,height=0.4\textwidth]{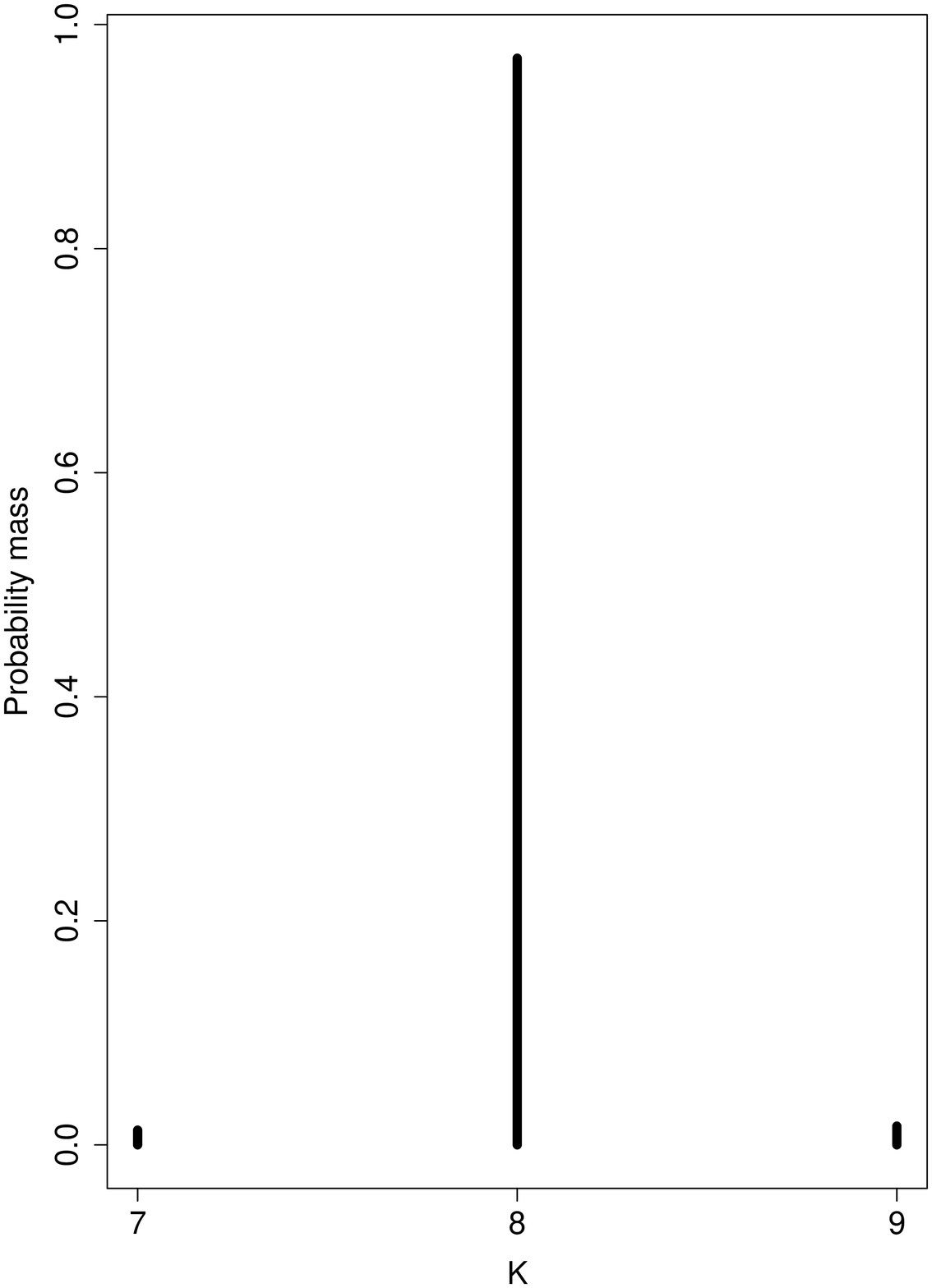}
       \includegraphics[width=0.4\textwidth,height=0.4\textwidth]{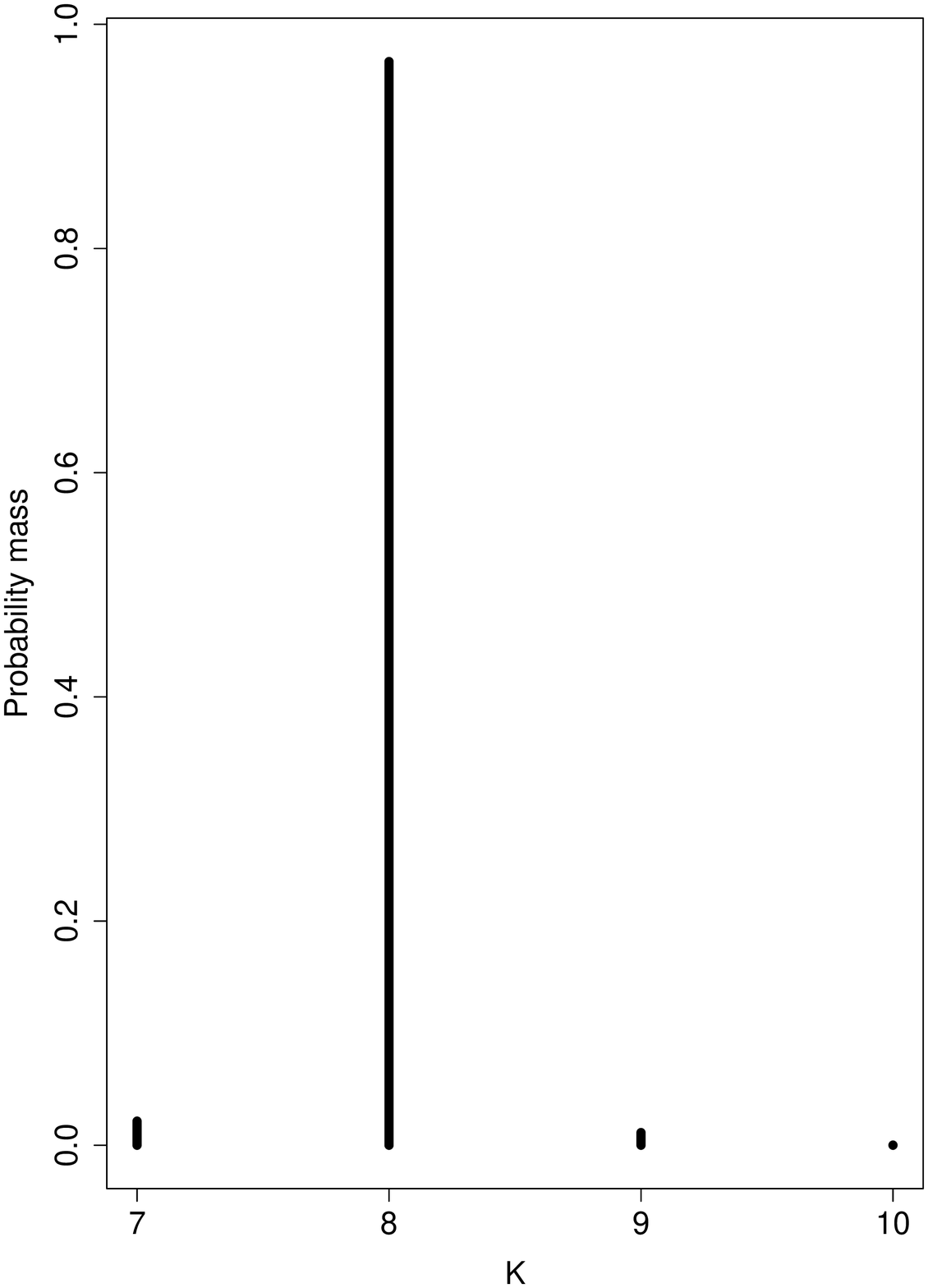}
  \end{center}
 \caption{Posterior distribution of $K$ for the simulated dataset from the mixture
of 8 components under Tests~$S_2$ (left) and $S_7$ (right) in
Table~\ref{tab:test_sim8}. }\label{fig:postk_8mixt}
\end{figure}

Table~\ref{tab:test_sim8} reports hyperparameters values for different
tests and posterior summaries  of interest, as well as prior mean and
variance of $K$. In particular, we show the posterior mean and variance
for the number of components $K$ (with which we assess the effectiveness
of the model for clustering), the mean squared error (MSE) and the
log-pseudo marginal likelihood (LPML) (that helps quantifying the
goodness-of-fit). In all cases we obtained a pretty satisfactory estimate
of the exact number of components, which is 8: the posterior is
concentrated around the true value with a very small variance. See also
Figure~\ref{fig:postk_8mixt}.

\begin{figure}[h!]
  \begin{center}
        \includegraphics[width=0.4\textwidth,height=0.45\textwidth]{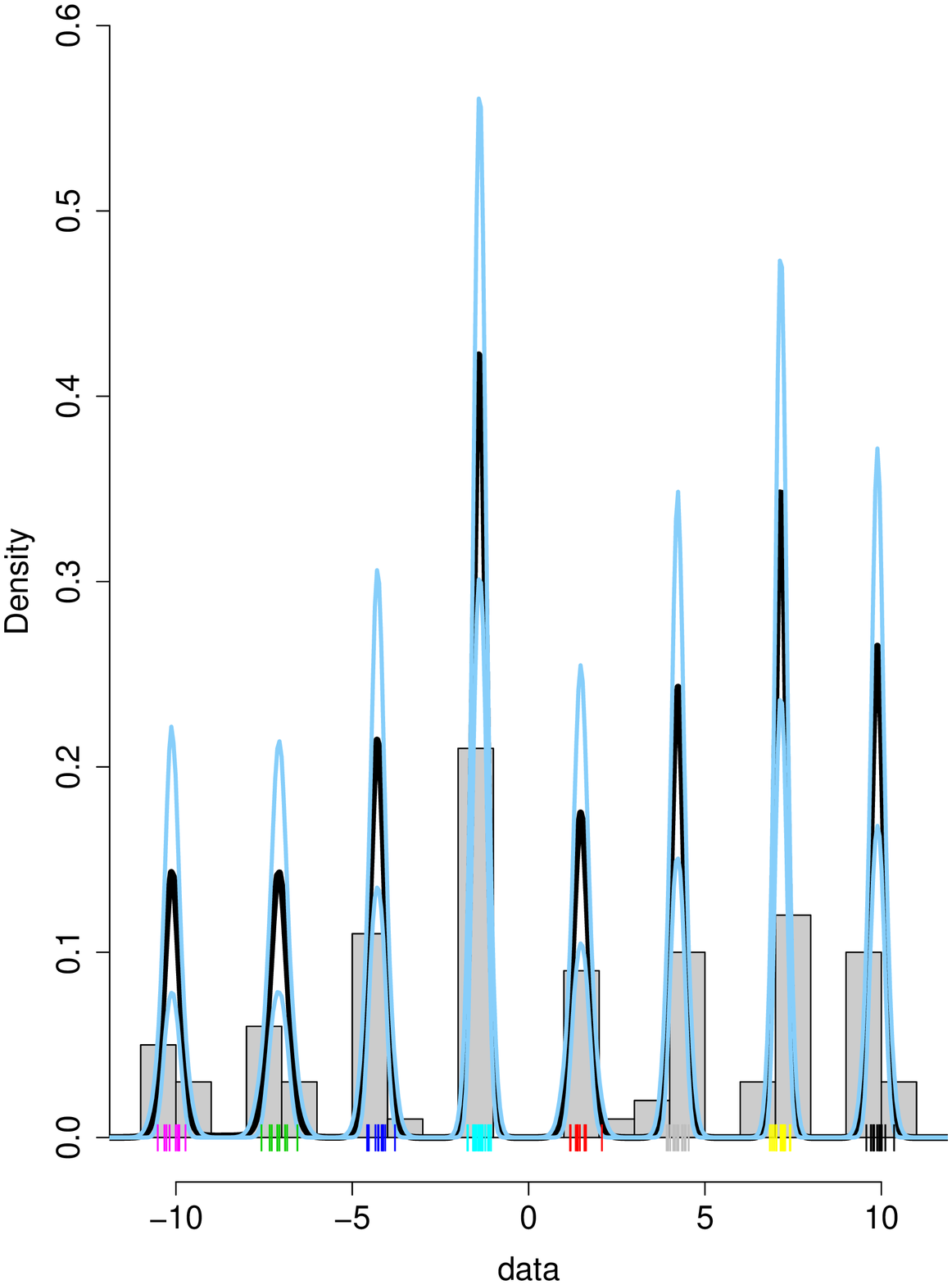}
        \includegraphics[width=0.4\textwidth,height=0.45\textwidth]{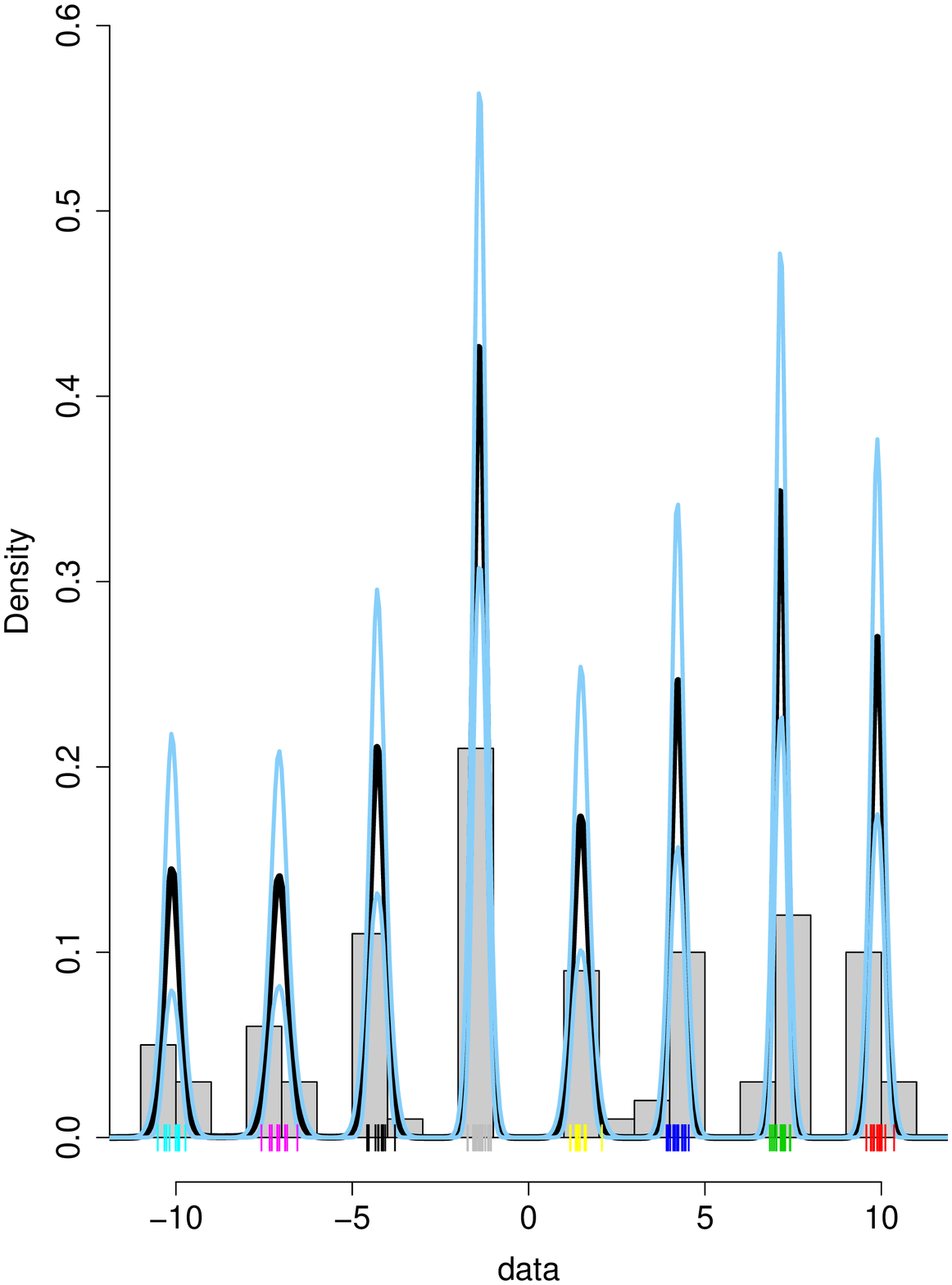}
    \end{center}
\caption{Density estimate and estimated partition for the simulated dataset from
the mixture of 8 components  under Tests~$S_2$ (left) and $S_7$ (right) in
Table~\ref{tab:test_sim8}. The points at the bottom of the density
estimate represent the data, and each color represents one of the eight
estimated clusters.} \label{fig:8mixt_dens}
\end{figure}

\begin{figure}[]
  \begin{center}
      \includegraphics[width=0.35\textwidth,height=0.4\textwidth]{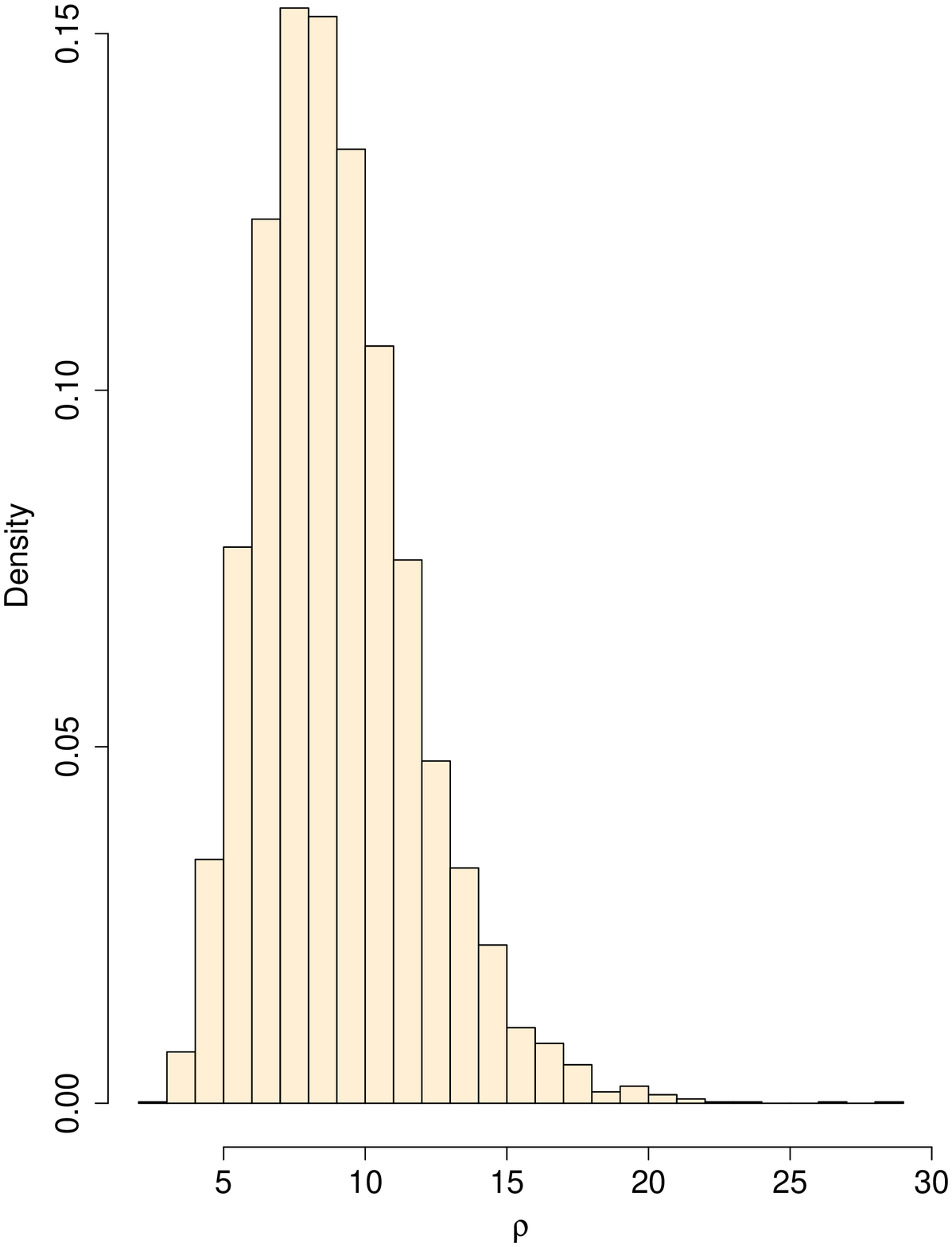}
     \includegraphics[width=0.35\textwidth,height=0.4\textwidth]{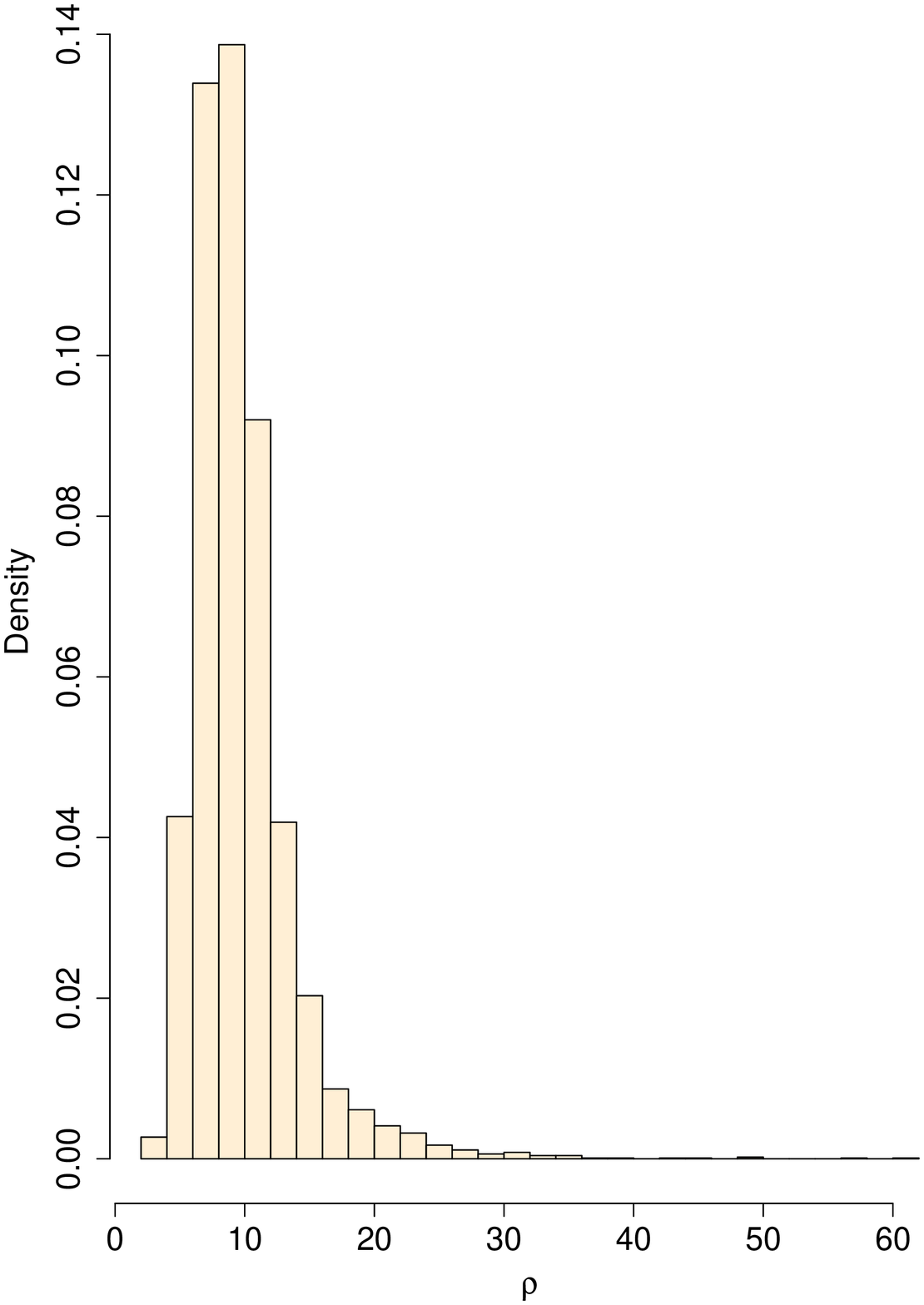}\\
    \end{center}
\caption{Posterior distribution of $\rho$  for the simulated dataset from the
mixture of 8 components  under Tests~$S_2$ (left) and $S_7$ (right) in
Table~\ref{tab:test_sim8}.} \label{fig:post_rho_8mixt}
\end{figure}

From the density estimation viewpoint, we have from
Table~\ref{tab:test_sim8} that both MSE and LPML are similar for all the
tests, thus indicating robustness with respect to the prior choice of
parameters $\rho$ and $\nu$. However, preferable tests seem to be $S_2$
and $S_7$; see  Figure~\ref{fig:8mixt_dens}, where density estimates and
estimated partitions for these two cases are displayed. The posterior
density of $\rho$ under Tests~$S_2$ and $S_7$ is shown in
Figure~\ref{fig:post_rho_8mixt}.

\section{Different spectral densities: application to the Galaxy dataset}
\label{supp:diff_spectral_dens}

We consider the proposed model with different spectral densities,
to check its robustness on the inference.
All the models presented in this paper are, as a matter of fact, general
and in principle any spectral density $\varphi(\cdot)$ satisfying the
conditions for the existence of the DPP process can be employed. The
choice of the spectral density in \eqref{eq:spectral_dens} is motivated by
its strong repulsiveness (see \citealp{Lav_etal2015}). However, in this
section we show inference on the Galaxy dataset obtained
when spectral representations other than the power spectral density, drive
the DPP.

We choose isotropic covariance functions that are well-known in
the spatial statistics literature: the Whittle-Mat\'ern and the
generalized Cauchy. Both densities depend on three parameters: intensity
$\rho>0$, scale $\alpha>0$ and shape $\nu>0$. In order to assure
$\varphi(x)<1$  for all $x$, $\rho$ must be smaller than
$\rho_{max}=\alpha^{-d}M$, where $M$ needs to be specified for each of the
two cases. For the Whittle-Mat\'ern we have
$$\varphi(x; \rho, \alpha, \nu) = \rho\dfrac{\Gamma(\nu+d/2)
 \left(2\alpha \sqrt{\pi} \right)^{d}}{\Gamma(\nu) \left(1+\| 2\pi\alpha x  \| \right)^{\nu+d/2}},
\qquad M = \dfrac{\Gamma(\nu)}{2^d\pi^{d/2} \Gamma(\nu+d/2) }$$
and for the generalized Cauchy
$$  \varphi(x; \rho, \alpha, \nu) = \rho\dfrac{2^{1-\nu}
\left(  2\sqrt\pi \right)^d}{\Gamma(\nu+d/2)}\| 2\pi\alpha x
\|^{d}\mathcal{K}_\nu(\| 2\pi\alpha x\|), \qquad M =
\dfrac{\Gamma(\nu+d/2)}{\Gamma(\nu)\pi^{d/2}}$$ where $d$ is the dimension
of the space where $x$ lives ($d=1$ in what follows) and
$\mathcal{K}_\nu(\cdot)$  is the modified Bessel function of the second
kind.

We fix $\rho=\dfrac{1}{2}\rho_{max}$ and $(\alpha, \nu)$ equal to:
$(i)$ (0.1,0.1), $(ii)$ (0.1,2), $(iii)$ (1,0.1) in the tests below. 
To fit the Galaxy data to the model in Section \ref{subsec:nocov_model},
the selected hyperparameter values are $\delta=1$ and $(a_0, b_0)=(3,3)$ 
(see \eqref{eq:w} and \eqref{eq:sigma}).

\begin{table}\begin{center}
   \begin{tabular}{ l c c c c  c c }
\hline
 ~&
\multicolumn{6}{ c}{Whittle-Mat\'ern}\\
\hline
Test&$\mathbb{E}(K)$ &Var$(K)$ & $\mathbb{E}(K\mid data)$ &Var$(K\mid data)$ & MSE & LPML\\
\hline
\rowcolor{black!20} $(i)$ & 10.21 & 17.29 &6.07 &1.09 & 73.67&-167.22 \\
$(ii)$ & 2.09 & 2.15&6.08&1.09 &73.89& -167.68  \\
\rowcolor{black!20}$(iii)$ & 3.53 & 9.87 &6.07 &1.10 & 75.80 &-167.33\\
\hline
\end{tabular}\medskip

   \begin{tabular}{ l c c c c c c  }
\hline
 ~&
\multicolumn{6}{ c}{Generalized Cauchy}\\
\hline
Test&$\mathbb{E}(K)$ &Var$(K)$ & $\mathbb{E}(K\mid data)$ &Var$(K\mid data)$ & MSE & LPML \\
\hline
\rowcolor{black!20} $(i)$ &5.65 & 14.49&6.09 & 1.09 &76.98&-166.60\\
$(ii)$ & 1.84 & 1.73&6.07 & 1.10& 75.75& -167.42 \\
\rowcolor{black!20}$(iii)$ & 0.25 & 0.06 &6.07  &1.12 & 80.66 &-169.84\\
\hline
\end{tabular}
\caption{Prior mean and variance of $K$ and posterior summaries for the
Galaxy dataset with Whittle-Mat\'ern (top) and generalized Cauchy (bottom)
spectral densities. }\label{tab:new_sd}
 \end{center}
\end{table}
Table \ref{tab:new_sd} displays posterior summaries for the two families
of spectral densities under hyperparameter settings $(i)$, $(ii)$ and
$(iii)$. Posterior summaries of the number of components $K$ and
goodness-of-fit values are close to those of Table~\ref{tab:test_galaxy}.
This gives evidence to robustness with respect to the choice of the
spectral density.


\section{Gibbs sampler in presence of covariates}\label{supp:cov_Gibbs}

The Gibbs sampler algorithm employed to carry out posterior inference for
model \eqref{eq:verx}-\eqref{eq:sigmagamma},
\eqref{eq:X}-\eqref{eq:rhonu}, \eqref{eq:clusterx}-\eqref{eq:beta} is
different from the one in Section~\ref{supp:nocov_Gibbs} except for the
full conditionals of $(\rho, \nu)$.  The sampling of labels
$\{s_i\}_{i=1}^n$ differs  from \eqref{eq:sample_s}, since now
$$ p(s_i=k\mid rest)\propto w_k(x_i) \mathcal{N}\left(y_i; \mu_k+x_i^T\gamma_k, \sigma^2_k \right)
\propto \exp\left(\beta^T_kx_i\right) \mathcal{N}\left(y_i;
\mu_k+x_i^T\gamma_k, \sigma^2_k \right).$$ The sampling of the
$\{\mu_k\}_{k=1}^K$ is similar as the same step in
Section~\ref{supp:nocov_Gibbs}, but now
$$  p(\mu_1, \dots, \mu_K\mid rest) \propto
det[\tilde{C} ](\{\tilde{\mu}_1, \dots, \tilde{\mu}_K\}; \rho,
\nu)\prod_{k=1}^K \prod_{i:\, s_i=k} \mathcal{N}\left(y_i - x_i^T\gamma_k;
\mu_k, \sigma^2_k \right). $$ However the substantial change from the
model without covariates to the model with covariates is due to the update
of $K$, the number of components in the mixture, and of  $\{ \beta_2,
\dots, \beta_K \}$ (recall that $\beta_1=0$ for identifiability reasons);
these are indeed complicated by the presence of the covariates. Moreover,
the update of $\{\sigma_k^2\}$  is now replaced  by
\begin{equation*}
 \begin{split}
&p(\gamma_k,\sigma_k^2 \mid rest)
\propto \prod_{i:\, s_i=k}\mathcal{N}\left(y_i; \mu_k + x_i^T \gamma_k, \sigma^2_k \right)
  \mathcal{N}_p\left(\gamma_k; 0,  \sigma^2_k \Lambda_0 \right)inv-gamma(\sigma^2_k;a_0,b_0)\\
  & \ \propto \dfrac{1}{(2\pi\sigma^2_k)^{n_k/2}}\exp\left( -\dfrac{1}{2\sigma^2_k}
  \sum_{i: \,  s_i=k} \left( y_i-\mu_k - x_i^T\gamma_k \right)^2  \right)
  \mathcal{N}_p\left(\gamma_k; 0,  \sigma^2_k \Lambda_0 \right)inv-gamma(\sigma^2_k;a_0,b_0)\\
 \end{split}
\end{equation*}
where  $n_k = \#\{i:\, s_i=k \}$;  here we assume the vector of the prior
mean of $\gamma_k$,  $\gamma_0$, to be equal to the 0-vector. This
full-conditional is the posterior of the standard conjugate normal
likelihood, normal - inverse gamma regression model.
In particular, we have that
$$  \gamma_k \mid\sigma^2_k, rest \sim \calN_p\left(m^*, \sigma^2_k \Lambda^* \right) $$
with $\Lambda^*= \left(\Lambda_0^{-1} + \sum_{i: \,s_i=k}x_i x_i^T
\right)^{-1}$ and $m^* = \Lambda^*\left( \sum_{i: \,  s_i=k}y_ix_i
\right)$. Moreover
$$ \sigma^2_k\mid rest \sim inv-gamma\left(a_0 + \dfrac{n_k}{2}, b_0 + \dfrac{1}{2}
\left(  \sum_{i: \,  s_i=k}y_i^2 - m^{*T}(\Lambda^{*})^{-1}m^*  \right)\right).$$

The full-conditional for the coefficients  $\beta_k$, $k=2,\dots,K$ is as
follows:
$$ p(\beta_2, \dots, \beta_K\mid rest) \propto \prod_{k=1}^K
 \left\{\prod_{i:\, s_i=k}\dfrac{\exp\left(\beta_k^Tx_i \right)}{\sum_{\ell=1}^K\exp\left(\beta_{\ell}^Tx_i \right)}
 \mathcal{N}_p(\beta_k; \beta_0, \Sigma_0)   \right\}$$
which has  no known  form. Therefore we resort to a Metropolis Hastings
step with a multivariate Gaussian proposal, centered in the current value
of the vector and with a diagonal covariance matrix, i.e.
$\zeta\mathbb{I}_{p \times p}$, where $\zeta$ is a tuning parameter chosen
to guarantee convergence of the chains.

On the other hand, the update of $K$ requires a Reversible Jump-type move.
However, the approach used in Section~\ref{supp:nocov_Gibbs} above is
difficult to implement when mixing weights depend on covariates, as in
this case,  so that we  need to find another way to define a transition
probability. Our approach is similar to that  of \cite{norets2015optimal},
with some differences that will be highlighted in the next lines.

As before, there are two available moves: split or combine. The
probability of proposing one of them is 0.5, except if $K=1$, when only
the split move can be proposed.

\medskip \medskip
\noindent \textbf{Split:} if this move is picked, $K^{prop} = K+1$, so
that we need to create a new group and its corresponding parameters (the
other parameters are kept fixed):
\begin{itemize}
\item[(i)] randomly pick one cluster, say $j$, containing at least two
    items
\item[(ii)]  randomly  divide data associated to this group,
    $\tilde{y}_j$, into two subgroups, $\tilde{y}_{j_1}$ and
    $\tilde{y}_{j_2}$;
\item[(iii)] set $\gamma_{j_1}=\gamma_j$, $\sigma^2_{j_1} =
    \sigma^2_j$, $\beta_{j_1} = \beta_j$, $\mu_{j_1} = \mu_j$. Now we
    need to choose a value for $\gamma_{j_2}$, $\sigma^2_{j_2}$,
    $\beta_{j_2}$ and $\mu_{j_2}$. In \cite{norets2015optimal}, this
    is done by sampling the new values from the posterior,
    conditioning also on the other parameters (even if, for practical
    purposes, Gaussian approximations for conditional posteriors are
    used in the implementation of the algorithm).
    Instead, we sample $\left(\mu_{j_2}, \gamma_{j_2}, \sigma^2_{j_2}
    \right)$ from the posterior of the following auxiliary model
    \begin{eqnarray*}
  \tilde{y}_{j_2}\mid \mu, \gamma, \sigma^2 &\stackrel{iid}{\sim}& \mathcal{N}(\mu+\tilde x_{j_2}^T\gamma, \sigma^2)   \\
\tilde{\gamma} = \begin{bmatrix}
         \mu \\
         \gamma \\
        \end{bmatrix}\mid \sigma^2 &\sim& \mathcal{N}_{p+1}(0, \sigma^2 \Gamma_0)\\
         \sigma^2 &\sim& inv-gamma(\xi_0, \nu_0)
   \end{eqnarray*}
    where $\tilde{x}_{j_2}$ and $\tilde{y}_{j_2}$ represent covariates
    and responses in the new group with label $j_2$, respectively.
    Parameter $\beta_{j_2}$ is sampled from a $p$-dimensional Gaussian
    distribution with mean $\beta^{mode}$ and variance covariance
    matrix $\Sigma^{mode}$. In particular, $\beta^{mode}$ is the
    argmax of the following expression
$$ \prod_{i:\, s_i=j_2}\dfrac{\exp\left( \beta_{j_2}^T x_i  \right)}{\exp\left( \beta_{j_2}^T x_i  \right)
+ \sum_{j \neq j_2}\mathbb{E}\left(
\exp\left( \beta_{j}^T x_i   \right)\right)}\mathcal{N}_p\left(\beta_{j_2};
\beta_0, \Sigma_0 \right),$$ which corresponds to  an approximation of
    the full-conditional of the $\beta_k$ (we dropped the dependence
    on the other $\beta_j$'s by considering the expected value in the
    denominator). Note that $\mathbb{E}\left(\exp\left( \beta_{j}^T
    x_i   \right)\right)$ is not other than the moment generating
    function, thus it is equal to $\exp\left(\beta_0^Tx_i + x_i^T
    \Sigma_0 x_i/2 \right)$.
\end{itemize}

\medskip \medskip
\noindent \textbf{Combine:} here $K^{prop} = K-1$, so that it suffices to
collapse two groups into one. Specifically, we randomly choose one group
to delete, say $j_1$, and remove the corresponding parameters $\beta_{j_1}
$, $\mu_{j_1}$ and $\sigma^2_{j_1}$. Then, we choose another group, $j_2$,
and assign all the data $\tilde{y}_{j_1}$ to it.

\medskip \medskip
\noindent \textbf{Acceptance rate:} this is simply given by
$$ \alpha(K \rightarrow K+1) = \dfrac{ p(y\mid K+1, \theta_{K+1})
\pi(K+1, \theta_{K+1}) }{p(y\mid K, \theta_{K}) \pi(K, \theta_{K})}
\dfrac{1}{f(\mu_{j_2}, \gamma_{j_2},\sigma^2_{j_2}, \beta_{j_2})}
\dfrac{p_{K+1}^S}{p_{K+1}^C} \dfrac{p_{c}(j_1, j_2)}{p_{s}(j)}$$
$$ \alpha(K \rightarrow K-1) = \dfrac{ p(y\mid K-1, \theta_{K-1})
\pi(K-1, \theta_{K-1}) }{p(y\mid K, \theta_{K}) \pi(K, \theta_{K})}
f(\mu_{j_1}, \gamma_{j_1}, \sigma^2_{j_1}, \beta_{j_1})
\dfrac{p_{K-1}^C}{p_{K}^S} \dfrac{p_{s}(j)}{p_{c}(j_1, j_2)}$$ where
$\theta_K = (\sigma^2_{1:K}, \gamma_{1:K}, \mu_{1:K}, \beta_{1:K})$.
Moreover, $p_{s}(j) $ is the probability of splitting component $j$ and
similarly for the other terms.


\section{Air quality index dataset}
\label{supp:airquality}

The Air Quality Index (AQI) is an index for reporting air quality, see for
instance \url{https://airnow.gov/index.cfm?action=aqibasics.aqi}. It
describes how clean or polluted the air is, and what associated health
effects might be a concern for the population. The Environmental
Protection Agency calculates the AQI for five major air pollutants
regulated by the Clean Air Act: ground-level ozone, particle pollution,
carbon monoxide, sulfur dioxide, and nitrogen dioxide. Data can be
obtained from several sources,  for instance,  from
\url{http://aqicn.org/}. For a real-time map, see
\url{https://clarity.io/airmap/}.

For the purpose of this illustration, we investigate the spatial relations
in measurements of the AQI made on September 13th, 2015, at 16pm.
We  consider 1147 locations scattered around North and South America (the values of AQI have been
standardized). 


We ran the MCMC algorithm to fit model
\eqref{eq:ver}-\eqref{eq:rhonu}, \eqref{eq:sigma}-\eqref{eq:beta}, with a
burn-in  of 10,000, a thinning of 10 and a final sample size of 5,000. As
before, $\beta_0 = 0$ and $\nu = 2$. Table \ref{tab:post_aqi} displays
different settings of the hyperparameters for which the prior mean on the
number of groups is 1.996 and the prior standard deviation is 1.290
(computed using a Monte Carlo approach). The different hyperparameter
settings differ for the specification of $\phi$, the scale hyperparameter
in the $g$-prior in \eqref{eq:beta}, and the prior mean $m$ and variance
$v$ of $\sigma^2_k$; see \eqref{eq:sigma}.

\begin{table}[h!]
  \begin{center}
    \begin{tabular}{ l  c  c   c c  c  c c }
    \hline
  Test  & $\phi$ & $m$ & $v$  &mean($K$) & sd($K$) & MSE & LPML\\
  \hline
\rowcolor{black!20}  $AQ_1$ &  1000 & 2 &1 & 6.999 & 1.469 & 861.048 &  -1101.013\\ 
  $AQ_2$ &  500&10 &$+\infty$& 5.192 & 1.143 & 870.689 & -1235.988\\ 
\rowcolor{black!20}  $AQ_3$ &  1000& 0.1& 1&  9.045 & 2.243 &  840.0685 & \textbf{-1071.931}\\ 
  $AQ_4$ &   500 & 5&$+\infty$& 7.811 & 2.665 & \textbf{835.596} &-1160.631 \\ 
  \hline
    \end{tabular}
\end{center}
\caption{Prior specification  for the Air quality index dataset. The scale
parameter $\phi$ appears in the $g$-prior specification of
\eqref{eq:beta}, while $m$ and  $v$ denote prior mean and variance,
respectively,  of $\sigma^2_k$ as in \eqref{eq:sigma}.}
\label{tab:post_aqi}
\end{table}

\begin{figure}[h!]
  \begin{center}
        \includegraphics[width=0.47\textwidth]{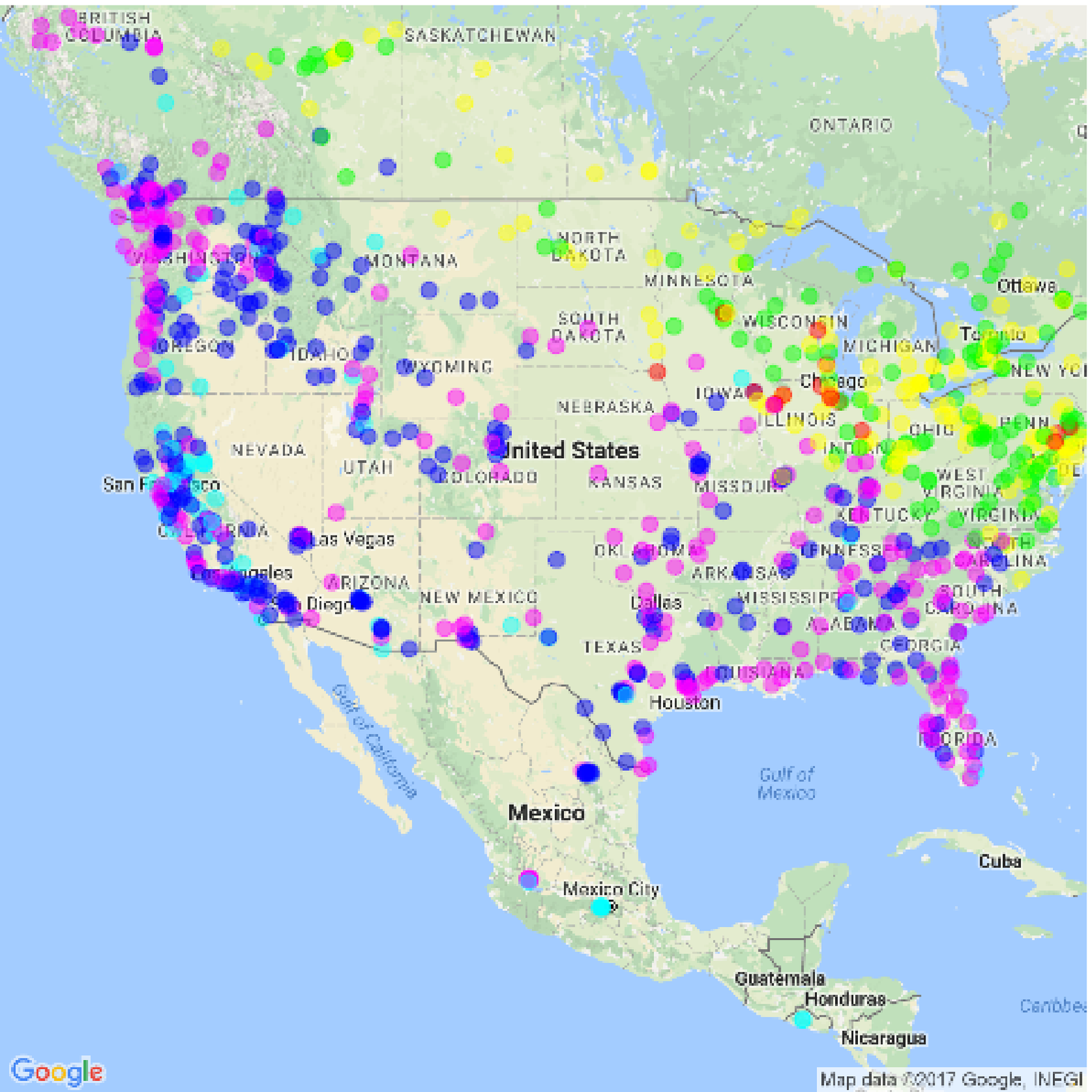}
        \includegraphics[width=0.47\textwidth]{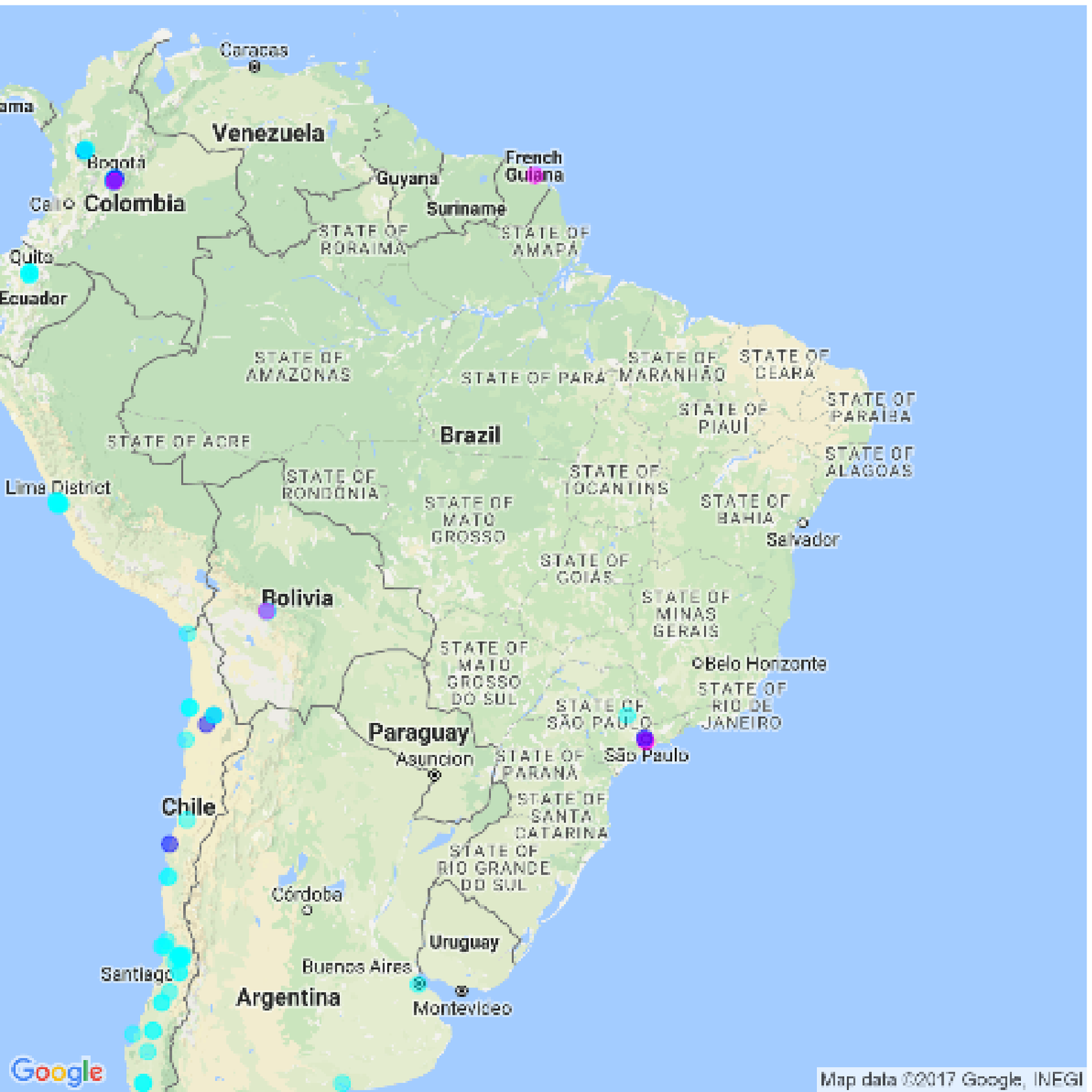}\\
   \includegraphics[width=0.47\textwidth, height = 0.55\textwidth,]{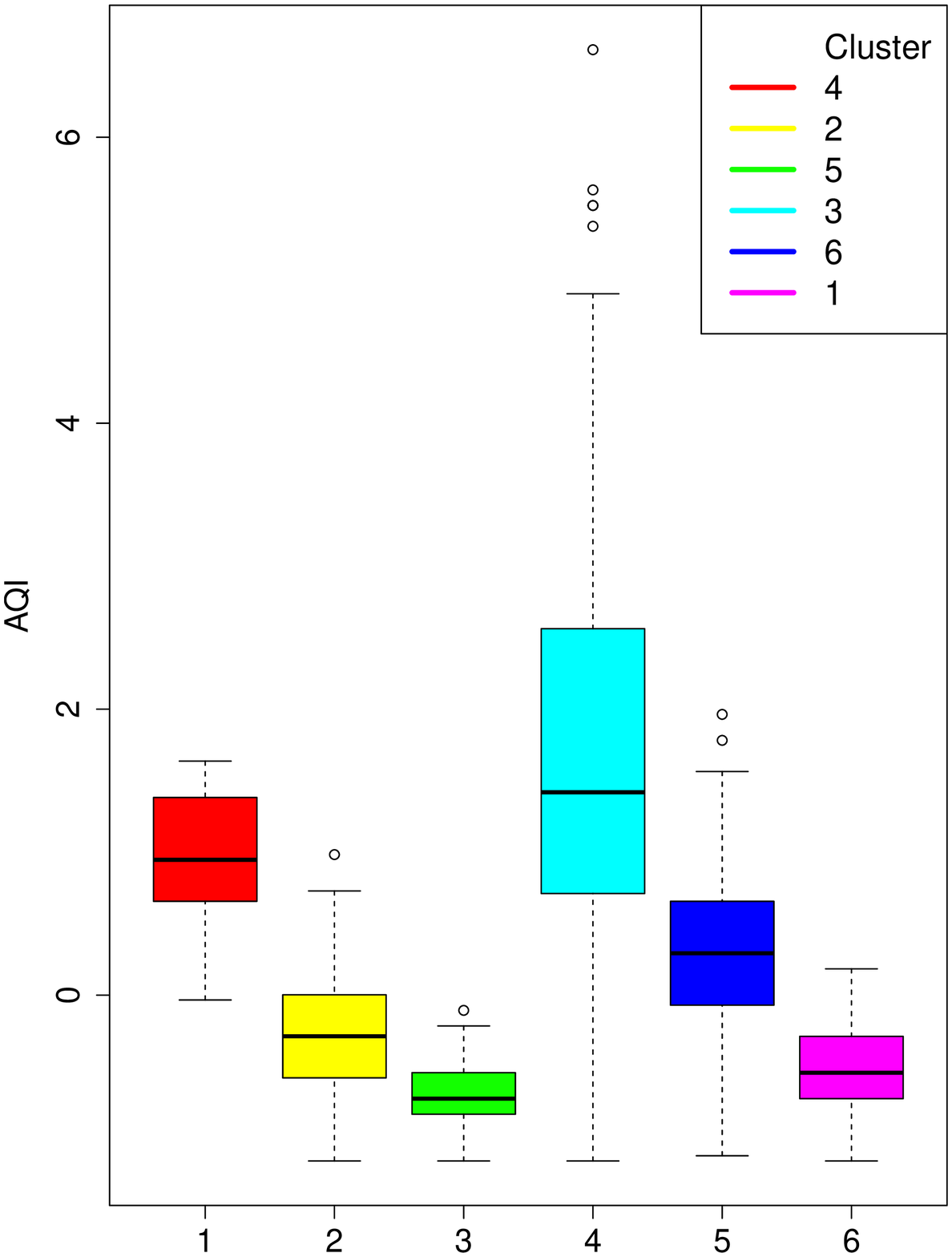}\\
\end{center}
\caption{Estimated partition of the Air Quality index dataset under hyperparameters
as those in  Test~$AQ_1$ in Table~\ref{tab:post_aqi}. The number of
estimated clusters is 6, each denoted by a different color, with  sizes
17, 221, 183, 136, 306, 284, respectively.} \label{fig:clu_aqi}
\end{figure}

Figure~\ref{fig:clu_aqi} shows the estimated clusters  obtained under
Test~$AQ_1$. The north - east coast seems to be associated with better
environmental  conditions, and it is clear that important urban sprawls
are generally grouped together. More in detail, the Binder loss function
method estimated 6 groups characterized by the following mean and standard
deviations of the AQI: (0.95, 0.44) in the red group, (-0.27, 0.45) in the
yellow, (-0.70, 0.21) in the green, (1.7,1.64) in the light blue, (0.28,
0.54) in the blue, (-0.51, 0.29) in the pink group; yellow, pink and green
points are associated to lower values of AQI, while red and light blue to
higher values. The boxplots of the AQI by cluster in
Figure~\ref{fig:clu_aqi} are clearly interpretable: the cluster depicted
in light blue gathers the polluted cities in south America and big cities
in the West coast of the U.S. (Las Vegas, Los Angeles, Seattle, for
instance). On the other hand, yellow and green points indicate
less dangerous environmental conditions that characterize the North-East
coast: however, the small red cluster contains the big cities that are
present in this area (Chicago, New York, Philadelphia, Boston).

\begin{figure}[]
  \begin{center}
        \includegraphics[width=0.32\textwidth]{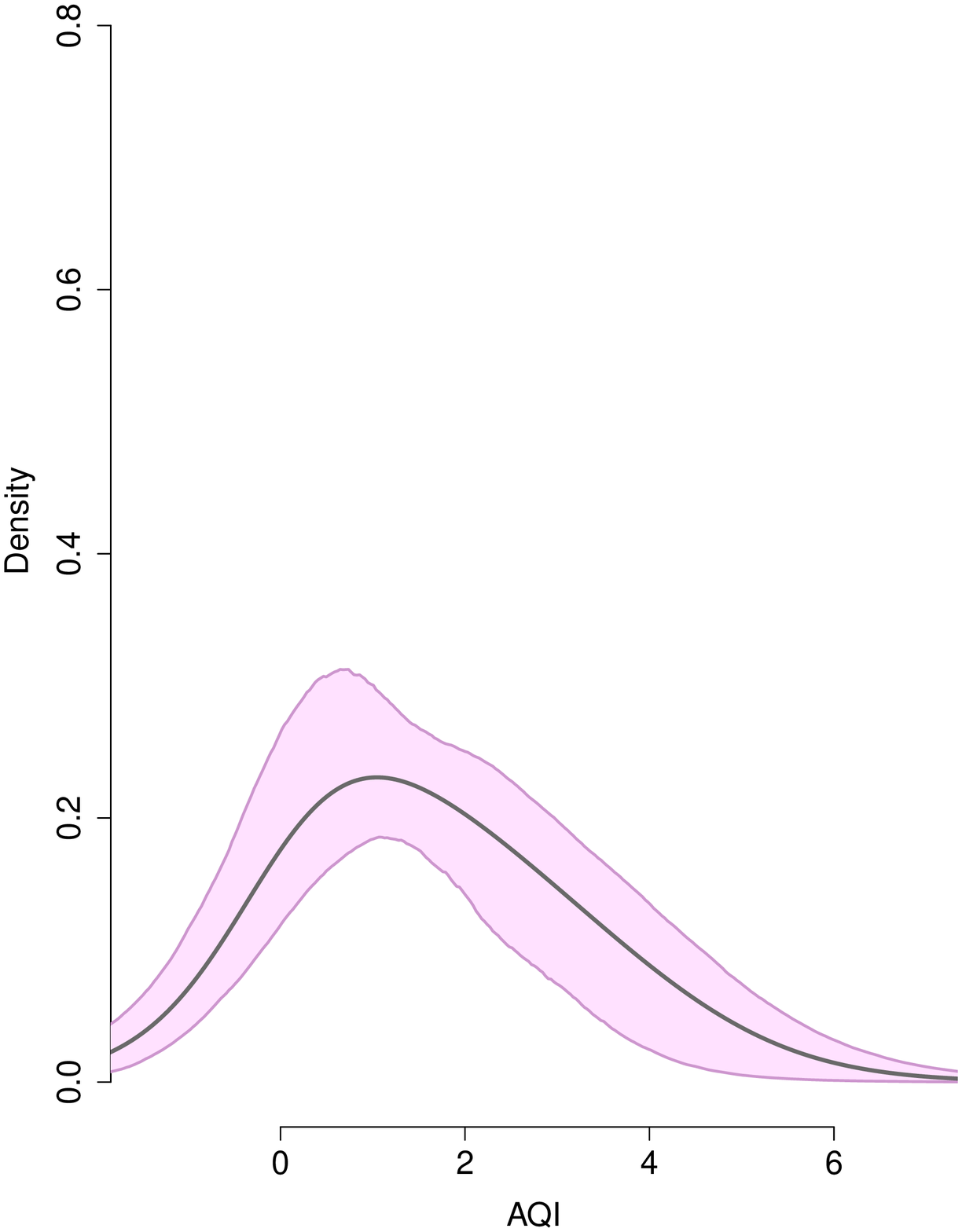}
        \includegraphics[width=0.32\textwidth]{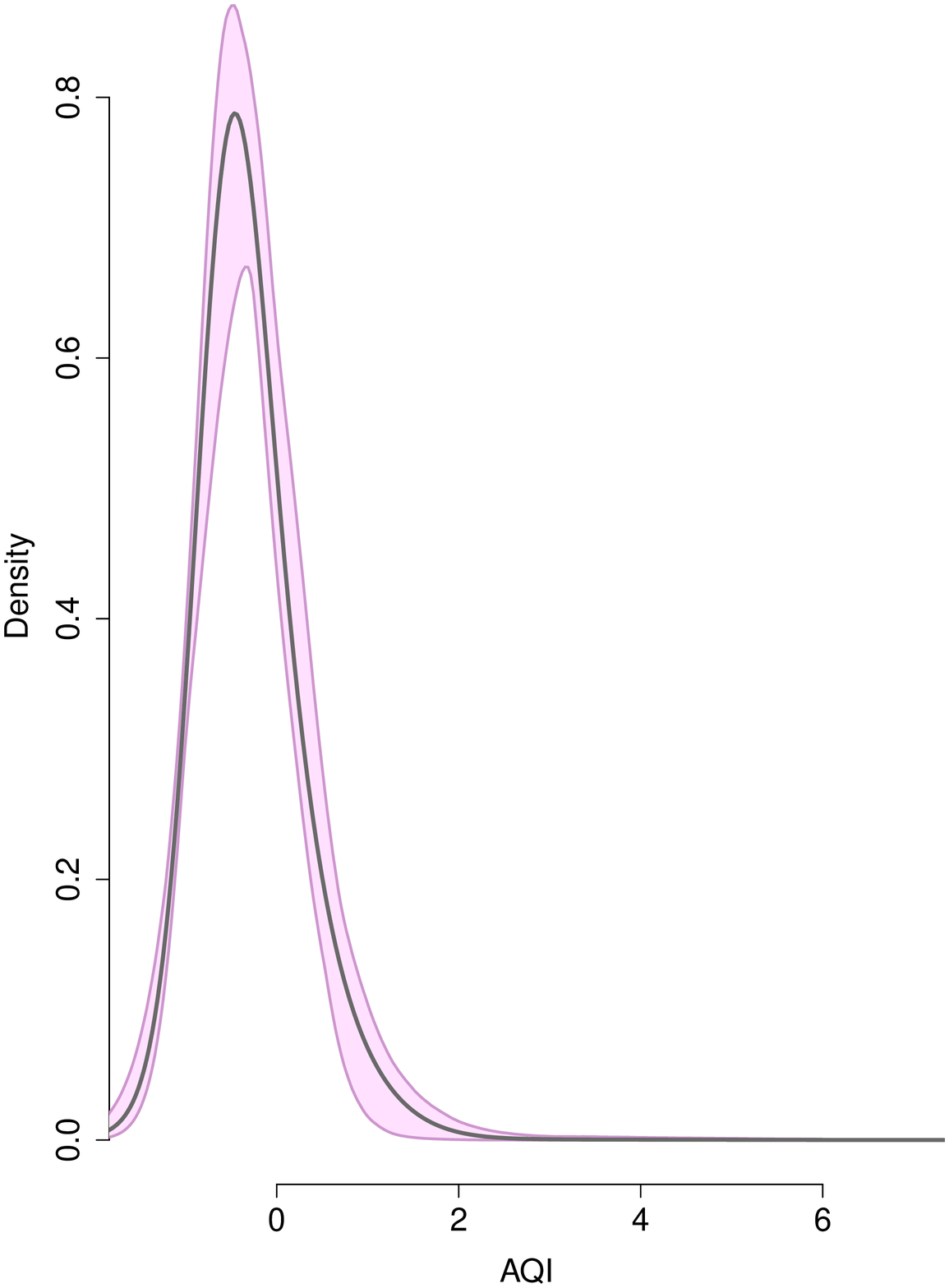}
        \includegraphics[width=0.32\textwidth]{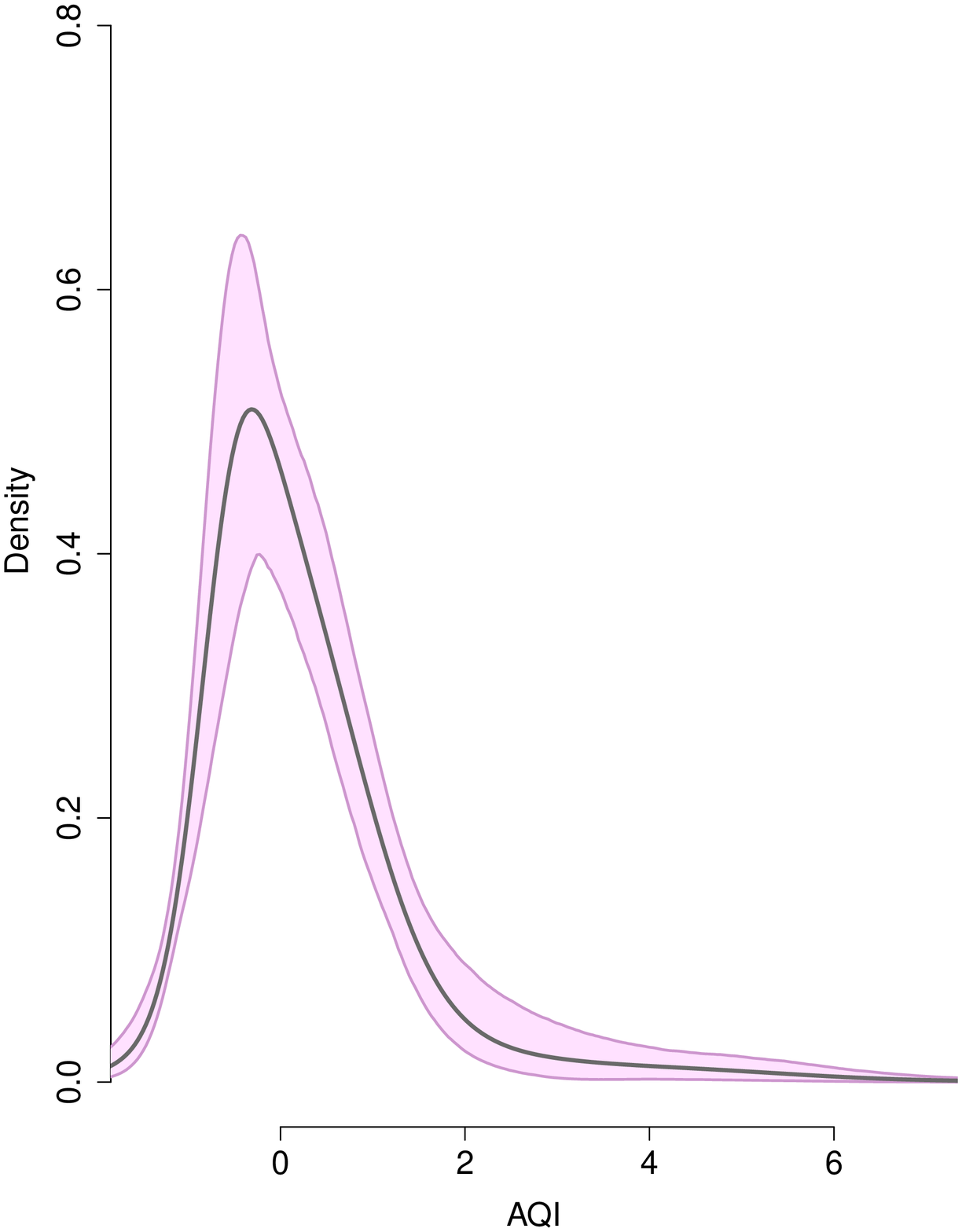}
\end{center}
\caption{Predictive distribution corresponding to 3 different locations
(New York, Sacramento, Monterrey) under Test~$AQ_4$ in
Table~\ref{tab:post_aqi} for the Air Quality Index dataset. }
\label{fig:pred_comp_aqi}
\end{figure}

Figure~\ref{fig:pred_comp_aqi} displays three different predictive laws
that correspond to different locations: Sacramento, which shows the lowest
predicted values of AQI, New York, where the environmental conditions are
worse, and Monterrey, that presents an intermediate situation.

Similarly as for the Biopics dataset, we compare the inference under our
model with the linear dependent Dirichlet process mixture model introduced
in \cite{de2004anova}. Prior information is fixed as follows:
$\alpha$ is distributed according to the $gamma(1,1)$ distribution for
Test~$AQ_5$, so that the prior mean and variance of $K$ are 7.15 and 36,
respectively, i.e. the prior of $K$ is vague. On the other hand, in Test
$AQ_6$ the mass parameter $\alpha$ of the Dirichlet process is set equal
to 0.15 such that $\mathbb{E}(K) = 2.09$ and $Var(K)=1.02$, which
approximately matches the prior information given on $K$ (mean 1.996
and variance 1.29). The baseline distribution is a multivariate Gaussian
with mean vector 0 and a random covariance matrix which is given a
non-informative prior and the hyperparameters of the inverse-gamma
distribution for the variances of the mixture components are such that
prior mean and variance are equal to 5 and 1, respectively. Posterior
summaries can be found in Table \ref{tab:aqi_lddp}.

\begin{table}[h!]\centering \small
\begin{tabular}{ l c c c c  }
\hline
Test& $\mathbb{E}(K\mid data)$ &Var$(K\mid data)$ & MSE & LPML \\
\hline \rowcolor{black!20}$AQ_{5}$ & 5.14  & 0.38 & 827.72 & -1100.73 \\
$AQ_6$ & 5.03 & 0.16 & 827.04  & -1100.06 \\
\hline
\end{tabular}
\caption{Posterior summaries for the Air Quality Index dataset under the
linear dependent Dirichlet process mixture for two different prior
specifications.}\label{tab:aqi_lddp}
\end{table}

\section{Additional plots}
\begin{figure}[h!]
  \begin{center}
      \includegraphics[width=0.4\textwidth, height=0.45\textwidth]{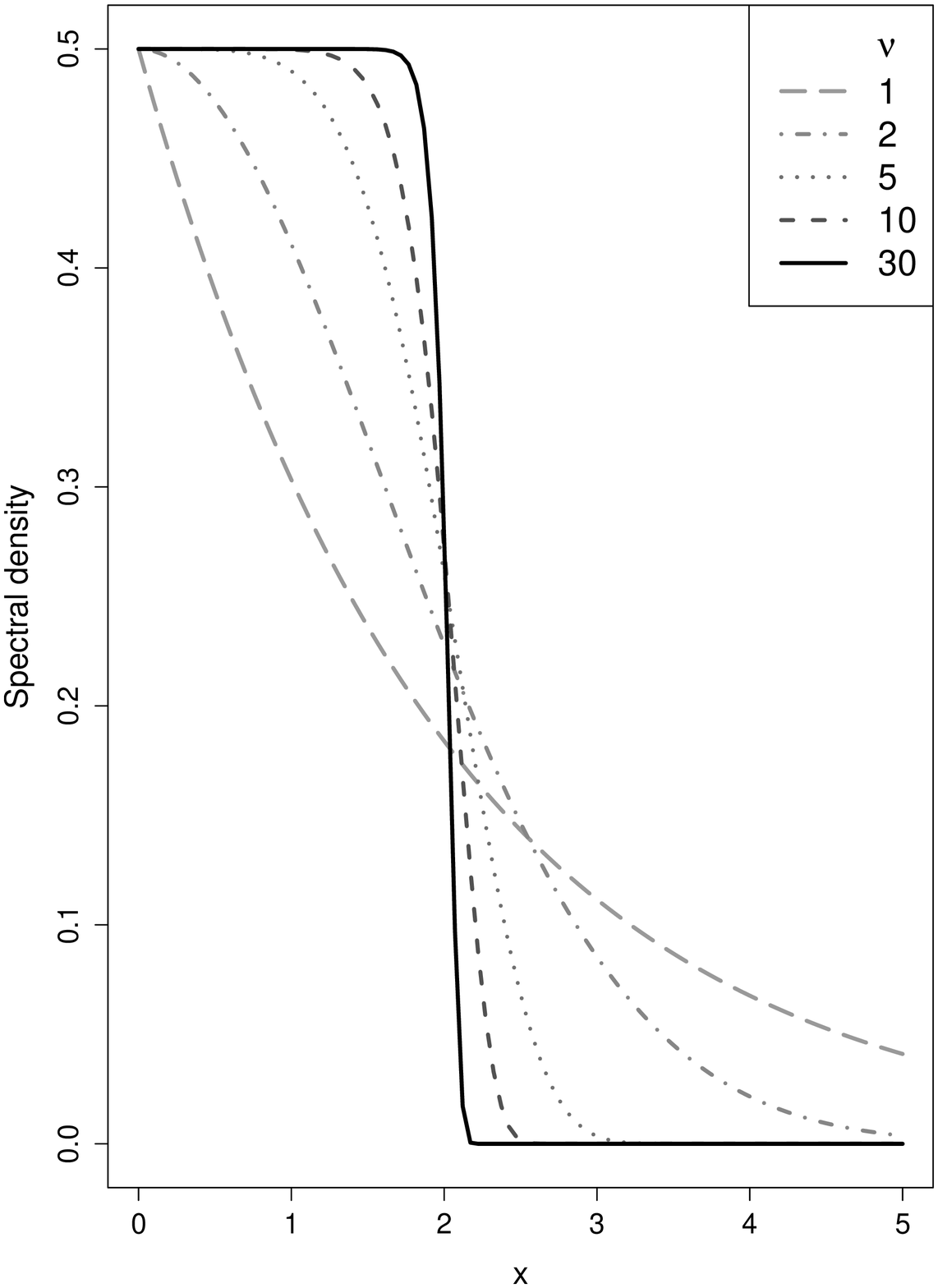}
      \includegraphics[width=0.4\textwidth,height=0.45\textwidth]{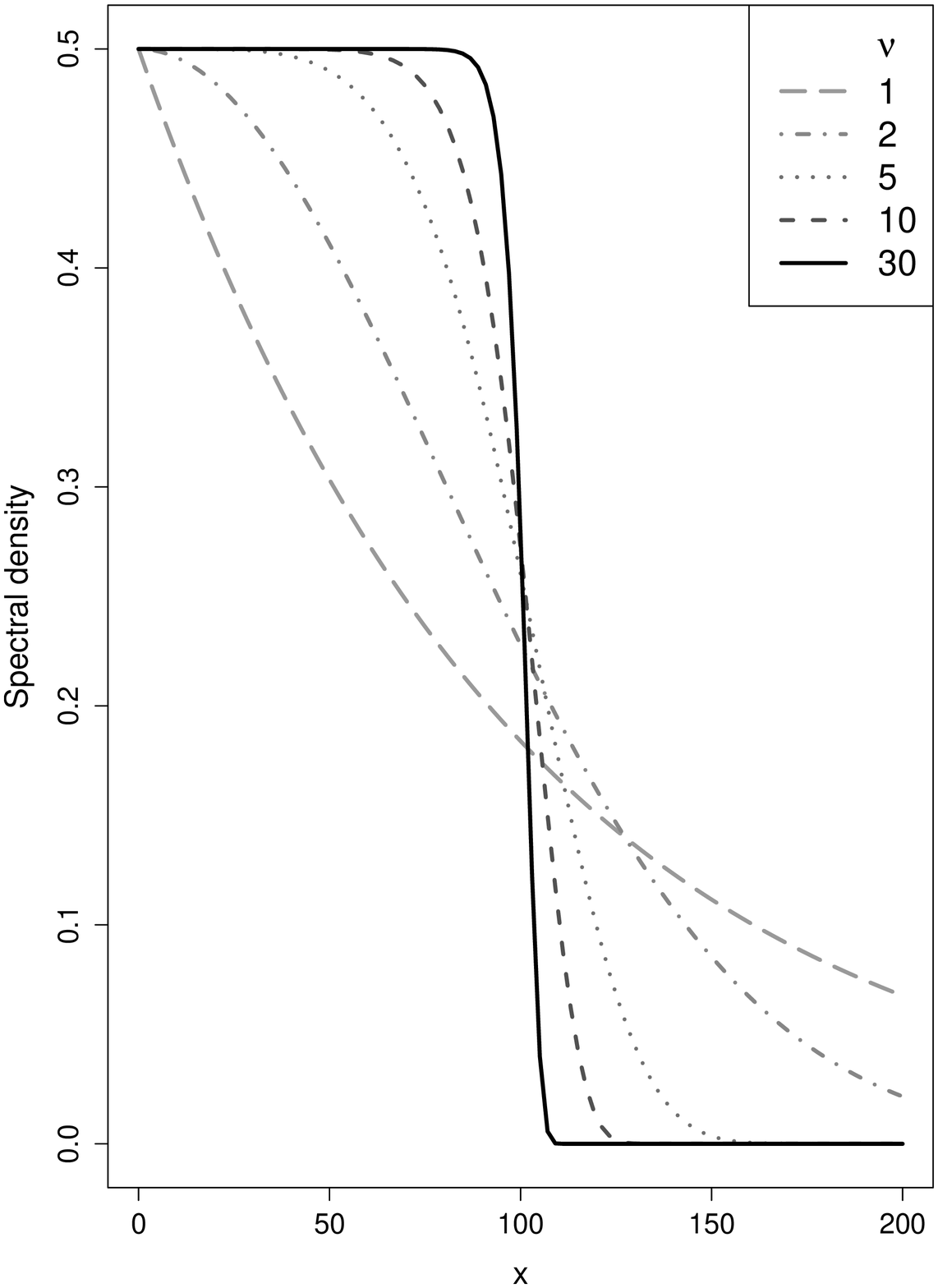}
\end{center}
\caption{Power exponential spectral density $\varphi(x; \rho, \nu)$ when $\rho$ is
2 (left) and 100 (right) and $\nu$ varies. }
 \label{fig:spectral_density}
 \end{figure}

\begin{figure}[]
\begin{center}
       \includegraphics[width=0.45\textwidth,height=0.5\textwidth]{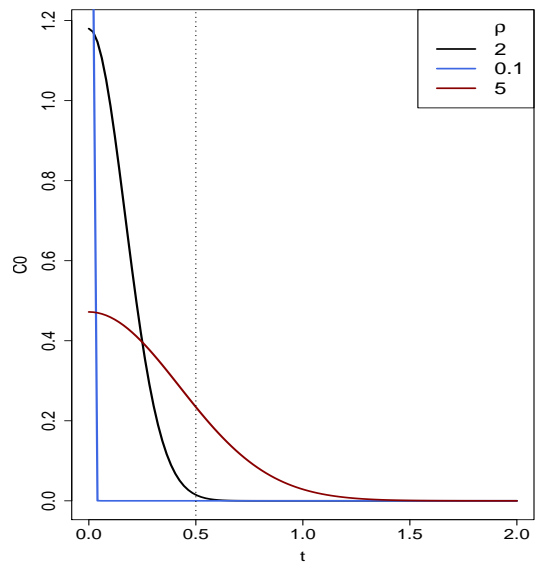}
\end{center}
\caption{Value of $C_0(t)$ corresponding to the Gaussian spectral density when $s=0.5$ and
$\rho$ is equal to  $0.1,2,5$.}
\label{fig:C0}
\end{figure}

\begin{figure}[]
  \begin{center}
        \includegraphics[width=0.32\textwidth,height=0.3\textwidth]{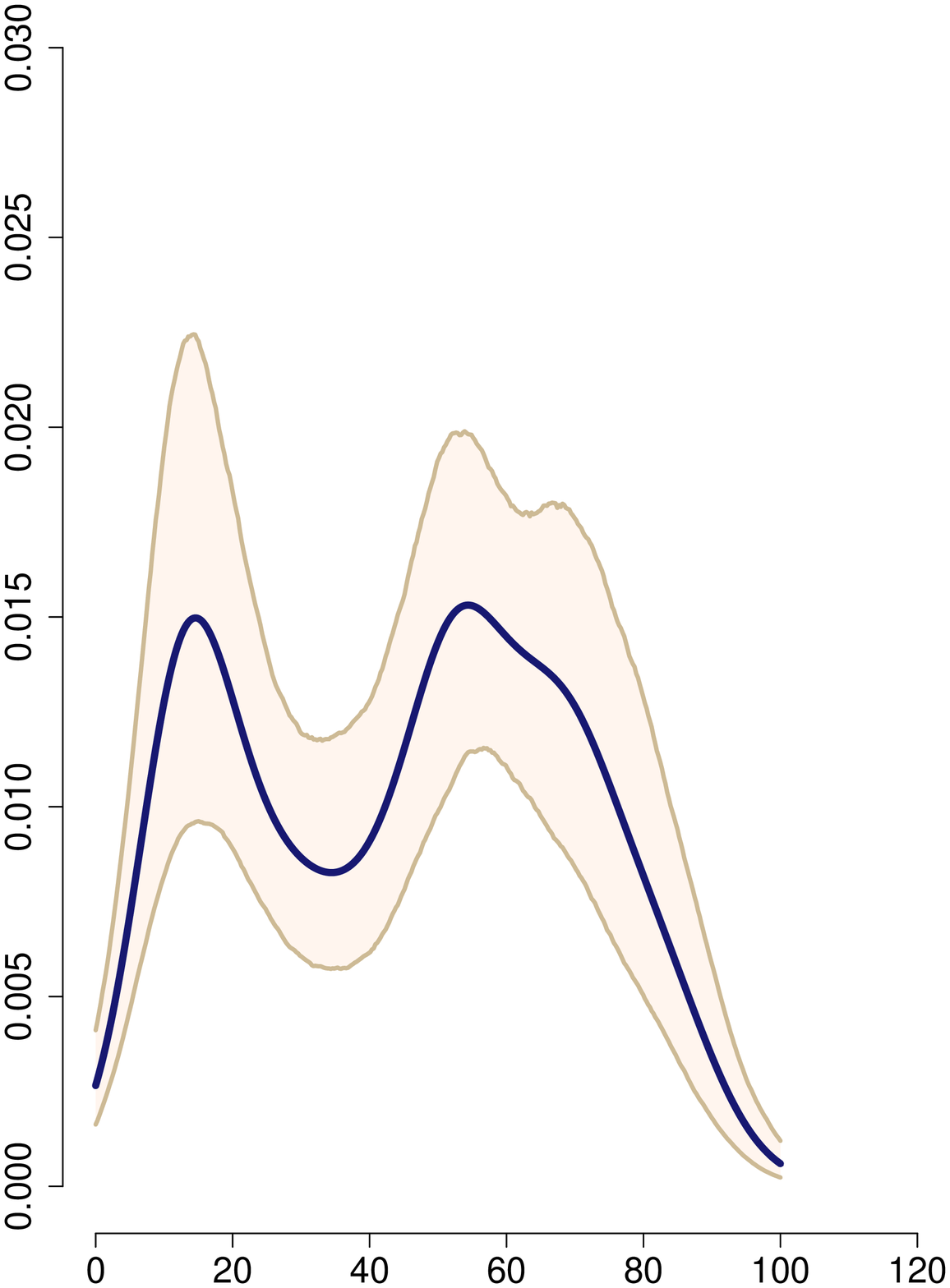}
       \includegraphics[width=0.32\textwidth,height=0.3\textwidth]{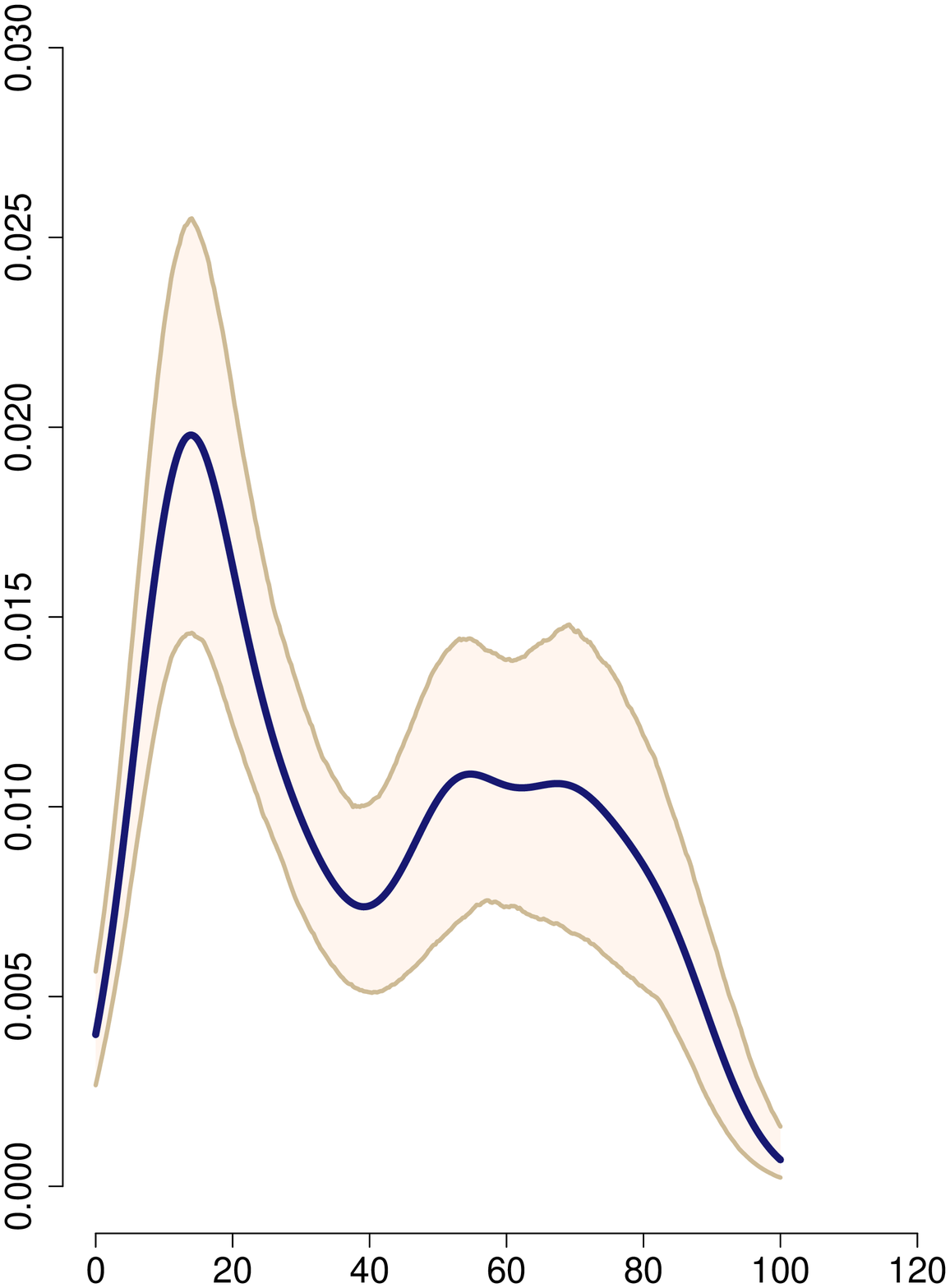}
        \includegraphics[width=0.32\textwidth,height=0.3\textwidth]{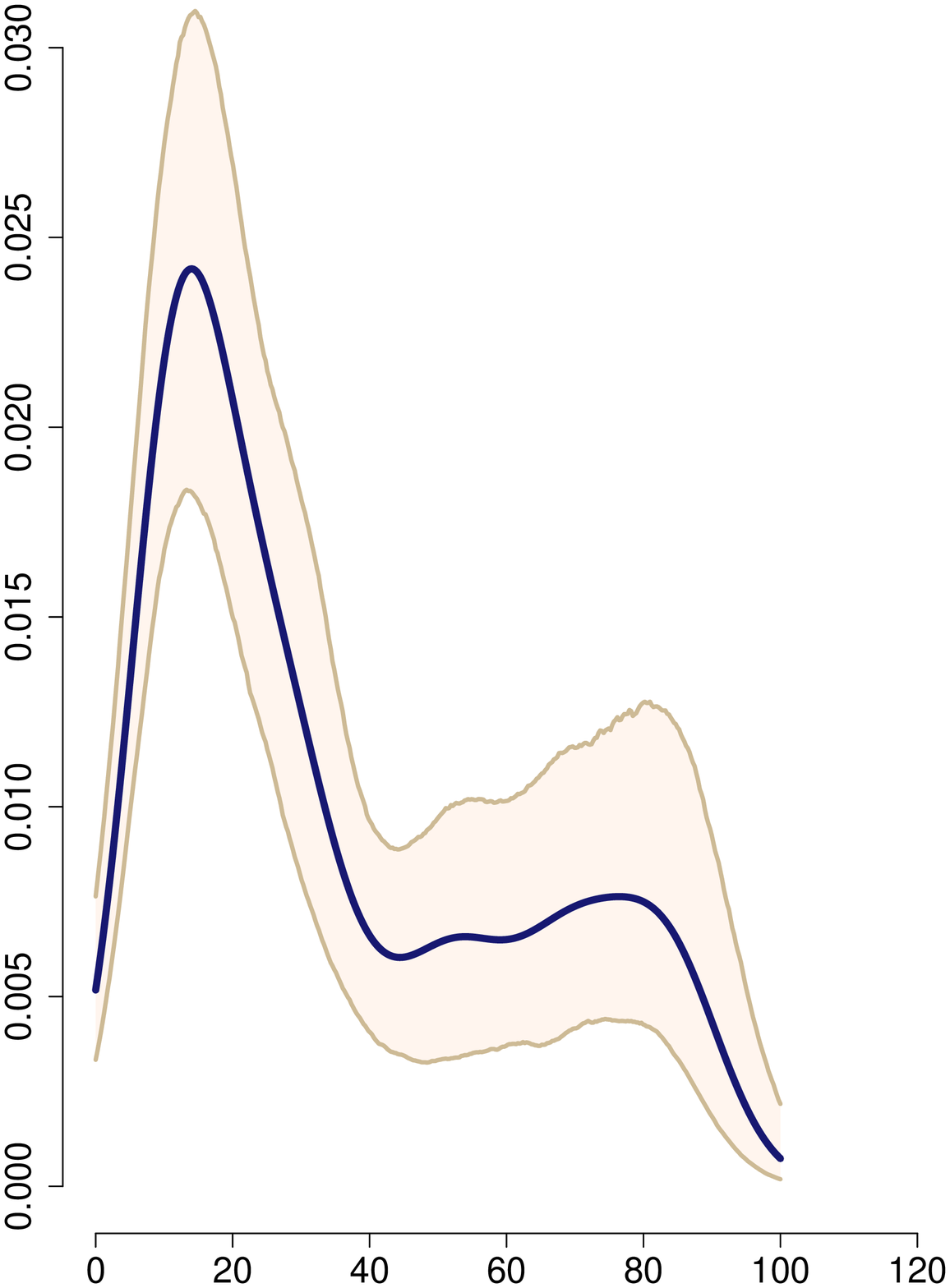}\\
        \includegraphics[width=0.32\textwidth,height=0.3\textwidth]{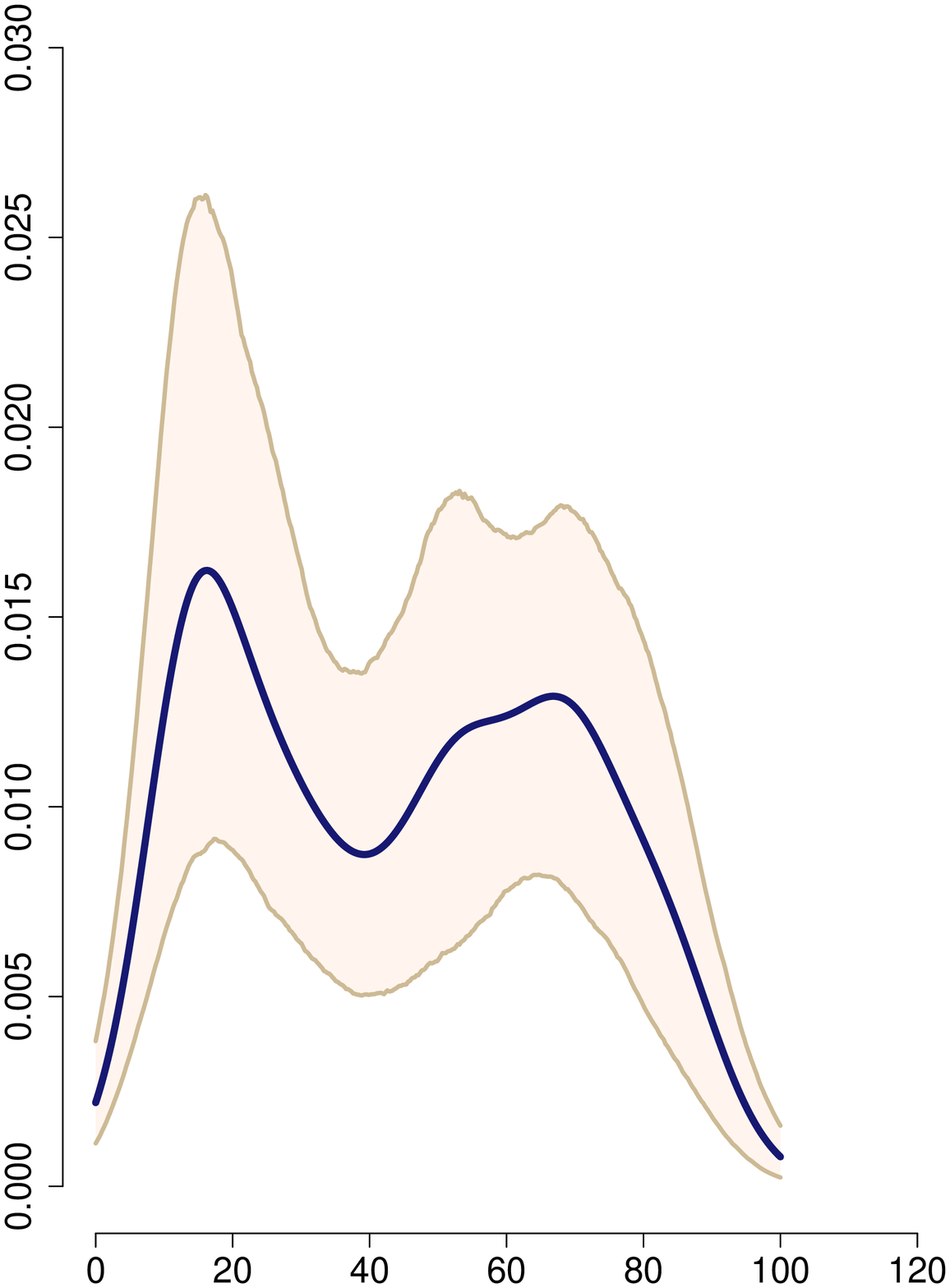}
       \includegraphics[width=0.32\textwidth,height=0.3\textwidth]{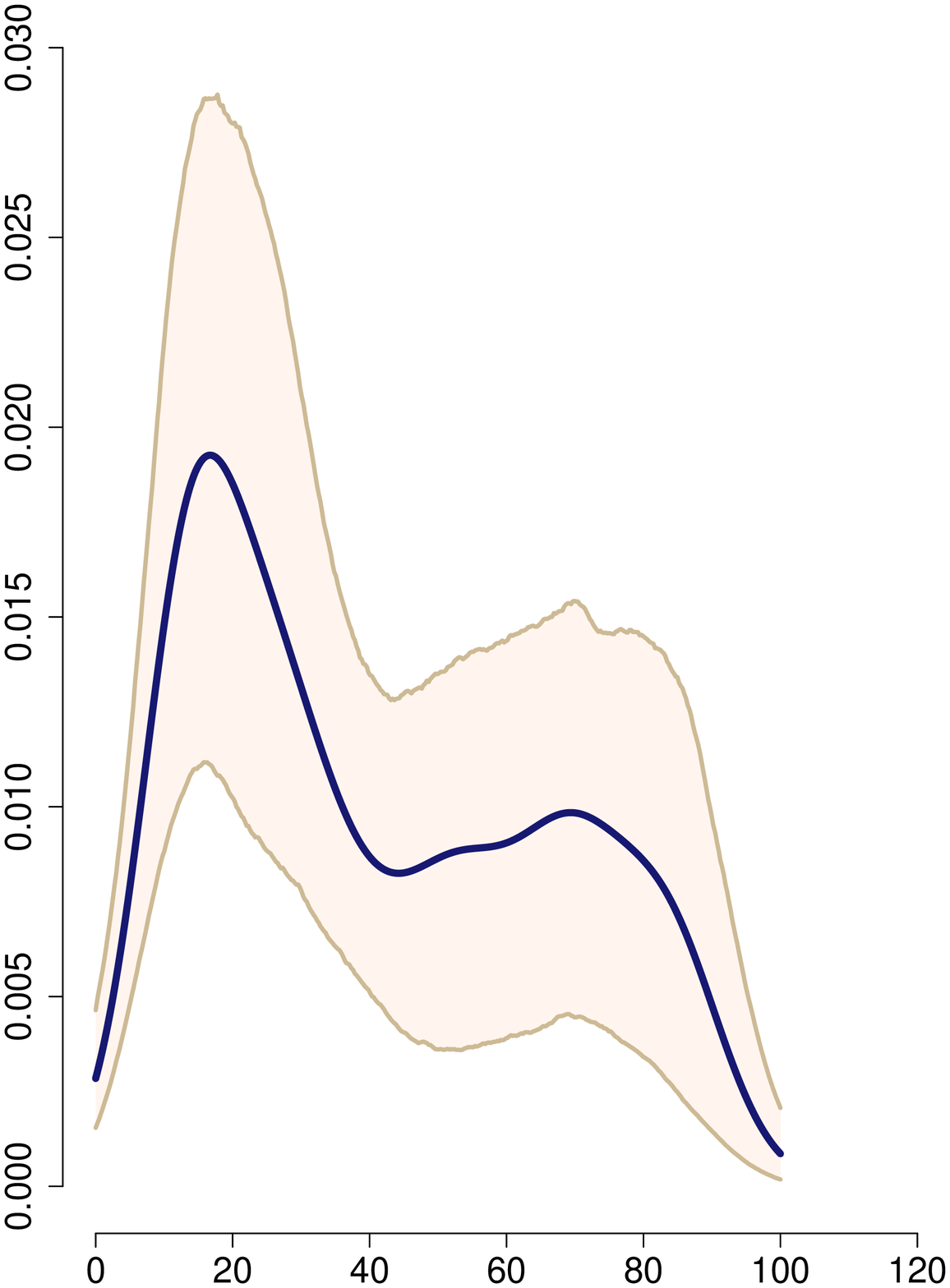}
        \includegraphics[width=0.32\textwidth,height=0.3\textwidth]{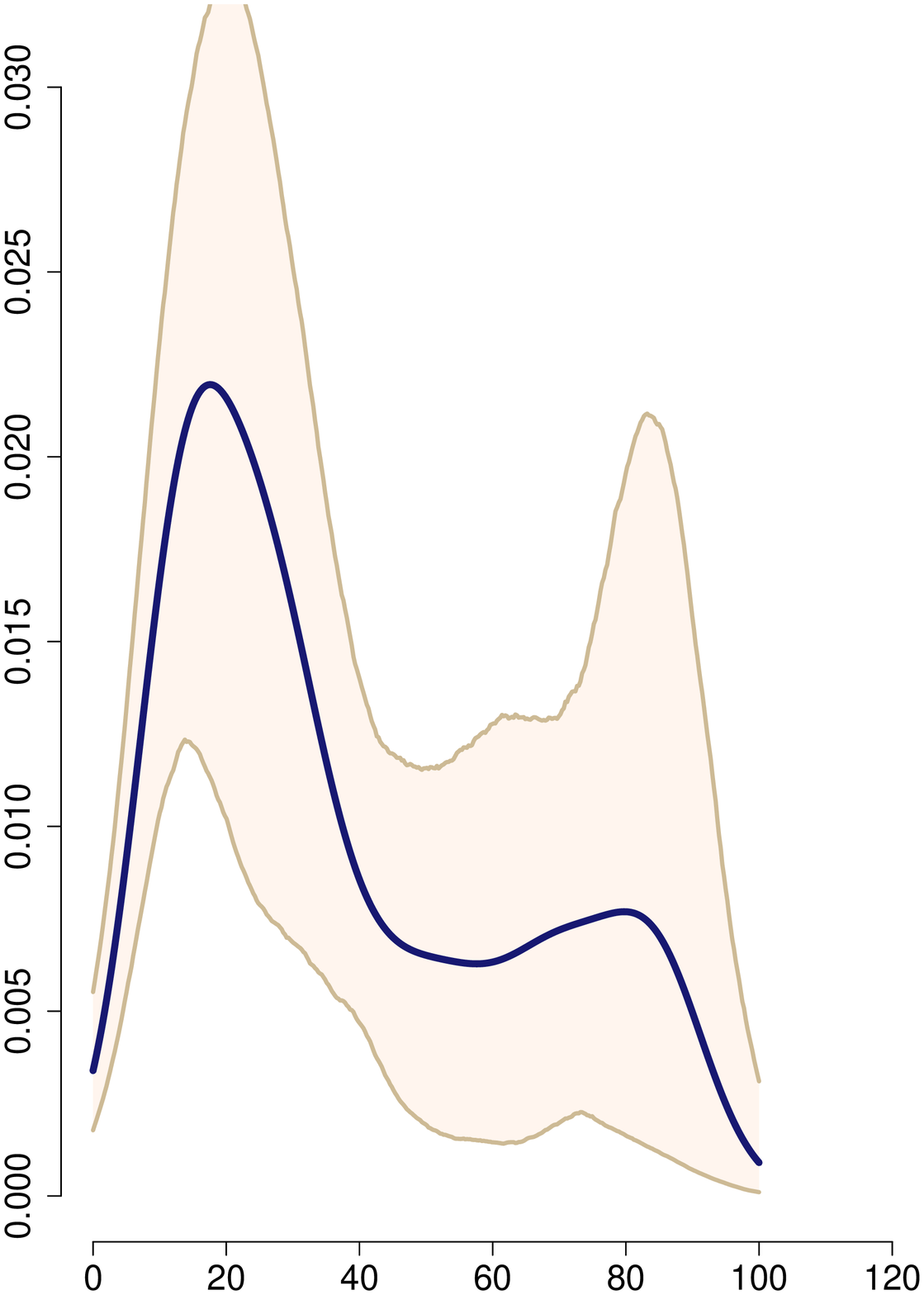}\\
        \includegraphics[width=0.32\textwidth,height=0.3\textwidth]{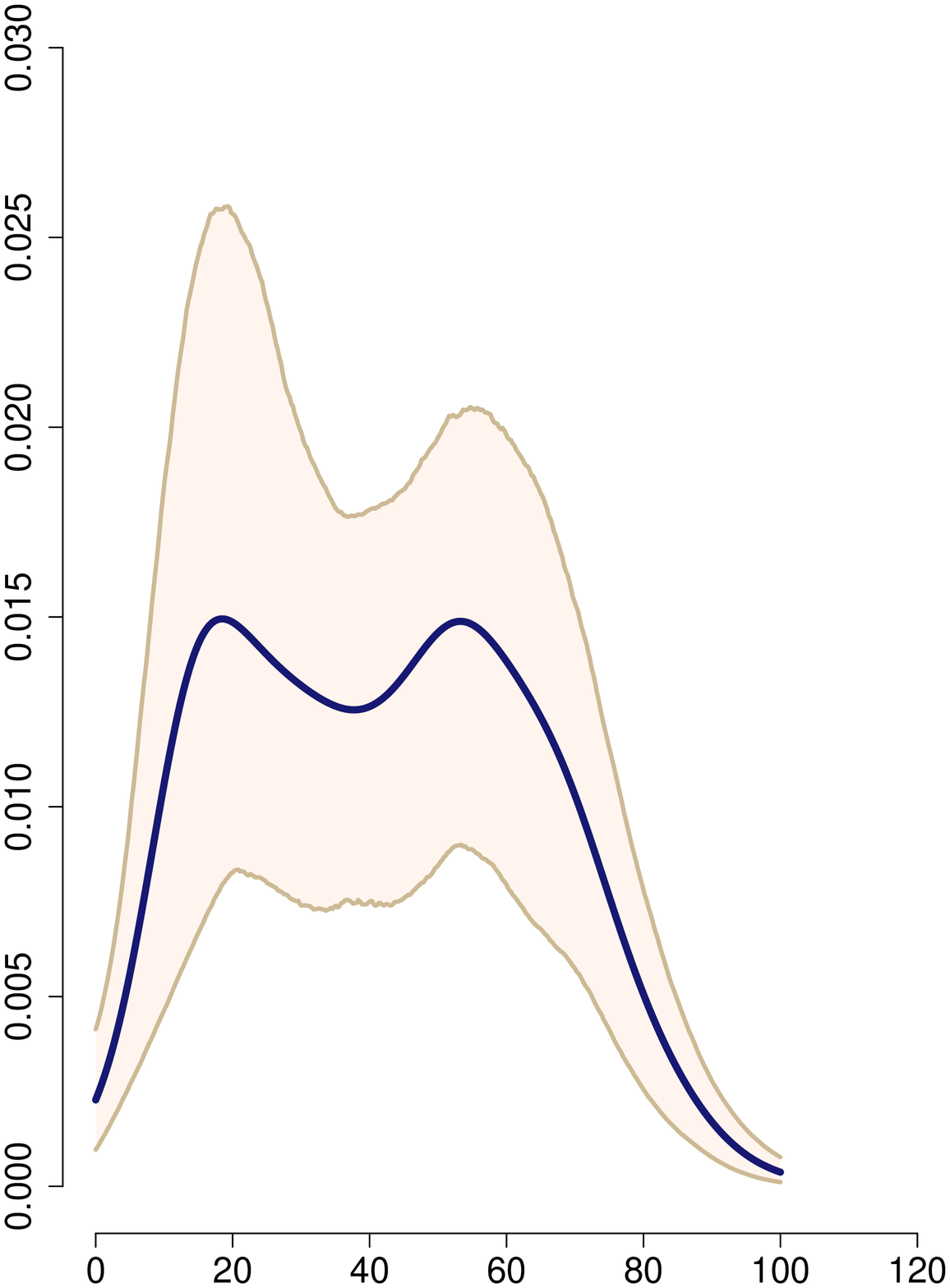}
       \includegraphics[width=0.32\textwidth,height=0.3\textwidth]{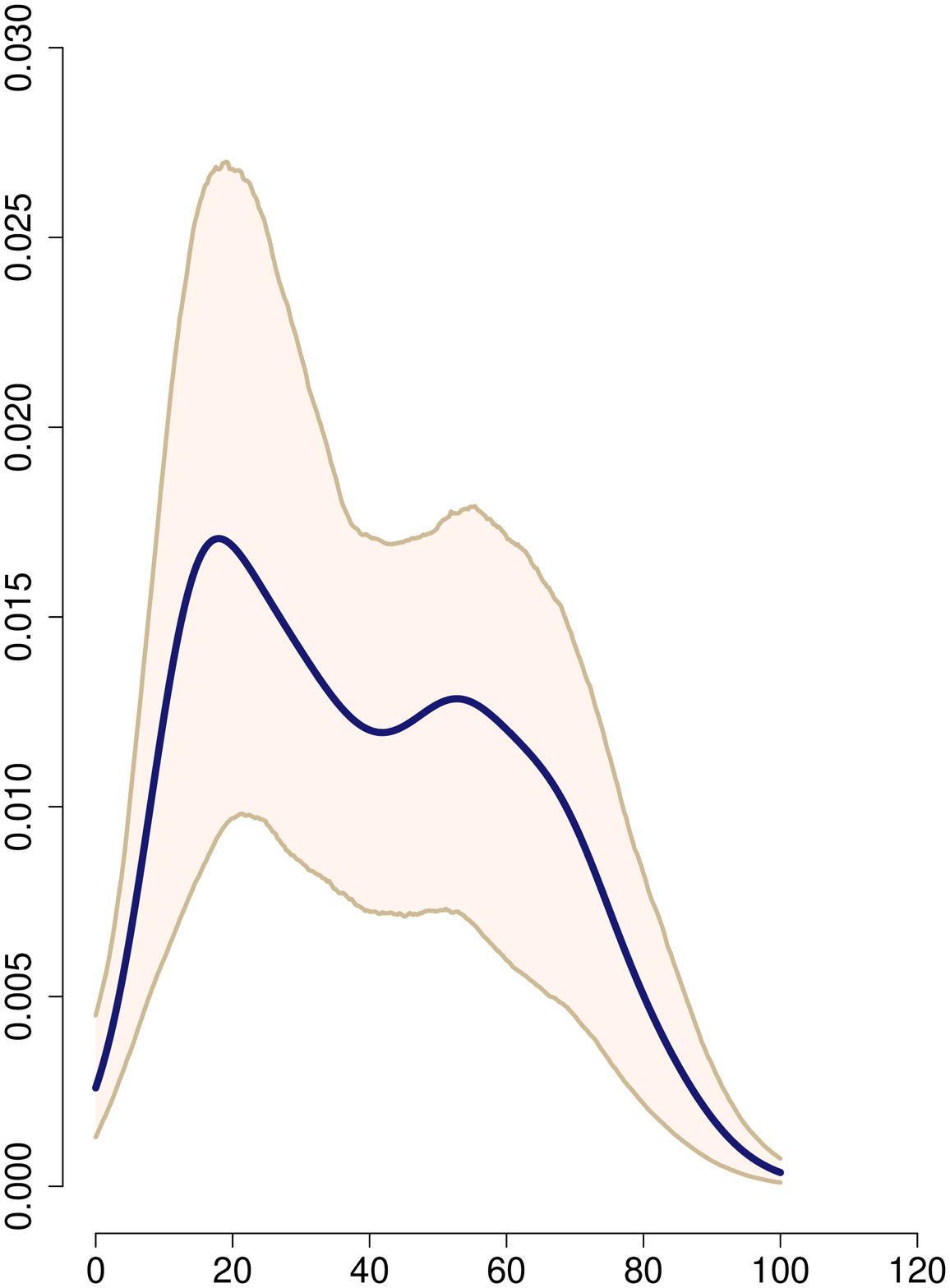}
        \includegraphics[width=0.32\textwidth,height=0.3\textwidth]{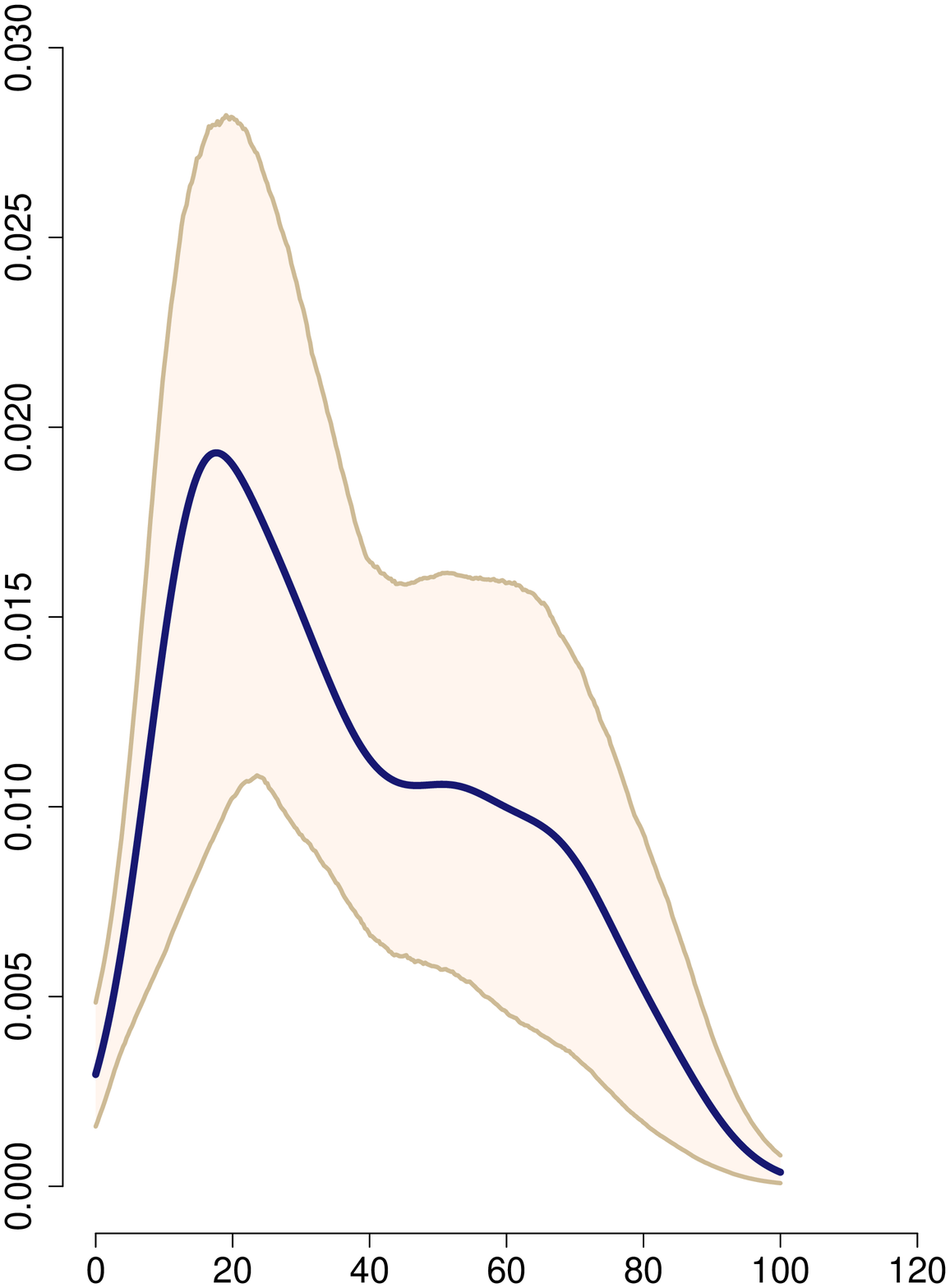}\\
        \includegraphics[width=0.32\textwidth,height=0.3\textwidth]{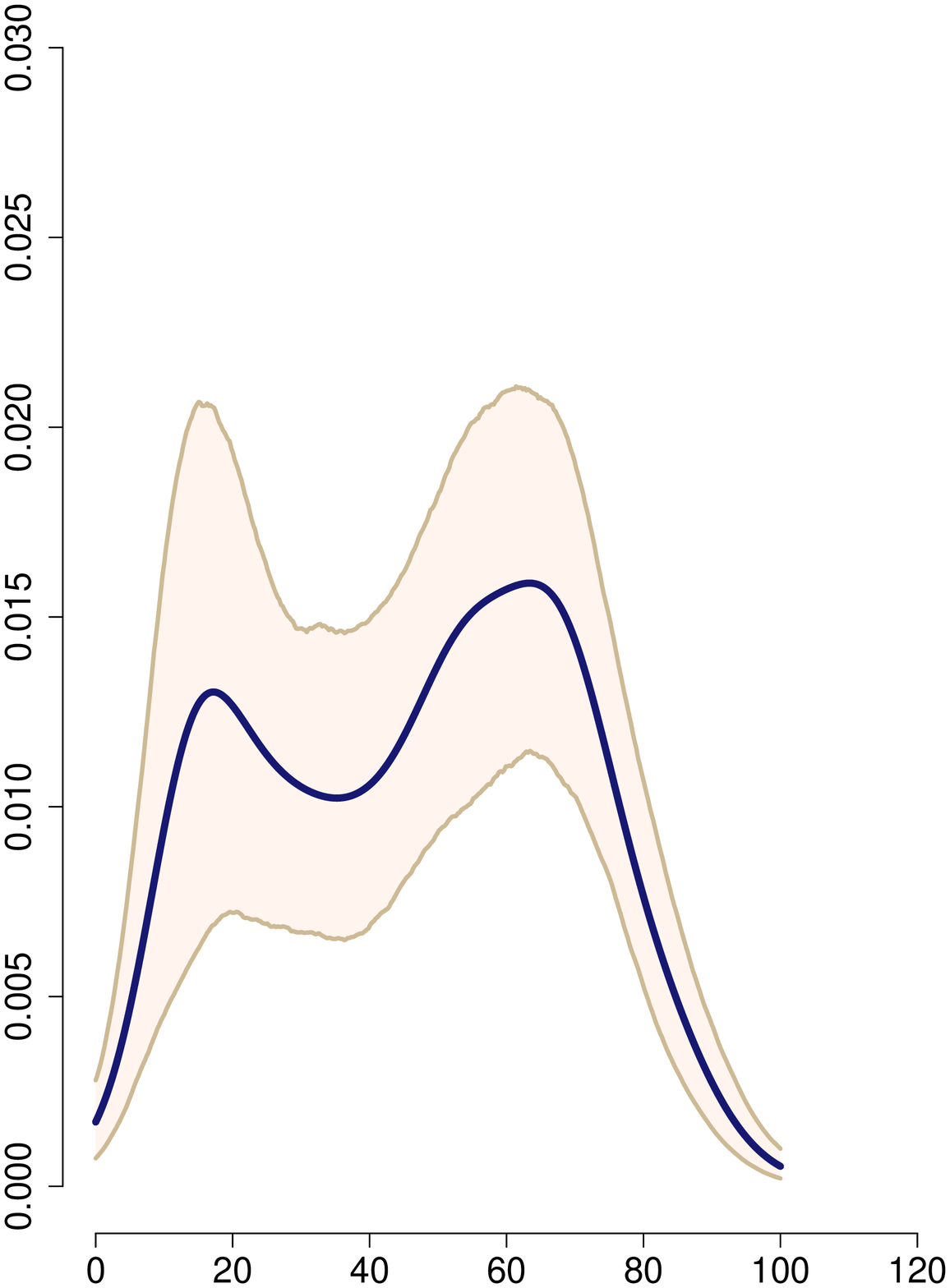}
       \includegraphics[width=0.32\textwidth,height=0.3\textwidth]{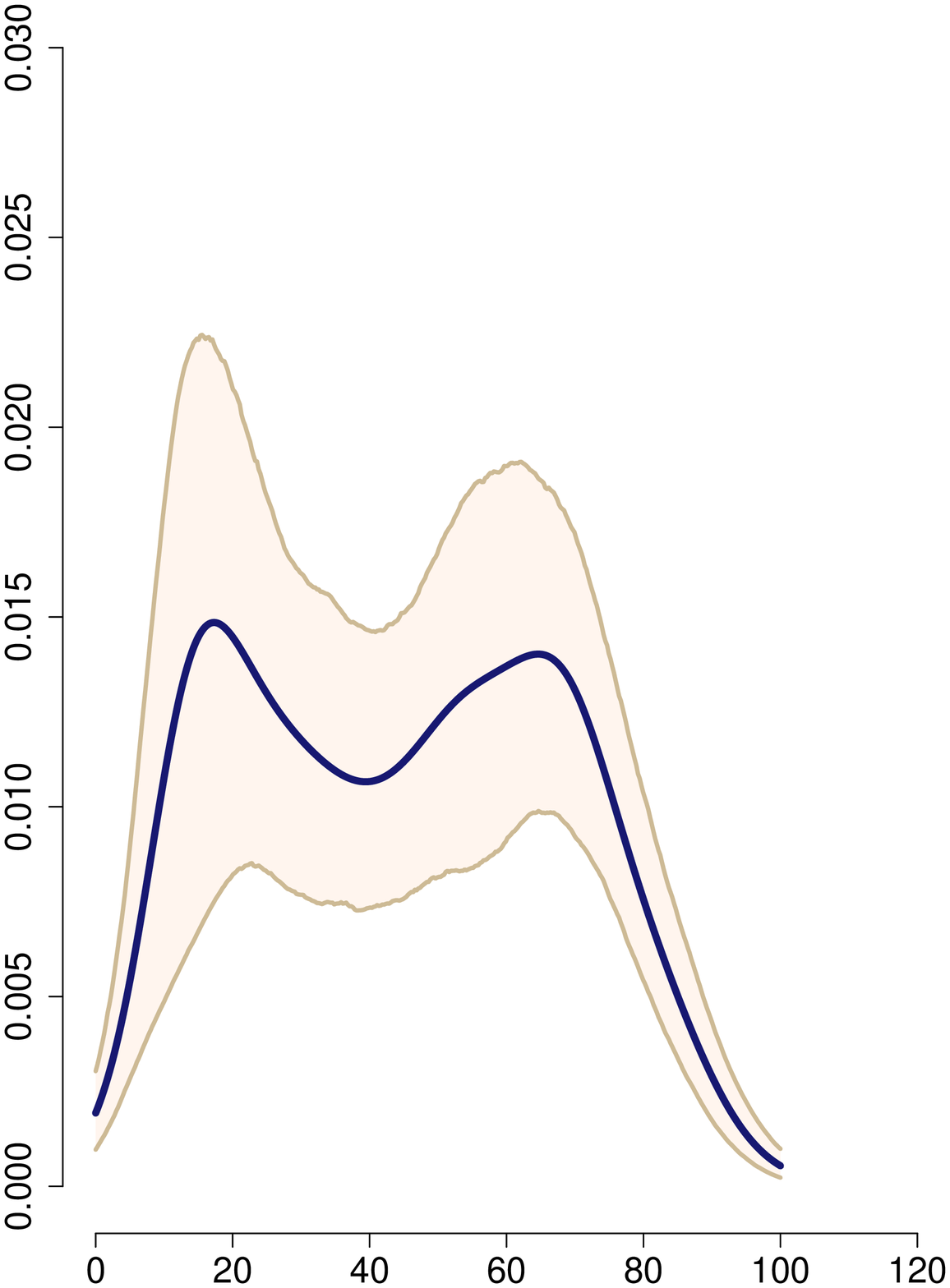}
        \includegraphics[width=0.32\textwidth,height=0.3\textwidth]{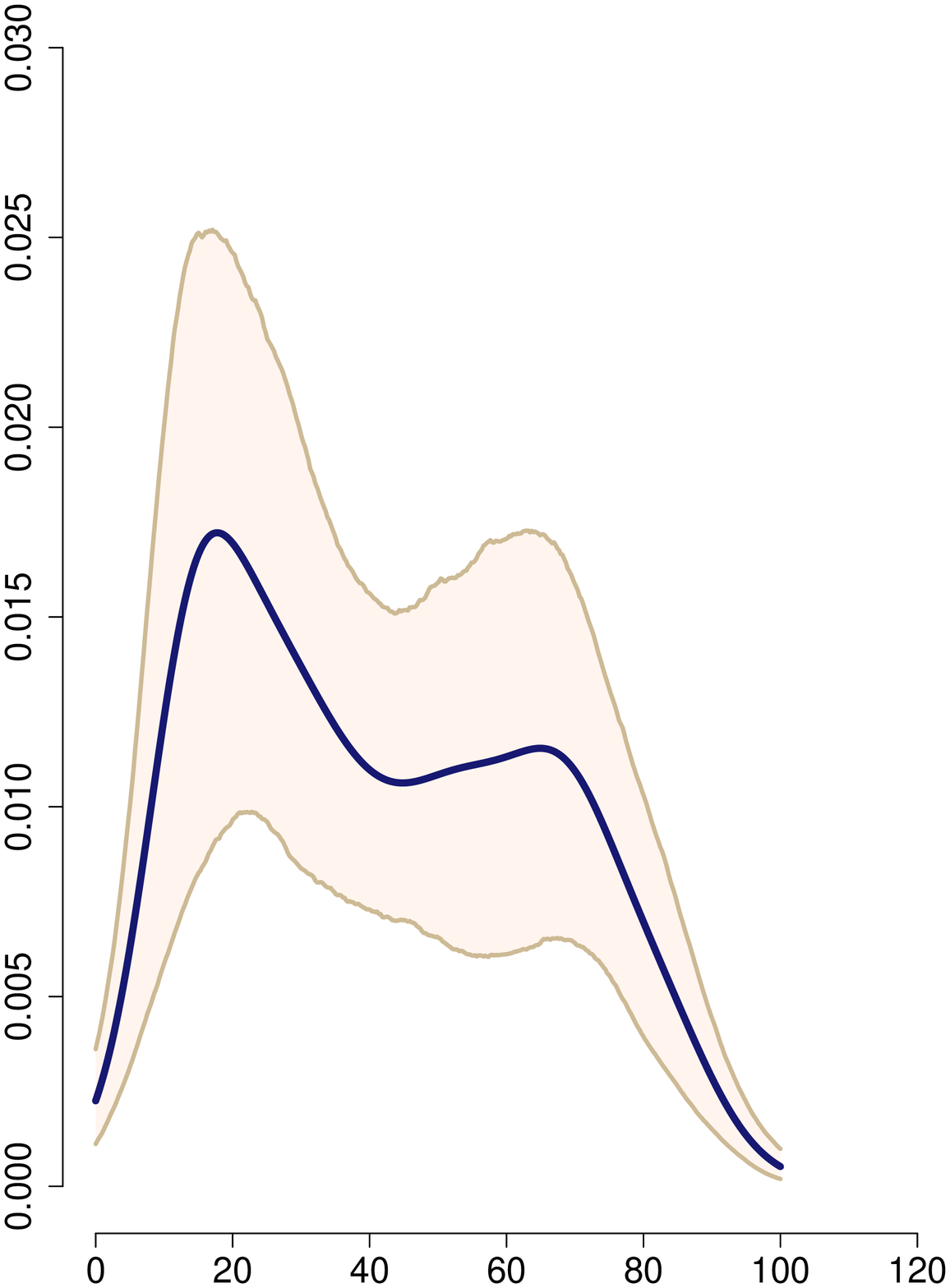}\\
\end{center}
\caption{Predictive distribution corresponding to the 12 different reference values
of the covariates. The simulation truth can be found in Figure~1 of
\cite{muller2011product}. }
\label{fig:pred_comp}
\end{figure}

\begin{figure}[h!]
  \begin{center}
    \includegraphics[width=0.55\textwidth, height=0.64\textwidth]{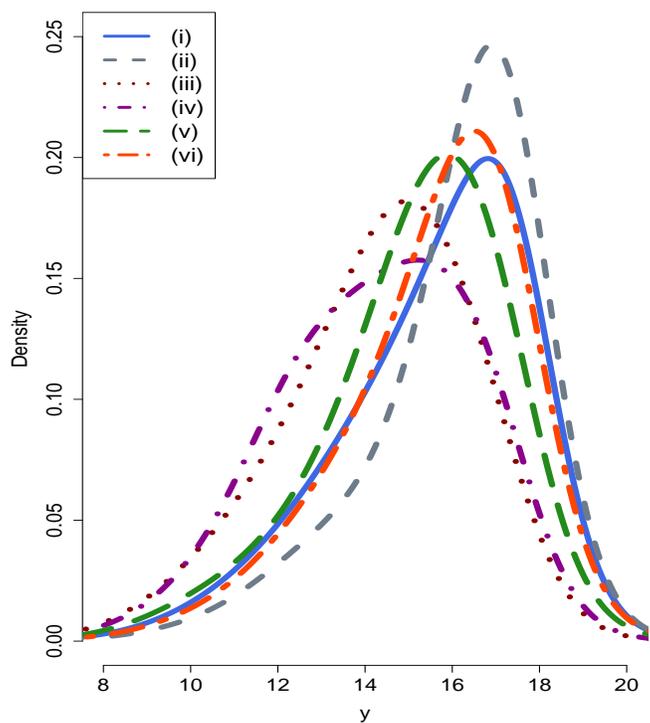}
\end{center}
 \caption{Predictive distribution for cases $(i)-(vi)$ under Test E in
Table~\ref{tab:par_biopics} for the Biopics dataset. }
 \label{suppfig:pred_bio}
 \end{figure}

\begin{figure}[h!]
  \begin{center}
      \includegraphics[width=0.32\textwidth]{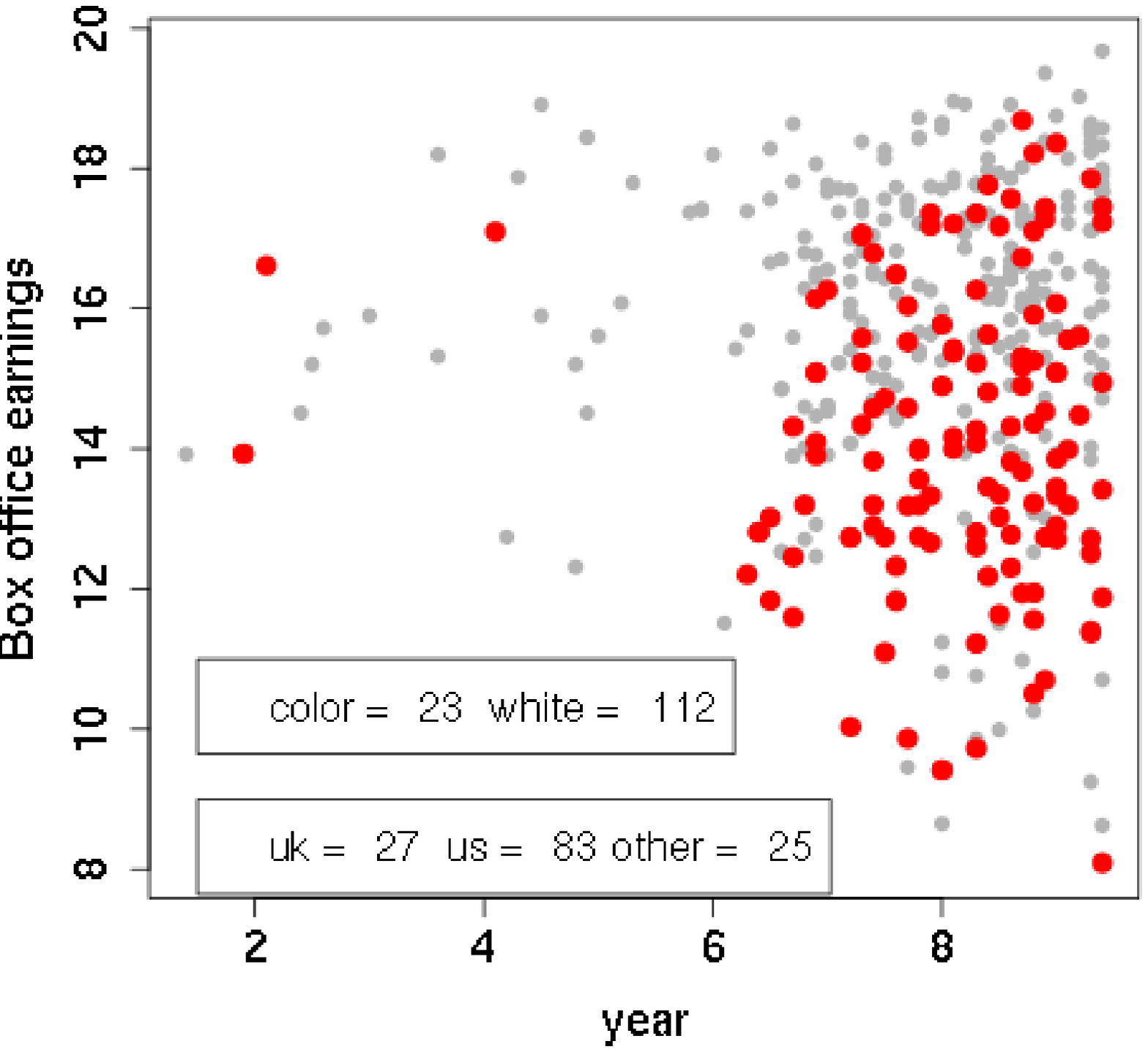}
      \includegraphics[width=0.32\textwidth]{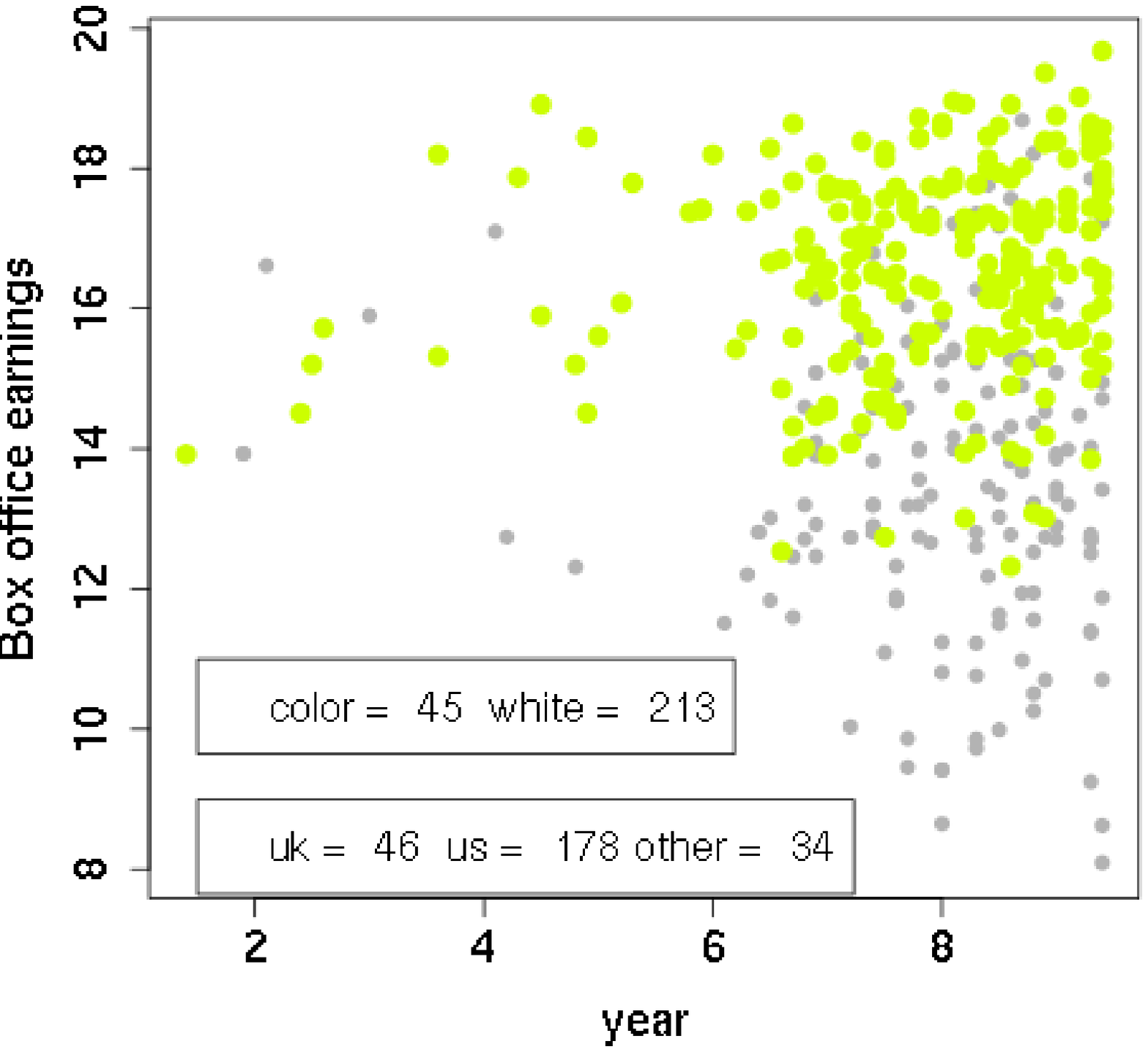}
      \includegraphics[width=0.32\textwidth]{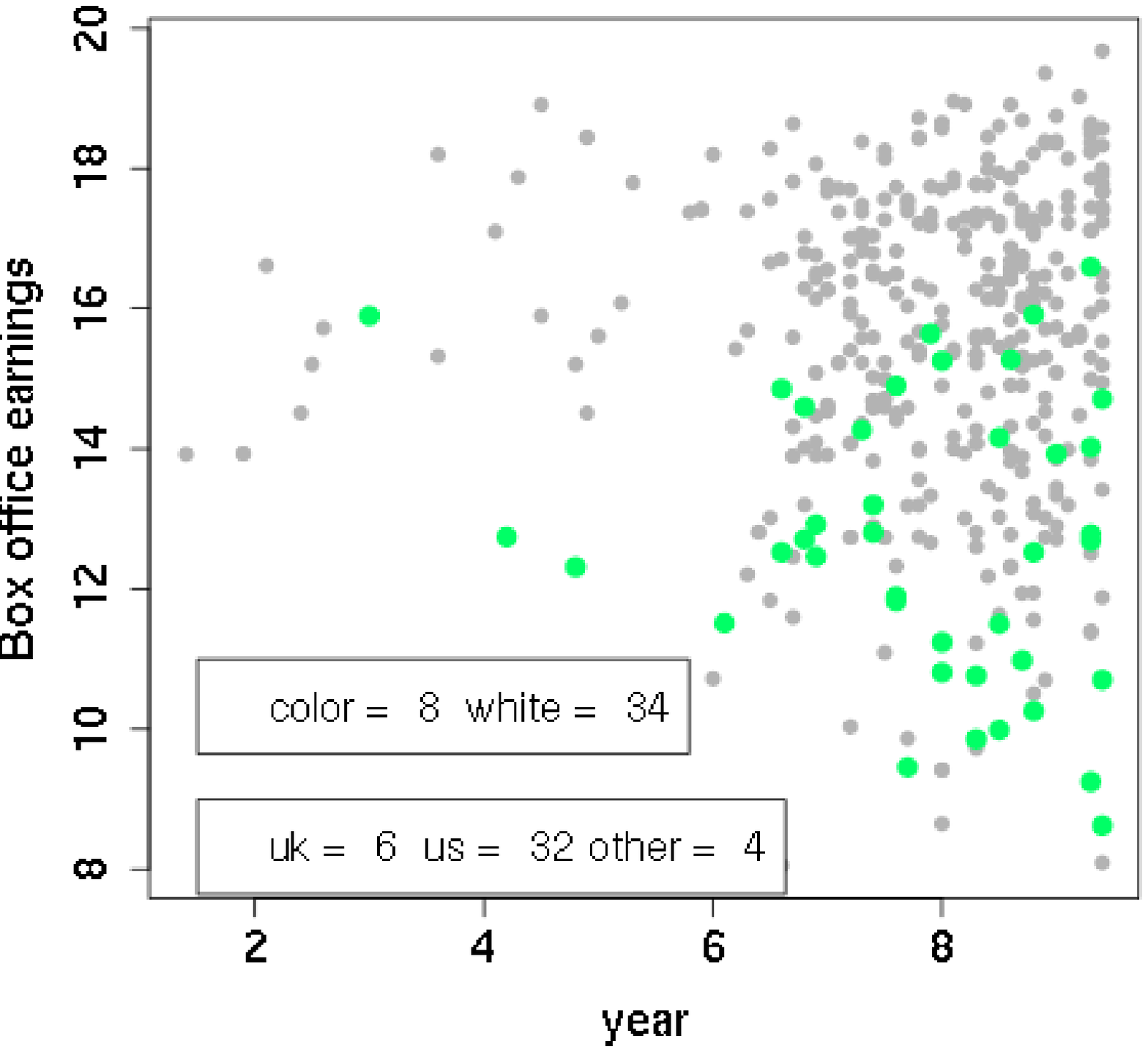}
\end{center}
\caption{Cluster estimate obtained under a linear dependent Dirichlet process model with
prior specification $G$ in Table \ref{tab:biopics_lddp} of the paper. }
 \label{fig:clu_est_bio_lddp}
 \end{figure}

\begin{itemize}
 \item Figure~\ref{fig:spectral_density} displays the graph of the
     power exponential spectral density $\varphi(x; \rho, \nu)$ in
     \eqref{eq:spectral_dens} when $\rho$ is  2 (left) and 100 (right)
     and $\nu$ varies.
 \item Figure~\ref{fig:C0} displays the value of $C_0(t)$
     corresponding to the Gaussian spectral density where $s=0.5$ and
     $\rho$ varies as in the legend. The vertical dashed line
     represents the upper limit of the set $S=\left(-\dfrac{1}{2},
     \dfrac{1}{2} \right)$. The approximation $C_0(t)\simeq 0$ for $t
     \notin S$ is perfectly adhered to when $\rho$ is small. The
     higher the $\rho$ is, the slower is the decay rate of the
     function $C_0(t)$.
\item Figure~\ref{fig:pred_comp} refers to
    Section~\ref{subsec:cov_simulated} of the paper, where we
    considered a simulated dataset with three covariates. Predictive
    distributions corresponding to the 12 different reference values
    of the covariates are shown. The simulation truth can be found in
    Figure~1 of \cite{muller2011product}.
\item Figure~\ref{suppfig:pred_bio} shows the predictive distribution
    for the response ``box-office earnings'' in the Biopics dataset
    application for cases $(i)-(vi)$ with parameter setting E in
    Table~\ref{tab:par_biopics}.
 \item Figure~\ref{fig:clu_est_bio_lddp} diplays the cluster estimate
     for the Biopics dataset obtained under a linear dependent
     Dirichlet process model with prior specification $G$ in Table
     \ref{tab:biopics_lddp} of the paper.
\end{itemize}



\end{document}